\documentclass[12pt,twoside]{article}
\usepackage[T1]{fontenc}
\usepackage[backend=bibtex,style=numeric-comp,autocite=plain,sorting=none,%
giveninits=true]{biblatex}
\usepackage{amsmath,amsfonts,amssymb,amsthm,mathabx}
\usepackage{times}
\usepackage{url}
\usepackage{tensor}
\usepackage{color}
\usepackage{caption}
\usepackage{enumitem}
\usepackage{graphicx}
\usepackage{array}
\usepackage{amsbsy}
\usepackage{appendix}
\usepackage{physics}
\usepackage{cancel}
\usepackage{yfonts}
\usepackage{chngcntr}
\usepackage{xcolor}
\usepackage{soul}
\usepackage{cancel}
\usepackage{setspace} 

\usepackage{hyperref}
\advance\textheight\topskip%
\numberwithin{equation}{section} \makeatletter
\@addtoreset{equation}{section}

\renewcommand{\cosh}{\operatorname{ch}}
\renewcommand{\sinh}{\operatorname{sh}}

 \def\scri{{\mathcal{J}}}%
 \def\scrip{\scri^{+}}%

\def\pd{{\partial}}

\newcommand\Yjms[2]{{{}_{#1}{Y}_{#2}}}
\newcommand\barYjms[2]{{{}_{#1}{\widebar{Y}}_{#2}}}
\newcommand\ThreeJ[6]{\begin{pmatrix}
		#1 & #3 & #5 \\
		#2 & #4 & #6
\end{pmatrix}}

\begin{document}

\title{Memory of Robinson-Trautman waves}

\author{Glenn Barnich, Ali Seraj}

%\date{}

\def\mytitle{Memory of Robinson-Trautman waves}

\pagestyle{myheadings} \markboth{\textsc{\small G.~Barnich, A.~Seraj}}
{\textsc{\small RT memory}}

\addtolength{\headsep}{4pt}

\begin{centering}

  \vspace{1cm}

  \textbf{\Large{\mytitle}}

    \vspace{1.5cm}

    {\large Glenn Barnich}

\vspace{.5cm}

\begin{minipage}{.9\textwidth}\small \it \begin{center}
    Physique Th\'eorique et Math\'ematique \\ Universit\'e libre de
    Bruxelles and International Solvay Institutes\\ Campus Plaine C.P. 231,
    B-1050 Bruxelles, Belgium \\
    E-mail: \href{mailto:Glenn.Barnich@ulb.be}{Glenn.Barnich@ulb.be}
    \end{center}
  \end{minipage}

\vspace{1cm}  

  {\large Ali Seraj}

  \vspace{.5cm}

\begin{minipage}{.9\textwidth}\small \it \begin{center}
    School of Quantum Physics and Matter\\ Institute for Research in
    Fundamental Sciences (IPM)
    \\ P.O.~Box 19395--5531, Tehran, Iran\\
    E-mail: \href{mailto:ali_seraj@ipm.ir}{ali\_seraj@ipm.ir}
    \end{center}
\end{minipage}

\end{centering}

\vspace{1cm}
  
\begin{center}
  \begin{minipage}{.9\textwidth} \textsc{Abstract}. The memory effect for
    Robinson-Trautman waves is explicitly worked out. In a first step, we
    construct the combined frame rotation and coordinate transformation in which
    Robinson-Trautman waves are manifestly locally asymptotically flat at future
    null infinity. This allows us to apply well-established results on how to
    derive the memory effect in this context. In a second step, we construct a
    suitably improved generalized mass aspect that provides a local Lyapunov
    function for the flow in the sense that it is manifestly positive. News-free
    solutions are studied in detail and shown to coincide with the vacuum sector
    of Euclidean Liouville theory. They correspond to boosted Schwarzschild
    black holes. As a by-product, we show that the displacement and non-linear
    memory effects in locally asymptotically flat spacetimes at future null
    infinity are invariant under supertranslations and covariant under
    $\mathrm{BMS}_4$ Lorentz transformations. We provide a novel interpretation
    of the gauge fixed modified flow that controls the low harmonics in terms of
    the instantaneous rest frame of the system.
 \end{minipage}
\end{center}

\vfill
\thispagestyle{empty}

\newpage

\begin{spacing}{0.85}
    \tableofcontents
\end{spacing}

\newpage

\section{Introduction}\label{sec:introduction}

In much the same spirit than three-dimensional or self-dual gravity are
interesting and tractable toy models for four-dimensional general relativity,
Robinson-Trautman (RT) waves~\cite{Robinson:1960zzb,Robinson:1962zz} constitute an analytically
controllable sub-sector of the full theory. They represent an isolated radiating
system that settles down to boosted Schwarzschild black holes by emission of
gravitational radiation. One should thus be able to explicitly compute the
displacement and non-linear memory effects~\cite{Zeldovich1974,Christodoulou:1991cr} (see
also~\cite{Blanchet:1990twn,Wiseman:1991ss,Thorne:1992sdb,Frauendiener1992,%
  Blanchet:1992br,Bieri:2013ada}) associated to these waves. The hope is
that the detailed solution to this theoretical exercise and the clarification of
the underlying conceptual issues may lead to valuable lessons for gravitational
wave emission in more realistic astrophysical processes.

We will focus on ``physical'' RT waves of type $\mathrm{II}$ with a strictly
positive constant mass parameter and spherical topology. In this context, one
considers ``half of a characteristic initial value problem'': given initial
conditions on a null hypersurface for the induced metric on a closed space-like
2-surface with the topology of a sphere, the problem consists in studying in
detail how the associated 4-dimensional solution to Einstein's equations settles
down to Schwarzschild black holes by emission of gravitational radiation
(\cite{robinson1964exact}, pp.~113--114). In conformally flat complex coordinates for the induced
metric, the dynamical problem reduces to the RT equation, a single fourth order
parabolic equation for the conformal factor.

The problem is tractable because RT spacetimes are algebraically special
solutions to the vacuum Einstein equations and admit a null geodesic congruence
that is shear-free, in line with the Goldberg-Sachs theorem~\cite{goldberg1962theorem}. The null
generator is in addition hypersurface-orthogonal and a gradient. In the
Newman-Penrose (NP) formalism~\cite{Newman:1961qr} with a suitable choice of frame and
coordinates, in which the outgoing vector $l$ is the repeated null direction,
the RT equation appears as the only non-trivial evolution equation for a
projected component of the Weyl tensor~\cite{Newman1962b}.

Whereas the late-time behavior involves algebraically special quasi-normal
modes\cite{Foster:1967,Couch:1973zc,Chandrasekhar:1984mgh,Qi:1993ey}, the non-linear convergence towards Schwarzschild
has been extensively studied in the late eighties and the beginning of the
nineties~\cite{lukacs1984lyapunov,Schmidt1988a,rendall1988existence,%
  RENDALL1989,Singleton_1990,Singleton2020,Chrusciel:1992cj}. The RT equation has been
interpreted~\cite{Tod1989} as a two-dimensional Calabi flow~\cite{Calabi1982} and important global
aspects have been clarified~\cite{Tod1989,Chrusciel:1992rv,Chrusciel:1992tj,mason1998asymptotic}. In the physics literature,
intriguing connections to integrability have been made for instance
in~\cite{Bakas:2005kv,Bakas:2005pc}. Relevant work for our purpose includes a recent analysis on
second order algebraically special perturbations of the Schwarzschild black
hole~\cite{BenAchour:2024skv}.

More recently, RT waves have been extensively studied from the viewpoint of the
AdS/CFT and the fluid/gravity correspondence~\cite{BernardideFreitas:2014eoi,Bakas:2014kfa,Skenderis:2017dnh,Ciambelli:2017wou} in
connection with non-equilibrium dynamics. As has been emphasized in these
references, RT waves are agnostic about the cosmological
constant~\cite{Bakas:2014kfa} in the sense that the effective dynamics, the
Calabi flow on the two-dimensional spatial surface, is the same for all values
of the cosmological constant $\Lambda$. Further recent work in this context
includes~\cite{Adami:2024mtu,Arenas-Henriquez:2025rpt,Poole:2025cmv}.

In the current work, we re-analyze these spacetimes using the tools developed in
the context of asymptotically flat spacetimes at future null infinity because a
standard derivation~\cite{Frauendiener1992} of the non-linear memory effect is available.

How to fit RT waves in the context of asymptotically flat spacetimes in the
sense of the Bondi-Sachs framework has been originally investigated by Hogan and
Trautman~\cite{hogan1987gravitational}, while the relation to Newman-Unti spacetimes has been clarified
in~\cite{Schmidt1988}. That there is no contradiction in having no shear and no news in the
algebraically special frame, while definite expressions may be computed in the
asympotically flat frame, for instance in terms of Penrose's news
function~\cite{Penrose:1986}, is explained in Singleton's thesis (\cite{Singleton2020}, pp.~79--80). In his
analysis, he states ``{\em Recently Hogan and Trautman (1987) and Hogan (1985)
  have attempted to generalize Bondi coordinates to explicitly include RT
  coordinates as a subcase. As pointed out by Schmidt (1988b) however, such
  coordinates are well known and due to Newman and Unti (1962). Their
  construction is also somewhat futile since they cannot cater for both the
  universal structure of J+ and the internal algebraically special nature of the
  spacetime — the coordinate freedom group (the Newman-Unti group) becomes much
  too large.}''

Two aspects of the problem, which warrant the current investigation, have
changed more recently: (i)~when suitably interpretated, the residual gauge
freedom in the Newman-Unti set-up reduces to the direct product of the BMS and
the (complex) Weyl group~\cite{Barnich:2009se,Barnich:2010eb,Barnich:2016lyg}, and (ii)~new connections between
gravitational memory effects, BMS symmetries and soft theorems have been
found~\cite{Strominger:2013jfa,StromingerZhiboedov2016,Pasterski:2015tva}, see also 
e.g.~\cite{Hollands:2016oma,Nichols:2017rqr,Mao_2019,Mitman:2020pbt,Seraj2023,%
  He:2023qha,Mitman:2024uss,} for further related recent
developments.

What seems to be novel in our analysis is
\begin{enumerate}[label= (\roman*)]
  \item a proof of supertranslation invariance as well as Lorentz covariance of the
        displacement and the non-linear memory effects in locally asymptotically
        flat spacetimes at future null infinity, complementary to the
        interpretation of~\cite{StromingerZhiboedov2016,Pasterski:2015tva},
  \item a systematic analysis of both the change of frame and the change of
        coordinates that provides all the asymptotically flat data for RT waves
        (see also~\cite{Mao_2019}),
  \item the manifestly positive local Lyapunov function for the RT flow based on an
        improved generalized mass aspect,
  \item the expressions for the displacement and the non-linear memory in
  natural RT coordinates (see also~\cite{Mao_2019}),
  
  \item a comprehensive study of the vacuum sector,

  \item the control of low harmonics through the instantaneous rest frame of the
        system,

  \item the explicit computation of the displacement and the non-linear memory
        effects to second order in perturbation theory.
\end{enumerate}

The paper is structured as follows. In Section~\ref{sec:rt-solutions}, we recall the description of
RT waves in the context of the NP formalism (see~\cite{Newman1962b,Foster:1967} and~\cite{newman:1980xx,newman_spin-coefficient_2009}
for reviews) and introduce notations and conventions. Volume preserving and
antipodally symmetric RT flows are briefly reviewed in Section~\ref{sec:gauged-fixed-flow}.

In Appendix~\ref{sec:locally-asympt-flat}, we briefly recall how Newman-Unti spacetimes encompass both RT
waves and locally asymptotically flat spacetimes at null infinity. Section~\ref{sec:non-linear-memory}
is devoted to reviewing Christodoulou's non linear memory effect~\cite{Christodoulou:1991cr} as
translated to locally asymptotically flat spacetimes at future null infinity by
Frauendiener~\cite{Frauendiener1992}. We show that in this context, both the displacement and the
non-linear memory effects are supertranslation invariant, and covariant under
Lorentz transformations. We also discuss the instantaneous rest frame of the
system.

In Section~\ref{sec:bms-data-rt} and Appendix~\ref{sec:asympt-frame-rott}, we use the BMS-Weyl
universal structure to systematically work out the asymptotically flat data for
RT waves, in the sense that we provide, in the asymptotically flat rather than
the algebraically special frame, expressions for the shear, the news,
generalized mass and angular momentum aspects as well as outgoing and incoming
radiation, as initiated in~\cite{Mao_2019}. We thus recover earlier results of~\cite{Singleton2020} for
the news, of~\cite{Adamo:2009vu} for the shear and incoming radiation from the class
$\mathrm{II}$ frame rotation, and of~\cite{Tafel2000} for the generalized mass aspect in
the Bondi-Sachs framework.

This allows us in Section~\ref{sec:memory-effect} to re-write the Robinson-Trautman equation in
terms of a mass-loss formula for the generalized mass aspect in natural
Robinson-Trautman coordinates. An additional step consists in applying a
suitable supertranslation, as proposed by Moreschi~\cite{Moreschi1986,Moreschi:1988pc}, that turns the
generalized mass aspect into a manifestly positive function that can never
increase. In other words, it provides a local Lyapunov function, rather than a
functional, for the flow. How one may work out the asymptotically flat frame,
without using Bondi coordinates in an intermediate step, is briefly sketched in
Appendix~\ref{sec:stay-natur-robins}.

In Section~\ref{sec:absence-news-1}, the vacuum sector, also called equilibrium solutions in~\cite{Singleton2020},
is analyzed in detail. Technical material on boosts, conformal transformations
of the sphere, and spin weighted spherical harmonics is provided in Appendices~\ref{sec:boosts-conf-fact}
and~\ref{sec:spin weighted harmonics}. RT vacua may be equivalently characterized by
\begin{enumerate}[label= (\roman*)]
    \item the absence of Bondi news,
  \item a conserved generalized mass aspect,
  \item the subset of solutions to the RT equation for which the quotient of the
        conformal factor to that of the round sphere is a constant linear
        combination of the first four real spherical harmonics,
  \item the homogeneous space
        $\mathbb R^*_+\times \mathrm{PSU}(2)\backslash \mathrm{PSL}(2,\mathbb C)$ of strictly
        positive real numbers times boosts,
  \item the locally asymptotically flat solutions at future null-infinity that can
        be obtained by applying boosts belonging to the $\mathrm{BMS}_4$ group
        at $\scrip$ to Schwarzschild black holes.
\end{enumerate}

\noindent We conclude the analysis of this sector by computing the conserved
$\mathrm{BMS}_4$ charges associated to the vacuum solutions using results on STF
harmonics reviewed in Appendix~\ref{app:STF Integrals}. In particular, we show that the not only the
total but also the angular momentum aspect of vacuum solutions vanishes.
      
In Section~\ref{sec:expand-around-vacu}, we show how the RT equation may be expanded around any of the
equivalent vacuum solutions, how the symmetry group acts on solutions, and how
solutions around an arbitrary vacuum may be generated by a scaling and a boost
of solutions around the round vacuum.

Section~\ref{sec:coll-coord} is devoted to how the RT flow appears when one allows for
time-dependent scalings and boosts. The dynamics of these collective coordinates
is fixed by the requirement that the system be in its instantaneous rest frame.

In Section~\ref{sec:settl-down-schw}, we recall how the algebraically special quasi-normal modes
appear in a linearized approximation~\cite{Foster:1967,Couch:1973zc,Chandrasekhar:1984mgh,Qi:1993ey}. We compute the
displacement memory to first order and show that there is no non-linear memory
to that order. We then work out the time-dependence to second order in
perturbation theory and briefly compare to the non-perturbative results of~\cite{Chrusciel:1992rv,Chrusciel:1992tj}
before computing displacement and non-linear memory to second order in
perturbation theory.

We will use mostly minus signature and $G=1=c$.

\section{RT waves in NP formalism}\label{sec:rt-solutions}

\subsection{Tetrad, metric, spin coefficients, Weyl tensor}

For RT waves, a NP tetrad and co-tetrad that is adapted to the fact that these
spacetimes are algebraically special,~i.e., such that $\Psi_0=0=\Psi_1$ in this frame,
is given in Bondi-like coordinates by
\begin{equation}
  \label{eq:9}
  \begin{split}
  l=\partial_r,\quad n=\partial_u+U\partial_r,\quad
    m=r^{-1}P\widebar\partial,\\
    l=du,\quad n=-Udu+dr,\quad m=-\frac{r}{P}d\zeta,
  \end{split}
\end{equation}
The associated line element is
\begin{equation}
  \label{eq:20}
  ds^2=2 l n-2m\widebar m=-2Udu^2+2dudr+r^2d\widebar s^2.
\end{equation}
Here, $d\widebar s^2$ is the line element on a spacelike $2$-surface
$\mathcal S$ written in conformally flat coordinates $x^A=(\zeta,\widebar\zeta)$
as
\begin{equation}
  \label{eq:14}
  d\widebar s^2=-2\frac{d\zeta d\widebar \zeta}{P^{2}}=-2e^{2\varphi}d\zeta d\widebar \zeta,
\end{equation}
while $\varphi$ is a real field of spin weight $0$ defined in terms of $P$ through
\begin{equation}
  \label{eq:4}
  P=e^{-\varphi}.
\end{equation}
If $\eta^s$ is a field of spin weight $s$, the $\eth$ operator and its complex
conjugate are defined by
\begin{equation}
  \label{eq:5}
    \eth \eta^{s}=P^{1-s}\widebar\partial ( P^s \eta^s),\quad
    \widebar\eth\eta^s=P^{1+s}\partial (P^{-s}\eta^s),
\end{equation}
with $\partial=\partial_\zeta,\widebar\partial=\partial_{\widebar\zeta}$. They raise and lower the spin weights by one
unit. The spin weight of the complex conjugate of a field is minus the spin
weight of the field. In particular,  
\begin{equation}
  \label{eq:44}
   \eth \varphi=P\widebar\partial\varphi,\quad \widebar\eth \varphi=P\partial\varphi,
\end{equation}
the scalar curvature is 
\begin{equation}
  \label{eq:15}
  R^{(2)}= 4\eth\widebar\eth \varphi=-4 P^{2}\partial\widebar\partial \ln P,
\end{equation}
the two-dimensional Laplacian is
\begin{equation}
  \label{eq:271}
  \Delta^{(2)}\eta^0=-\frac{1}{\sqrt{|g^{(2)}|}}
  \partial_A(\sqrt{|g^{(2)}|}g^{AB}\partial_B\eta^0)=2\eth\widebar\eth \eta^0=2P^2\partial\widebar\partial \eta^0,
\end{equation}
and 
\begin{equation}
  \label{eq:10}
  [\eth,\widebar\eth]\eta^s=\frac{s}{2}R^{(2)}\eta^s=2s\eth\widebar\eth\varphi\eta^s.
\end{equation}
Define
\begin{equation}
  \begin{split}
  \label{eq:13}
    & \alpha^0=-\frac 12\widebar\eth \varphi=\frac 12  P\partial \ln P,\quad\gamma^0=\frac 12\partial_{u}\varphi=-\frac 12
      \partial_u \ln P,\\ &\nu^0=2\widebar\eth \gamma^0
                                                  =\widebar\eth\partial_u\varphi,\quad
  \mu^0=\eth\widebar\eth\varphi=\frac 14 R^{(2)}=P^{2}\partial\widebar\partial\varphi,
  \end{split}
\end{equation}
where, in this context, the spin weights of $\gamma^0,\nu^0,\mu^0$ are $0,-1,0$. 
It follows that
\begin{equation}
  \label{eq:41}
  \begin{split}
  [\eth,\partial_u]\eta^s&=2(\gamma^0\eth+s\eth\gamma^0)\eta^s=(\partial_{u}\varphi\eth+s\eth\partial_{u}\varphi)\eta^{s},\\ [\widebar\eth,\partial_u]\eta^s
      &=2(\gamma^0\widebar \eth-s\widebar \eth\gamma^0)\eta^s=(\partial_{u}\varphi\widebar\eth-s\widebar\eth\partial_{u}\varphi)\eta^{s}.
  \end{split}
\end{equation}
For RT waves, the non-vanishing Weyl scalars are
\begin{equation}
  \label{eq:11}
  \Psi_2=r^{-3}\Psi^0_2,\quad \Psi_3=r^{-2}\Psi^0_3,\quad
  \Psi_4=r^{-1}\Psi^0_4-r^{-2}\widebar\eth\Psi^0_3, 
\end{equation}
with spin weights of $\Psi^0_2,\Psi^0_3,\Psi^0_4$ given by $0,-1,-2$,
and
\begin{equation}
  \label{eq:17}
  \Psi^0_2=-1,\quad \Psi^0_3=\widebar\eth\mu^0=\widebar\eth\eth\widebar\eth \varphi,\quad \Psi^0_4=\widebar\eth
  \nu^0=\widebar\eth^2\partial_u\varphi.
\end{equation}
Note that in some standard presentations of RT spacetimes, $\Psi^0_2=-m(u)$.
We have used here the available coordinate freedom to put $m(u)$ to $1$, see
e.g.~section 28.1 of~\cite{Stephani:2003tm}. There is no loss of generality as
the mass parameter will re-appear when working out the general solution to the
RT equation below. The non-vanishing spin coefficients are
\begin{equation}
  \label{eq:12}
  \begin{split}
    & \rho=-r^{-1},\\ & \alpha=r^{-1}\alpha^0=-\frac 12 r^{-1}\widebar\eth \varphi,\\ &
    \beta=-r^{-1}\widebar\alpha^0=\frac 12 r^{-1}\eth \varphi =-\widebar\alpha,\\ & \gamma=\gamma^0-\frac 12 r^{-2}
                                                             \Psi^0_2=\frac 12 \partial_{u}\varphi+\frac 12 r^{-2}=\widebar\gamma,\\ & \mu=r^{-1}\mu^0-r^{-2}\Psi^0_2
                                                                                                                  =r^{-1}\eth\widebar\eth\varphi+r^{-2}=\widebar\mu,\\
    & \nu=\nu^0-r^{-1}\Psi^0_3 =\widebar\eth\partial_{u}\varphi-r^{-1}\widebar\eth\eth\widebar\eth \varphi.
\end{split}
\end{equation}
In these terms,
\begin{equation}
  \label{eq:18}
  U=-r(\gamma^0+\widebar\gamma^0)+\mu^0-\frac 12
  r^{-1}(\Psi^0_2+\widebar \Psi^0_2)=-r\partial_u\varphi+\eth\widebar\eth\varphi+r^{-1}.
\end{equation}

\subsection{RT equation}
The only remaining equation to be satisfied in order to have a vacuum solution
to Einstein's equations is the evolution equation
\begin{equation}
  \label{eq:19} (\partial_u+3\gamma^0+3\widebar\gamma^0)\Psi^0_2=\eth\Psi^0_3,
\end{equation}
When using~\eqref{eq:17} and
$\eth\Psi^0_3={(\eth\widebar\eth)}^{2}\varphi=-P^{2}\partial \widebar \partial \big(P^{2}\partial \widebar \partial \ln P\big)$,
this equation is the RT equation
\begin{equation}
  \label{eq:36}
  \boxed{3\partial_u\varphi=-{(\eth\widebar\eth)}^2 \varphi\iff 3 \partial_u P=P^3|\partial^2 P|^2-P^4\partial^2\widebar\partial^2 P}.
\end{equation}
Two others forms of this equation turn out to be useful,
\begin{equation}
  \label{eq:270}
  \partial_u P^{-3}=\partial^2\widebar\partial^2 P-P^{-1}|\partial^2 P|^2,      
\end{equation}
and 
\begin{equation}
  \label{eq:275}
  \frac{3}{2}\partial_u P^{-2}=P \partial^2\widebar\partial^2 P - |\partial^2 P|^2.        
\end{equation}
The identification of equation~\eqref{eq:275} as a particular case of a two dimensional
Calabi flow, 
\begin{equation}
  \label{eq:125}
  \partial_u g^{(2)}_{AB}=-\frac{1}{12}(\Delta^{(2)} R^{(2)}) g^{(2)}_{AB},
\end{equation}
expressed in conformally flat coordinates is due to Tod~\cite{Tod1989}.

When the mass parameter is set to $1$, the only transformations that are left
(out of the transformations (28.9) discussed in~\cite{Stephani:2003tm}) that preserve the
line element and the solution to the RT equation are,
\begin{equation}
  \label{eq:160}
u'=u+C,\quad r'=r,\quad  \zeta'=\zeta'(\zeta),\quad \widebar\zeta'=\widebar\zeta' (\widebar\zeta),
\end{equation}
with $C$ constant, under which
\begin{equation}
  \label{eq:165}
  P'(u',x^{\prime})=P(u,x)|\frac{\partial\zeta' }{\partial\zeta}|\iff    \varphi'(u',x^\prime)=\varphi(u,x)- \ln |\frac{\partial \zeta' }{\partial\zeta}|.
\end{equation}
Indeed, when using
\begin{equation}
  \label{eq:152}
  \eth\widebar\eth \varphi=e^{-2\varphi}\partial\widebar\partial\varphi,
\end{equation}
it follows that
\begin{equation}
  \label{eq:149}
  \eth' \widebar\eth' \varphi' =\eth \widebar\eth \varphi,\quad \partial_{u'}\varphi'=\partial_{u}\varphi,
\end{equation}
so that the RT equation is equivalent to
\begin{equation}
  \label{eq:134}
  3\partial_{u'}\varphi'=- {(\eth' \widebar\eth')}^{2} \varphi'.
\end{equation}
Even though they have the same action on $\zeta,\widebar\zeta$, the
transformations~\eqref{eq:160} are not superrotations because there is no
associated change of the coordinates $u,r$.

\subsection{Spherical topology and background structure} When
$\widebar ds^2_{\mathfrak k}$ is the metric of the round sphere with unit radius
(${\mathfrak k}=1$), the complex plane (${\mathfrak k}=0$), or the open unit
disk (${\mathfrak k}=-1$),
\begin{equation}
  \label{eq:16}
  P_{\mathfrak k}=\frac{{\mathfrak k} |\zeta|^{2}+1}{\sqrt 2}\iff \varphi_{\mathfrak k}=
  \frac 12\ln 2-\ln ({\mathfrak k} |\zeta|^{2}+1),
\end{equation}
it follows that
\begin{equation}
  \label{eq:21}
  R^{(2)}_{\mathfrak k}=-2{\mathfrak k},\quad \mu^0_{\mathfrak k}=-\frac{{\mathfrak k}}{2}
  =\eth_{\mathfrak k}\widebar\eth_{\mathfrak k}\varphi_{\mathfrak k},\quad \gamma^0_{\mathfrak k}=0=\nu^0_{\mathfrak k},\quad
  \alpha^0_{\mathfrak k}=\frac{{\mathfrak k} \widebar\zeta}{\sqrt 2},
\end{equation}
while the only non-vanishing Weyl scalar is $\Psi_2=-r^{-3}$. For
${\mathfrak k}=1$, the solution reduces to a Schwarzschild black hole with mass
$M=1$, while for ${\mathfrak k}=0$, it is a special Kasner metric (see
e.g.~section 28.1 of~\cite{Stephani:2003tm}).

For definiteness, let us concentrate on the case of the sphere, $\mathfrak k=1$.
For later convenience and agreement with the literature, quantities that refer
to the round sphere with $\mathfrak k=1$ will be denoted in the following with a
subscript $\circ$ rather than a subscript $1$. The following relations will be used
throughout,
\begin{equation}
  \label{eq:62}
  \eth_{\circ}^{2}\varphi_{\circ}+{(\eth_{\circ}\varphi_{\circ})}^{2}=0,
  \quad \eth_{\circ}\widebar\eth_{\circ}\varphi_{\circ}=-\frac 12.
\end{equation}
One may then consider the deviation of $\varphi$ from $\varphi_{\circ}$ or the quotient of $P$
and $P_\circ=\frac{1}{\sqrt 2}(|\zeta|^2+1)$,
\begin{equation}
  \label{eq:47}
V_\circ=\frac{P}{P_\circ}=e^{-\Phi_\circ},\quad  \Phi_\circ=\varphi-\varphi_{\circ}.
\end{equation}
At this stage, our assumption is that the positive function $V_\circ>0$ and $\Phi_\circ$
may be expanded in spherical harmonics. The following relations are useful,
\begin{equation}
  \label{eq:123}
  \begin{split}
    \eth\eta^{s}&=V^{1-s}_\circ\eth_\circ (V_\circ^s\eta^s)=e^{-\Phi_\circ}[\eth_{\circ}\eta^{s}-s\eth_{\circ}\Phi_\circ\eta^{s}], \\
    \widebar \eth\eta^{s}& =V^{1+s}_\circ\widebar\eth_\circ (V_\circ^{-s}\eta^s)
                     =e^{-\Phi_\circ}[\widebar \eth_{\circ}\eta^{s}+s\widebar \eth_{\circ}\Phi_\circ\eta^{s}],\\
                     [\eth_\circ,\widebar\eth_\circ]\eta^s&=-s\eta^s,\\ \eth_\circ\widebar\eth_\circ\eta^0&=\frac 12 \Delta_\circ \eta^0,\\   \eth_\circ^2\widebar\eth_\circ^2\eta^0&=
  \big[{(\eth_\circ\widebar\eth_\circ)}^2+\eth_\circ\widebar\eth_\circ\big]\eta^0=\frac 14 \Delta_\circ(\Delta_\circ +2)\eta^0.
  \end{split}
\end{equation}
In these terms, the RT evolution equation~\eqref{eq:36} becomes an equation for the
deviation,
\begin{equation}
  \label{eq:27}
  \boxed{3 \partial_u V_\circ=
    -V^4_\circ\eth_\circ^2\widebar\eth_\circ^2V_\circ+V^3_\circ|\eth_\circ^2V_\circ|^2},
\end{equation}
cf.~\cite{Chrusciel:1992tj}, equation~(2.11), or equivalently
\begin{equation}
  \label{eq:302}
  \frac{3}{n}\partial_u V^n_\circ=-V_\circ^{3+n}\eth_\circ^2\widebar\eth_\circ^2V_\circ+V_\circ^{2+n}|\eth_\circ^2V_\circ|^2,
\end{equation}
where besides $n=1$, $n=-2,n=-3$ will also be useful, and finally,
\begin{multline}
  \label{eq:52}
  3e^{4\Phi_\circ}\, \partial_{u}\Phi_\circ=-\eth_\circ^2\widebar\eth_\circ^2\Phi_\circ
   +2\widebar\eth_{\circ}\Phi_\circ \eth_{\circ}\widebar\eth_{\circ}\eth_{\circ} \Phi_\circ
  +  2\eth_{\circ}\Phi_\circ\widebar\eth_{\circ}\eth_{\circ}
  \widebar\eth_{\circ}\Phi_\circ\\ +2|\eth_{\circ}\Phi_\circ|^2  +2{(\eth_{\circ}\widebar\eth_{\circ}\Phi_\circ)}^2 
  -4|\eth_{\circ}\Phi_\circ|^2\eth_{\circ}\widebar\eth_{\circ}\Phi_\circ.
\end{multline}
In these terms,
\begin{equation}
  \label{eq:124}
  \begin{split}
    \gamma^{0}&=-\frac 12 V_\circ^{-1}\partial_u V_\circ=\frac 12 \partial_{u}\Phi_\circ,\quad \alpha^{0}
           =\frac 12(\widebar\eth_\circ V_\circ-V_\circ\widebar\eth_0\varphi_\circ)
           =-\frac 12 e^{-\Phi_\circ}(\widebar\eth_{\circ}\varphi_{\circ}+\widebar\eth_{\circ}\Phi_\circ),\\
    \nu^{0}&=V_\circ^{-1}\widebar\eth_\circ V_\circ\partial_u V_\circ-\widebar\eth_\circ \partial_u V_\circ
           =e^{-\Phi_\circ}\widebar\eth_{\circ}\partial_{u}\Phi_\circ,\\ \mu^{0}&
                                                    =\eth_\circ V_\circ\widebar\eth_\circ V_\circ -V_\circ \eth_\circ\widebar\eth_\circ V_\circ
                                                    -\frac 12 V_\circ^2=e^{-2\Phi_\circ}(\eth_{\circ}\widebar\eth_{\circ}\Phi_\circ-\frac 12),\\
    \Psi^{0}_{3}&=V_\circ\eth_\circ V_\circ\widebar\eth_\circ^2V_\circ -V_\circ^2\eth_\circ\widebar\eth^2_\circ V_\circ=e^{-3\Phi_\circ}[\eth_{\circ}\widebar\eth^2_{\circ}\Phi_\circ
               -2\widebar\eth_{\circ}\Phi_\circ\eth_{\circ}\widebar\eth_{\circ}\Phi_\circ],\\
    \Psi^{0}_{4}&=\widebar\eth^2_\circ V_\circ\partial_u V_\circ -V_\circ\widebar\eth_\circ^2\partial_u V_\circ =e^{-2\Phi_\circ}[\widebar\eth_{\circ}^{2}\partial_{u}\Phi_\circ
               -2\widebar\eth_{\circ}\Phi_\circ\widebar\eth_{\circ} \partial_{u}\Phi_\circ].
  \end{split}
\end{equation}
Besides the constant time shift, out of the transformations~\eqref{eq:160} that
preserve RT solutions, globally well-defined ones are the fractional linear
transformations. If 
\begin{equation}
  \label{eq:159}
\mathrm{SL}(2,\mathbb C)\ni   g=\begin{pmatrix} a& b\\ c & d 
  \end{pmatrix},\quad ad-bc =1, 
\end{equation}
they acts as
\begin{equation}
  \label{eq:147a}
  x^\prime_{g}(x):\  \zeta' =\frac{ a\zeta+  b }{ c\zeta+ d},\ \widebar \zeta' =\frac{ \widebar{ a} \widebar\zeta+ \widebar{ b} }{\widebar{ c}\widebar\zeta+ \widebar{ d}}\iff
  x_{ g^{-1}}(x^\prime):\  \zeta =\frac{  d\zeta^\prime-  b }{- c\zeta^\prime+  a},\ \widebar \zeta =\frac{ \widebar{ d} \widebar\zeta^\prime- \widebar{ b} }{-\widebar{ c}\widebar\zeta^\prime+ \widebar{ a}},
\end{equation}
the transformation law~\eqref{eq:165} becomes
\begin{equation}
  \label{eq:148a}
  P'_{ g}(u,x^\prime_{ g}(x))=\frac{P(u,x)}{| c\zeta+ d|^{2}}\iff
  \varphi'_{ g}(u,x^\prime_{ g}(x))=\varphi(u,x)+\ln | c\zeta + d|^{2}.
\end{equation}

\section{Restricted flows}\label{sec:gauged-fixed-flow}

\subsection{Volume}\label{sec:volume}

Let $\mathcal S$ be a closed surface of genus $0$ diffeomorphic to $\mathbb S^2$
with metric $g$. In conformally flat coordinates, with $g$ defined in terms of
$P$ as in equation~\eqref{eq:14}, the volume $\mathrm{Vol}(g)$ is given by 
\begin{equation}
  \label{eq:206}
  \mathrm{Vol}(g)=\int_{\mathbb S^2}\frac{d\zeta\wedge d\widebar\zeta}{\imath P^{2}}
  =\int_{\mathbb S^2}\frac{d\zeta\wedge d \widebar\zeta}{\imath P^{2}_{\circ}}V_\circ^{-2}.
\end{equation}
The RT flow is volume preserving on account of Stokes' theorem. To see this
explicitly on the second form, one uses that the right hand side of the RT
equation with $n=-2$ in equation~\eqref{eq:302} is a total $\eth_\circ$ (or
$\widebar\eth_\circ$) derivative
\begin{equation}
  \label{eq:120}
  \begin{split}
     |\eth^2_\circ V_\circ|^2-V_\circ \eth_\circ^2\widebar\eth_\circ^2V_\circ
    &=\eth_\circ\big(\eth_\circ V_\circ\widebar\eth_\circ^2V_\circ               -V_\circ \eth_\circ \widebar\eth_\circ^2V_\circ                                                              \big)\\
    &=\widebar\eth_\circ\big(\widebar\eth_\circ V_\circ\eth_\circ^2V_\circ -V_\circ \widebar\eth_\circ \eth_\circ^2V_\circ\big).
  \end{split}
\end{equation}
This allows one to re-write the RT equation as
\begin{equation}
  \label{eq:293}
  -\frac 32 \partial_u V^{-2}_\circ=\frac 12 \eth_\circ\big(\eth_\circ V_\circ\widebar\eth_\circ^2V_\circ -V_\circ \eth_\circ \widebar\eth_\circ^2V_\circ      \big)+\frac 12 \widebar\eth_\circ\big(\widebar\eth_\circ V_\circ\eth_\circ^2V_\circ  -V_\circ \widebar\eth_\circ \eth_\circ^2V_\circ  \big).
\end{equation}
As a consequence, the normalization condition on the volume,
\begin{equation}
  \label{eq:137}
  \mathrm{Vol}(g)=4\pi,
\end{equation}
is compatible with the flow. Note that this translates into a normalization
condition for the zero mode of $V_\circ^{-2}$: if
${(V_\circ)}^{-2}=b^{jm} {}_0Y_{jm}$ the condition requires that
$b^{00}=\frac{1}{\sqrt{4\pi}}$. When imposing this condition on the initial
data, it will be maintained at all times for solutions to the RT equation. 

In order to exploit this property, the RT equation has been reformulated in terms of a function $f>-1$ defined through
\begin{equation}
  \label{eq:81}
V^{-2}_\circ =1+f \iff V_\circ={(1+f)}^{-\frac 12},   
\end{equation}
where one assumes that, for all times that
\begin{equation}
  \label{eq:219}
  f=f_{j\geq 1}.
\end{equation}
When using
\begin{equation}
  \label{eq:220}
  \eth_\circ V_\circ=-\frac 12 V_\circ^3\eth_\circ f,\quad
  \eth^2_\circ V_\circ=-\frac 12V_\circ^3\big[\eth_\circ^2 f
  -\frac 32V_\circ^2{(\eth_\circ f)}^2\big], 
\end{equation}
the RT equation~\eqref{eq:26} becomes
\begin{equation}
\begin{split}
  \label{eq:207}
  \partial_u f=&-\frac 13(1+f)^{-2} \eth_\circ^2\widebar\eth_\circ^2 f\\
  & + (1+f)^{-3}\big[\eth_\circ f\eth_\circ\widebar\eth_\circ^2 f+\widebar\eth_\circ f\eth_\circ^2\widebar\eth_\circ f +(\eth_\circ\widebar\eth_\circ f)^2
  +\frac 13|\eth^2_\circ f|^2\big]\\
  & -(1+f)^{-4}\big[\eth_\circ^2 f(\widebar\eth_\circ f)^2
  +\widebar\eth^2_\circ f(\eth_\circ f)^2
  +5\eth_\circ f\widebar\eth_\circ f\eth_\circ \widebar\eth_\circ f \big]\\
               &+(1+f)^{-5}4|\eth_\circ f|^4.
\end{split}
\end{equation}
In order to discuss non-linear convergence, this equation is written as
\begin{equation}
  \label{eq:249}
  3\partial_u f+ \eth_\circ^2\widebar\eth_\circ^2 f=3g(f),
\end{equation}
where $g(f)$ denotes terms that are at least quadratic in $f$ on the right hand
side of~\eqref{eq:207}, cf.~\cite{rendall1988existence,Singleton_1990},~\cite{Singleton2020} equation (4.21),~\cite{Chrusciel:1992cj}
equation (5.4),~\cite{Chrusciel:1992rv} equations (4.1), (4.3).

\subsection{Gauss-Bonnet theorem}\label{sec:gauss-bonnet}

Independently of the dynamics, there is also the integral invariant,
\begin{equation}
  \label{eq:203}
   \int_{\mathbb S^2}\frac{d\zeta\wedge d\widebar\zeta}{\imath P^{2}}(-\frac 12 R^{(2)})=4\pi,
\end{equation}
according to the Gauss-Bonnet theorem at genus $0$. In this context, this can be
explicitly shown by using that
\begin{equation}
  \label{eq:145}
  \eth\widebar\eth\varphi=e^{-2\Phi_\circ}[\eth_{\circ}\widebar\eth_{\circ}\Phi-\frac 12],
\end{equation}
which implies that
\begin{equation}
  \label{eq:144}
 \int_{\mathbb S^2}\frac{d\zeta\wedge d\widebar\zeta}{\imath P^{2}}(-\frac 12 R^{(2)})=  \int_{\mathbb S^2}\frac{d\zeta\wedge d\widebar\zeta}{\imath P^{2}}(-2\eth\widebar\eth\varphi)=
\int_{\mathbb S^2}\frac{d\zeta\wedge d\widebar\zeta}{\imath P^{2}_{\circ}}(-2\eth_{\circ}\widebar\eth_{\circ}\Phi+1)=
  4\pi,
\end{equation}
on account of Stokes' theorem. In terms of $V_\circ$, this follows from
\begin{equation}
  \label{eq:170}
  \eth\widebar\eth\varphi=\eth_\circ V_\circ\widebar\eth_\circ V_\circ -V_\circ \eth_\circ\widebar\eth_\circ V_\circ-\frac 12 V_\circ^2,
\end{equation}
so that
\begin{equation}
  \label{eq:180}
  \int_{\mathbb S^2}\frac{d\zeta\wedge d\widebar\zeta}{\imath P^{2}}(-\frac 12 R^{(2)})=
  (-2)\int_{\mathbb S^2}\frac{d\zeta\wedge d\widebar\zeta}{\imath P^{2}_\circ }V_\circ^{-2}[\eth_\circ V_\circ\widebar\eth_\circ V_\circ -V_\circ \eth_\circ\widebar\eth_\circ V_\circ-\frac 12 V_\circ^2].
\end{equation}
Since
\begin{equation}
  \label{eq:201}
  V_\circ^{-2}\eth_\circ V_\circ\widebar\eth_\circ V_\circ-V_\circ^{-1}\eth_\circ\widebar\eth_\circ V_\circ =-\eth_\circ(V_\circ^{-1}\widebar\eth_\circ V_\circ)=-\widebar\eth_\circ(V_\circ^{-1}\eth_\circ V_\circ),
\end{equation}
Stokes' theorem now implies that the contribution of the first two terms
vanishes, so that the result comes from the last term which reduces to the
volume of the sphere.

The first considerations on Lyapunov functionals in this context have been based
on this topological invariant~\cite{lukacs1984lyapunov} (see also~\cite{Singleton2020}, section
4.3 for a comprehensive discussion).

\subsection{Antipodally symmetric solutions}\label{sec:antip-symm-solut}

As shown in~\cite{rendall1988existence}, one may restrict oneself consistently to antipodally
symmetric solutions, in which case the problem simplifies because the zero modes
at $j=1$ are absent.

More explicitly, under the antipodal transformation $A$ described by
\begin{equation}
  \label{eq:178}
A(x)\ :\  \zeta^A=-\frac{1}{\widebar \zeta},\quad \widebar\zeta^A=-\frac{1}{ \zeta},
\end{equation}
the conformal factor transforms as 
\begin{equation}
  \label{eq:208}
  P^A(u,x)=|\zeta|^2P(u,A(x)),
\end{equation}
and, in particular, the conformal factor for the sphere is invariant,
$P^A_\circ(\zeta,\widebar\zeta)=P_\circ(\zeta,\widebar\zeta)$.

Furthermore, on the level of $\mathrm{SL}(2,\mathbb C)$ transformations, we have
\begin{equation}
  \label{eq:131}
x_{g}(A(x^\prime))=  A(x_{{g^{-1}}^\dagger}(x^\prime)). 
\end{equation}
This means that rotations commute with the antipodal map, while for boosts, the
boost or rapidity vectors change signs, cf.~Appendix~\ref{sec:boosts-conf-fact}.
Spin and boost weighted functions $\eta^{s,w}$ are mapped to functions $\eta^{s,w A}$
of weights $(-s,w)$ according to
\begin{equation}
  \label{eq:209}
  \eta^{s,w A}(u,x)
  ={(-)}^s\big{(\frac{\widebar\zeta}{\zeta}\big)}^s
  \eta^{s,w}(u,A(x)),
\end{equation}
so that
\begin{equation}
  \label{eq:213}
  {(\eta^{\prime s,w}_{;g^{-1}})}^A(u,x^\prime)={\eta^{s,w A}}^\prime_{;g^\dagger}(u,x^\prime).
\end{equation}
It follows that
\begin{equation}
  \label{eq:210}
  \eth_A\, \eta^{s,w A}=-{\big(\widebar\eth \eta^{s,w}\big)}^A,\quad \eth_\circ  \eta^{s,w A}=-{\big(\widebar\eth_\circ \eta^{s,w}\big)}^A. 
\end{equation}
In particular, the RT equations~\eqref{eq:27},~\eqref{eq:302} are invariant
under the antipodal map.

Restricting the problem to antipodally symmetric conformal factors that satisfy 
\begin{equation}
  \label{eq:211}
  P^A(u,\zeta,\widebar\zeta)=P(u,\zeta,\widebar\zeta),
\end{equation}
is consistent with the RT equation. Since antipodally symmetric spherical
harmonics ${}_{0,1}Y_{jm}$ (cf.~Appendix~\ref{sec:spin weighted harmonics}) form
a subalgebra composed of those with even $j$, antipodally symmetric solutions
$V^{S}_\circ$ admit an expansion of the form $V^{S}_\circ=c+g^{S}$, $g^S > -c$,
and $g^S$ only involves even harmonics with $j\geq 2$.

When imposing in addition the normalization condition on the volume and working
with the variable $f$ defined in~\eqref{eq:81}, the reduced flow may be analyzed in terms
of the variable
\begin{equation}
  \label{eq:226}
  {(V^S_\circ)}^{-2}=1+f^S,\quad f^S=d^{jm}\,{}_{0,-2} Y_{jm}\big\rvert{j\geq 2},
\end{equation}
and its expansion in real even spherical harmonics with $j\geq 2$. In the RT
equation~\eqref{eq:207} for $f^S$, the highest term in derivatives involving
$\eth^2_\circ\widebar\eth^2_\circ f^S$ no longer has zero-modes. As a consequence exponential
decay of $f^S$, and thus to the Schwarzschild solution with $M=1$, may be proved
under suitable assumptions~\cite{rendall1988existence}.

Note however that the restriction to even spherical harmonics of spin and boost
weight $(0,1)$ is not preserved under boosts.

\section{Memory in asymptotically flat spacetimes}\label{sec:non-linear-memory}

\subsection{Displacement and non-linear memory}\label{sec:total-non-linear}

In the closely related study of locally asymptotically flat spacetimes at future
null infinity\footnote{A standard (and easily controllable) difference between the Newman-Penrose-Unti and the BMS definitions of locally asymptotically flat spacetimes consists in using a different radial coordinate, an affine parameter for null geodesics versus the ``luminosity'' distance.}~\cite{Newman:1961qr,Newman:1962cia,Exton:1969im}, the first element of the Newman-Penrose
tetrad, $l'$, is taken to be the generator of an affinely parametrized null
geodesic, $l'=\partial_{r'}$, that is twist-free, hypersurface orthogonal and a
gradient, $l'=du'$, but does generically admit a non-vanishing shear $\sigma'$. In
terms of the affine parameter $r'$, all remaining spin-coefficients that have
not been set to zero by the above conditions vanish at large $r'$, while to
leading order the conformally flat 2-dimensional line element
$d\widebar s^{\prime 2}$ is that of the round sphere,
$m'=-\frac{r'}{P'_{\circ}}+\mathcal O(r^{\prime -1})$,
$P'_{\circ}=\frac{1}{\sqrt 2}(|\zeta'|^2+1)$.

In this asymptotically flat solution space, the relevant equations are  
\begin{equation}
  \label{eq:56}
  \lambda'^{0}=\partial_{u'}\widebar\sigma'^{0},\quad \Psi'^{0}_{3}=-\eth^\prime_{\circ}\partial_{u'}\widebar\sigma^{\prime 0},\quad
  \Psi^{\prime 0}_{4}=-\partial_{u'}^{2}\widebar\sigma^{\prime 0},
\end{equation}
and the evolution equation
\begin{equation}
  \label{eq:23}
  \partial_{u'}\Psi^{\prime 0}_{2}=\eth^{\prime}_{\circ}\Psi^{\prime 0}_{3}+\sigma^{\prime 0}\Psi^{\prime 0}_{4},
\end{equation}
in terms of Bondi time $u'$. A further equation relating these quantities is the
reality condition
\begin{equation}
  \label{eq:6}
  \Psi^{\prime 0}_{2}
  +\sigma^{\prime 0}\lambda'^{0}+\eth^{\prime 2}_{\circ}\widebar\sigma^{\prime 0} -\mathrm{c.c.}=0.
\end{equation}
Introducing the notation
$\Delta'f' (u',\zeta',\widebar\zeta')=f'(u'_{f},\zeta',\widebar\zeta')-f'(u'_{i},\zeta',\widebar\zeta')$, the
displacement (or total) memory is obtained by integrating the first equation
of~\eqref{eq:56} between some initial and some final time,
\begin{equation}
  \label{eq:250}
 \boxed{ \Delta^\prime \widebar\sigma^{\prime 0}=\int_{u'_i}^{u'_f}du' \lambda^{\prime 0}}. 
\end{equation}

How Christodoulou's non linear memory effect~\cite{Christodoulou:1991cr} appears in this context has
been shown concisely by Frauendiener~\cite{Frauendiener1992}. Paraphrasing his derivation along
the lines of Section 6.5 of~\cite{Barnich:2016lyg}, one may start from the total mass aspect
\begin{equation}
  \label{eq:68}
  4\pi \mathcal M'_{T}=-(\Psi^{\prime 0}_{2}
  +\sigma^{\prime 0}\lambda'^{0}+\eth^{\prime 2}_{\circ}\widebar\sigma^{\prime 0})+\int^{u'}_{u'_{i}}du'\, \partial_{v'}\sigma^{\prime 0}\partial_{v'}\widebar\sigma^{\prime 0}.
\end{equation}
The total mass aspect is real on account of~\eqref{eq:6}, and conserved
$\partial_{u' }\mathcal M'_{T}=0$ on account of the evolution equations so that
$\Delta'\mathcal M'_{T}=0$ or, more explicitly,
\begin{equation}
  \label{eq:58}
  \eth_{\circ}^{\prime 2}\Delta'\widebar\sigma^{\prime 0}=-\Delta'(\Psi'^{0}_{2}+\sigma^{\prime 0}\lambda'^{0})
    +\int^{u'_{f}}_{u'_{i}}\,dv' |\lambda'^{0}|^{2},
\end{equation}
where one recognizes the combination that appears in the Bondi mass aspect
$\mathcal M'_{B}$ on the right hand side,
\begin{equation}
  \label{eq:59}
  8\pi \mathcal M'_{B}=-(\Psi^{\prime 0}_{2}+\sigma^{\prime 0}\lambda'^{0}+\mathrm{c.c.}),
\end{equation}
as well as the energy that is radiated away to infinity per unit solid angle between $u'_{i}$ and $u'_{f}$,
\begin{equation}
  \label{eq:91}
  4\pi \mathcal F_{[u'_{i},u'_{f}]}=\int^{u'_{f}}_{u'_{i}}\,du' |\lambda'^{0}|^{2}.
\end{equation}
This is the relevant quantity in the non-linear memory effect that we will
compute for the Robinson-Trautman spacetimes. The above equations encode how
this quantity is relevant for the geodesic deviation equation with respect to
the asymptotic time-like geodesic with tangent vector
$\frac{1}{\sqrt 2}(l'+n')$.

The asymptotic part of the shear $\widebar\sigma^{\prime 0}$ has spin-weight
$-2$ and may be expanded in terms of the appropriate spin-weighted spherical
harmonics,
\begin{equation}
  \label{eq:60}
  \widebar\sigma^{\prime 0}=\sum_{j\geq 2}\sum_{m=-j}^{j}\widebar\sigma^{\prime 0}_{jm}(u')\ {}_{-2}Y_{jm}(\zeta',\widebar\zeta').
\end{equation}
The left hand side of~\eqref{eq:58} only involves harmonics with $j\geq 2$. When
projecting on the four lowest harmonics, one gets the flux balance laws
\begin{equation}
  \label{eq:61}
  \Delta'\int_{\mathbb S^{2}} d^{2}\Omega^{\prime} (\Psi^{\prime 0}_{2}+\sigma^{\prime 0}\lambda'^{0})\,\widebar{{}_{0}Y_{jm}}
  =\int^{u'_{f}}_{u'_{i}}\,du' \int_{\mathbb S^{2}} d^{2}\Omega\,
  |\lambda'^{0}|^{2}\widebar{{}_{0}Y_{jm}},\quad j\leq 1,
\end{equation}
where
$d^{2}\Omega^{\prime}=\frac{2d\zeta'd\widebar\zeta'}{i{(1+\zeta'\widebar\zeta')}^{2}}=\sin\theta' d\theta' d\phi'$,
while for the higher order harmonics, 
one gets
\begin{multline}
  \label{eq:63}
  \Delta'\widebar\sigma^{\prime 0}_{jm}=\frac{1}{\sqrt{s_j}}
  \Big[-\Delta'(\int_{\mathbb S^{2}} d^{2}\Omega'
  (\Psi^{\prime 0}_{2}+\sigma^{\prime 0}\lambda'^{0})\,\widebar{{}_{0}Y_{jm}})\\
  +\int^{u'_{f}}_{u'_{i}}\,du' \int_{\mathbb S^{2}} d^{2}\Omega'\,
  |\lambda'^{0}|^{2}
  \widebar{{}_{0}Y_{jm}}\Big],\quad j\geq 2,
\end{multline}
with $s_j$ given in~\eqref{eq:259}. The quadratic terms in the shear on the
right hand side involve the decomposition of the product of two spin-weighted
spherical harmonics ${}_{-2}Y_{j_{1}m_{1}}{}_{2}Y_{j_{2}m_{2}}$ into standard
spherical harmonics ${}_{0}Y_{jm}$ using the Wigner $3j$-symbols (see e.g.~\cite{Beyer:2013loa}
for details).

Note that a slightly different interpretation of equation~\eqref{eq:58} that encodes the
local version of the non-linear memory effect is possible: if one uses the
generalized, real, mass aspect that includes the linear term in the shear,
cf.~\cite{walker1983positivity}, equation~(35),
\begin{equation}
  \label{eq:75}
4\pi \mathcal M'_{G}= \Psi',\quad  \Psi' =-[\Psi^{\prime 0}_{2}
  +\sigma^{\prime 0}\lambda'^{0}+\eth^{\prime 2}_{\circ}\widebar\sigma^{\prime 0}],
\end{equation}
equation~\eqref{eq:58} reduces to a local mass loss formula for the generalized mass
aspect,
\begin{equation}
  \label{eq:76}
\boxed{\partial_{u'}\Psi'= -|\lambda'^{0}|^{2}\iff  \Delta'\Psi'=-\int^{u'_{f}}_{u'_{i}}\,du' |\lambda'^{0}|^{2}}.
\end{equation}
In other words, the generalized mass aspect $\Psi'(u',\zeta',\widebar\zeta')$ is a
decreasing function of $u'$. In what follows we will refer to~\eqref{eq:250} or the
associated evolution equation, the first of~\eqref{eq:56}, as the displacement or total
memory effect and to~\eqref{eq:76} as the non-linear memory effect.

For later use, let us also note that the initial angular momentum aspect is given
by~\cite{Dray1984},
\begin{equation}
  \label{eq:28}
  \widebar\Psi^\prime_J=-[\Psi^{\prime 0}_1+\frac 32 \sigma^{\prime 0}\eth^\prime_\circ\widebar\sigma^{\prime 0}+\frac 12\eth^\prime_\circ\sigma^{\prime 0}\widebar\sigma^{\prime 0}].
\end{equation}
If $\mathcal Y^\prime(x^\prime)$ is of spin weight $-1$, and satisfies
$\widebar\eth^\prime_\circ \mathcal Y^\prime=0$, so that
\begin{equation}
  \label{eq:80}
  \mathcal Y^\prime(x^\prime)=y^{1m}\, {}_{-1} Y_{1m}=y^{1m}\widebar\eth^\prime_\circ\, {}_{0} Y_{1m},\quad y^{1m}\in \mathbb C,
  \end{equation}
  and the associated real solutions describe conformal Killing vectors of the
  sphere~\cite{Held1970}, the total angular momentum of RT waves is given by
\begin{equation}
  \label{eq:34}
  8\pi Q_{\mathcal Y^\prime,\widebar{\mathcal Y}^\prime}
  =\int_{\mathbb S^2}d^2\Omega^\prime\,  [\mathcal Y^\prime \widebar \Psi^\prime_{J}+ \widebar{\mathcal Y} {\Psi^\prime_J}+ u^\prime(\eth^\prime_\circ\mathcal Y^\prime+\widebar\eth^\prime_\circ \widebar{\mathcal Y}^\prime)\Psi^\prime],
\end{equation}
while the time-dependent angular momentum aspect is 
\begin{equation}
  \label{eq:312}
  \widebar \Psi^\prime_{J}(u^\prime)=\widebar \Psi^\prime_{J}-u^\prime\mathcal \eth^\prime_\circ\Psi^\prime.
\end{equation}

\subsection{Covariance of memory effects}\label{sec:covar-memory-effect}

Applying a standard supertranslation in the asymptotically
flat Bondi frame means that one considers the asymptotic change of
coordinates\footnote{We omit a subscript indicating that this change of
  coordinates is only valid to leading order in the asymptotic expansion.}
\begin{equation}
  \label{eq:8a}
  \tilde u=u'+\alpha'(x^\prime),\quad \tilde \zeta=\zeta',\quad \tilde{\widebar\zeta}=\widebar\zeta' \iff \tilde x^A=x^{\prime A}
\end{equation}
If 
\begin{equation}
  \label{eq:309}
  \eta^{\prime s}(u^\prime,x^\prime)|=\eta^{\prime s}(\tilde u-\alpha',\tilde x),
\end{equation}
it follows that
\begin{equation}
  \label{eq:22}
  \partial_{\tilde u}[\eta^{\prime s}|]=[\partial_{u^\prime}\eta^{\prime s}]|,\quad \tilde\eth_{\circ}[\eta^{\prime s}|]=[\eth'_{\circ}\eta^{\prime s}-\eth'_{\circ}\alpha'\partial_{u^\prime}\eta^{\prime s}]|.
\end{equation}
The results of~\cite{Barnich:2016lyg} in the particular case of a (pure)
supertranslation then imply that
\begin{equation}
  \label{eq:24}
  \begin{split}
    \tilde\sigma^{0}&=[\sigma^{\prime 0}+\eth_{\circ}^{\prime 2}\alpha']|,\\
    \tilde \lambda^{0}&=\partial_{\tilde u}\tilde{\widebar \sigma}^{0}=[\partial_{u^\prime}\widebar\sigma^{\prime 0}]|\\
\tilde\Psi^{0}_{2}&=[\Psi^{\prime 0}_{2}-2\eth_{\circ}^{\prime}\alpha'\Psi^{\prime 0}_{3}+{(\eth_{\circ}^{\prime}\alpha')}^{2}\Psi^{\prime 0}_{4}]|,
  \end{split}
\end{equation}
where the quantities on the left hand sides are functions of
$\tilde u,\tilde x^A$, and where~\eqref{eq:56} holds. It then follows that
\begin{equation}
  \label{eq:57a}
  \tilde\eth_\circ\tilde{\widebar\sigma}^{0}=\big[\eth^{\prime}_{\circ}\widebar\sigma^{0}+\eth^{\prime}_{\circ}\widebar\eth^{\prime 2}_{\circ}\alpha'  -\eth^{\prime}_{\circ}\alpha' 
  \partial_{u'}\widebar\sigma^{\prime 0}\big]\big\rvert,
\end{equation}
\begin{equation}
  \label{eq:71a}
  \tilde\eth^{2}_\circ\tilde{\widebar\sigma}^{0}=\big[\eth^{\prime 2}_{\circ}\widebar\sigma^{0}+\eth^{\prime 2}_{\circ}\widebar\eth^{\prime 2}_{\circ}\alpha'  -\eth^{\prime 2}_{\circ}\alpha' 
  \partial_{u'}\widebar\sigma^{\prime 0} -2\eth^{\prime }_{\circ}\alpha' \eth^{\prime}_{\circ}\partial_{u'}\widebar\sigma^{\prime 0}+ {(\eth^{\prime}_{\circ}\alpha)}^{2}\partial_{u'}^{2}
  \widebar\sigma^{\prime 0} \big]\big\rvert.
\end{equation}
As a consequence, the generalized mass aspect in equation~\eqref{eq:75} transforms simply
as
\begin{equation}
  \label{eq:26}
  \boxed{\tilde\Psi=\Psi'\big\rvert-\tilde\eth^{2}_{\circ}\tilde{\widebar\eth}^{2}_{\circ}\tilde\alpha},\quad \tilde\alpha=\alpha^\prime|, 
\end{equation}
cf.~\cite{Moreschi1986,Moreschi:1988pc} (see also~\cite{RignonBret2025} for further considerations).

These transformation laws mean in particular that the displacement and the
non-linear memory effects are supertranslation invariant. In addition, since
ordinary translation correspond to the first for real harmonics in an expansion
of $\alpha^\prime$, which are in the kernel of $\eth^{\prime 2}_{\circ}\widebar\eth^{\prime 2}_{\circ}$, the
generalized mass aspect is translation invariant. Finally, since
$\tilde\eth^2_\circ\tilde\widebar\eth^2_\circ$ is invertible on
$\alpha_{j\geq 2}$, proper supertranslation with parameters that do not contain
standard translations and whose expansion contains spherical harmonics with
$j\geq 2$ can be used to set $\Psi_{j\geq 2}$ to zero at a fixed time
$u$~\cite{Moreschi1986,Moreschi:1988pc}.

Applying a Lorentz transformation $g$ inside the $\mathrm{BMS}_4$ group in the
asymptotically flat Bondi frame means that one considers the asymptotic change
of coordinates
\begin{equation}
  \label{eq:308}
  \tilde x_g(x^\prime), \quad \tilde u=e^{E^g_R}(x^\prime)u^\prime,
\end{equation}
where
\begin{equation}
  \label{eq:306}
  e^{E^g_R}(x^\prime)=\frac{|\zeta^\prime|^2+1}{X_g(x^\prime)},\quad X_{g}(x^\prime)= |a\zeta^\prime+b|^{2}+|c\zeta^\prime +d|^{2}, \quad
  e^{\imath E^g_I}(x^\prime)=\frac{\widebar c\widebar\zeta^\prime+\widebar d}{c\zeta^\prime+d}, 
\end{equation}
with $E^g=E^g_R+\imath E^g_I$, $\eth^{\prime 2}_\circ E^g_R-{(\eth^{\prime}_\circ E^g_R)}^2=0$, and the
understanding that $E^g_R,E^g_I$ are of spin weights zero. The inverse
transformation is
\begin{equation}
  \label{eq:339}
      x^\prime_{g^{-1}}(\tilde x),\quad  u^\prime =
    e^{-E^g_R}(x^\prime_{g^{-1}}(\tilde x))\tilde u=e^{E^{g^{-1}}_R}(\tilde x)\tilde u,
  \end{equation}
where the last expression for $u'$ has been obtained by taking into account~\eqref{eq:132}.

Under~\eqref{eq:308}, a field $\eta^{\prime s,w}(u^\prime,x^\prime)$ of spin and boost weights
$(s,w)$ transforms as
\begin{equation}
  \label{eq:366}
  \tilde\eta^{s,w}_{;g}(\tilde u,\tilde x)=\big(e^{w E_R^g+s\imath E_I^g}\eta^{\prime s,w}\big)\big\rvert,
\end{equation}
with
\begin{equation}
  \label{eq:310}
  f(u^\prime,x^\prime)|=f(e^{-E^g_R}(x^\prime_{g^{-1}}(\tilde x))
  \tilde u ,x^\prime_{g^{-1}}(\tilde x)).
\end{equation}
As a consequence,
\begin{equation}
  \label{eq:2}
  \partial_{\tilde u} \tilde\eta^{s,w}_{;g}=\big(e^{(w-1)E^g_R+s\imath E^g_I}\partial_{u^\prime}\eta^{\prime s,w}\big)\big\rvert,
\end{equation}
so that $\partial_{u^\prime}$ lowers the boost weight by $1$. Furthermore, let
$\tilde\eth_{\circ}$ denote the eth operator in $\tilde x$ variables that involves
$P_\circ(\tilde x)$ and $\eth^\prime_\circ$ denote the eth operator in $x^\prime$ variables
that involves $P_\circ(x^\prime)$. When taking into account that
\begin{equation}
  \label{eq:85}
  P_\circ(\tilde x)\frac{\partial\widebar\zeta^\prime}{\partial\widebar{\tilde\zeta}}=
  \big(P_\circ e^{-E^g+\imath E^I_g}(x^\prime)\big)\big\rvert,
\end{equation}
a somewhat involved computation yields
\begin{equation}
  \label{eq:7}
  \tilde\eth_{\circ} \tilde\eta^{s,w}_{;g}=\big( e^{(w-1)E^g_R+(s+1)\imath E^g_I}[\eth_\circ^\prime+(w-s)\eth_\circ^\prime E^g_R-\eth_\circ^\prime E^g_R u^\prime\partial_{u^\prime}]\eta^{\prime s,w}\big)\big\rvert.
\end{equation}

It follows for instance from~\cite{Barnich:2016lyg} that, in the particular case of a Lorentz
transformations inside the $\mathrm{BMS}_4$ group,
\begin{equation}
  \label{eq:305}
  \begin{split}
  \tilde\sigma^0&=\big(e^{-E^g_R+2\imath E^g_I}
                                \sigma^{\prime 0}\big)\big\rvert,\\
    \tilde\lambda^0&=\big(e^{-2E^g}\lambda^{\prime 0}\big)\big\rvert,\\
    \tilde\Psi^0_2&=\big(e^{-3E^g_R}\big[\Psi^{\prime 0}_2-2\eth^\prime_\circ E^g_R u^\prime\Psi^{\prime 0}_3
+{(\eth^\prime_\circ E^g_R u^\prime)}^2\Psi^{\prime 0}_4\big]\big)\big\rvert.
  \end{split}
\end{equation}
In particular, $\sigma^0,\lambda^0$ are of spin and boost weights $(2,-1)$ and $(-2,-2)$.
As a consequence,
$\partial_{\tilde u}\tilde{\widebar\sigma}^0 =\big(e^{-2E^g}\partial_{u^\prime}\widebar\sigma^{\prime 0}\big)\big\rvert$
and transforms like $\tilde\lambda^{0}$. Similarly,
\begin{equation}
  \label{eq:307}
  \begin{split}
    \tilde\sigma^0\tilde\lambda^0&=\big(e^{-3E^g_R}\sigma^{\prime 0}\lambda^{\prime 0}\big)\big\rvert,\\
    \tilde\eth_{\circ}\tilde{\widebar\sigma}^0&= 
\big(e^{-2 E^g_R-\imath E_I^g}\big[\eth_\circ^\prime+\eth_\circ^\prime E^g_R-\eth_\circ^\prime E^g_R u^\prime\partial_{u^\prime} \big]\widebar\sigma^{\prime 0}\big)\big\rvert,\\
    \tilde\eth^2_{\circ}\tilde{\widebar\sigma}^0&=\big(e^{-3E^g_R}\big[\eth^{\prime 2}_\circ-2\eth^\prime_\circ E^g_R u^\prime\eth_\circ^\prime\partial_{u^\prime}+{(\eth^\prime_\circ E^g_R u^\prime)}^2\partial^2_{u^\prime}\big]\widebar\sigma^{\prime 0}\big)\big\rvert.
  \end{split}
\end{equation}
When using~\eqref{eq:56}, the generalized mass aspect transforms homogeneously, 
\begin{equation}
  \label{eq:370}
  \boxed{\tilde\Psi=\big(e^{-3E^g_R}\Psi^\prime\big)\big\rvert},
\end{equation}
and is thus of spin and boost weights $(0,-3)$. Since
\begin{equation}
  \label{eq:100}
  |\tilde\lambda^0|^2=\big(e^{-4E_R^g}|\lambda^\prime_0|^2\big)\big\rvert,\quad \partial_{\tilde u}
  =e^{-E^g_R}\partial_{u^\prime},\quad d\tilde u= e^{E^g_R}du^\prime,
\end{equation}
both the displacement and the non-linear memory effects are covariant under
Lorentz transformations inside the $\mathrm{BMS}_4$ group.

\subsection{Bondi supermomentum and rest frame}\label{sec:bondi-superm-inst}

(Generalized) Bondi supermomentum is defined in
terms of the generalized mass aspect and a real supertranslation field $T^\prime(x^\prime)$ of weights $(0,1)$ through
\begin{equation}
  \label{eq:354}
4\pi  P^\prime(u^\prime)=\int_{\mathbb S^2}\,d^2\Omega^\prime \, \Psi^\prime  \, T^\prime,
\end{equation}
where we have taken into account that under Lorentz
transformations,
\begin{equation}
  \label{eq:155}
  d^2\tilde\Omega=\big(d^2\Omega^\prime  e^{2 E^g_R}\big)\big\rvert,
\end{equation}
i.e., the integration measure is of spin and boost weights $(0,2)$. As a
consequence, Bondi supermomentum transforms simply as
\begin{equation}
  \label{eq:369}
 4\pi\tilde P(\tilde u)=\int_{\mathbb S^2}\,d^2\tilde\Omega \, \tilde \Psi\, \tilde T=4\pi P^\prime\rvert. 
\end{equation}
Both the supertranslation field and the generalized mass aspect are expanded in
terms of spin and boost weighted spherical harmonics~(cf.~Appendix~\ref{sec:spin weighted harmonics}),
\begin{equation}
  \label{eq:378}
  T^\prime=  T^{\prime jm}\, {}_{01} Y_{jm},\quad \Psi^\prime= \Psi^{\prime jm} \, {}_{0,-3} Y_{jm},
\end{equation}
so that the components of BMS supermomentum are given by
\begin{equation}
  \label{eq:375}
  4\pi P^{\prime jm}{(u^\prime)}=\Psi^{\prime jm}. 
\end{equation}

Standard translations are defined by the condition
\begin{equation}
  \label{eq:374}
  \eth^{\prime 2}_\circ T^\prime=0,
\end{equation}
and its complex conjugate, which select the span of the four lowest real
spherical harmonics ${}_{01} Y_{jm},j\leq 1$ and form the vector representation
(see~\cite{Sachs1962,Newman1966,Held1970} and~\cite{Penrose:1984}, section 4.15) of the Lorentz group, which
appear here as a finite dimensional Lorentz invariant sub-representation of a
real fields of spin and boost weights $0,1$. The dual co-vector representation,
$D_{-2,-2}/F_{-2,-2}$ in the terminology of~\cite{gelfand1966generalized,Held1970}, is described by the
quotient of a field of spin and boost weight $(0,-3)$ by fields of the form
$\eth^2_\circ\widebar\eth^2_\circ \eta^{0,1}$ and may be represented by the span of the first
four spherical harmonics ${}_{0,-3}Y_{jm}, j\leq 1$.

As will be explicitly shown below in the current context, where
$P^{\prime 00}(u^\prime)>0$, one may always perform a suitable boost belonging to the
${\rm BMS}_4$ group at $\scrip$ so that $P^{\prime 1m}(u^\prime)=0$ at a given time $u^\prime$,
which we take to be the initial time $u^\prime=0$. This defines the initial rest
frame for the system. Furthermore, by a suitable scaling, one may always assume
that $4\pi P^{\prime 00}(0)=M^\prime$ for a given normalization condition
determined by $M^\prime$.

\section{Asymptotically flat data for RT waves}\label{sec:bms-data-rt}

The coordinate transformation that is necessary to put RT metrics with ``natural''
time $u$ in asymptotically flat form with Bondi time $u'$ has been discussed for
instance in~\cite{Tafel2000} in the metric approach and in~\cite{Adamo:2009vu} in the NP framework. We
will follow here the approach of~\cite{Barnich:2016lyg} as applied to the
current problem in~\cite{Mao_2019} in order to get more complete results.

A rather direct generalization by Newman and Unti~\cite{Newman:1962cia} (NU) of locally
asymptotically flat spacetimes at $\scrip$ in the sense of NP, that encompasses
RT spacetimes as described above, consists in leaving the conformal factor in
front of the spherical metric associated to a cut of $\scrip$ arbitrary, see
Appendix~\ref{sec:locally-asympt-flat}. In this case, the spin coefficients $\gamma$ and $\nu$ do no longer
vanish asymptotically,
\begin{equation}
  \label{eq:101}
  \gamma=\gamma^{0}+\mathcal O(r^{{-2}}),\quad \nu=\nu^{0}+\mathcal O(r^{-1}),
\end{equation}
with $\gamma^{0},\nu^{0}$ (and also $\alpha^{0},\mu^{0}$) as in~\eqref{eq:13}. For later use, some of
the equations of NU solution space that are relevant for us below are
\begin{equation}
  \label{eq:102}
  \lambda^{0}=(\partial_{u}+2\gamma^{0})\widebar\sigma^{0},\quad \Psi^{0}_{3}=-\eth\lambda^{0}+\widebar\eth\mu^{0},
  \quad \Psi^{0}_{4}=\widebar\eth\nu^{0}-(\partial_{u}+4\gamma^{0})\lambda^{0},
\end{equation}
the reality condition\footnote{In the description of RT waves in Section~\ref{sec:rt-solutions}, it
  trivially holds since $\Psi^{0}_{2}$ is real while all other quantities vanish.}
\begin{equation}
  \label{eq:66}
  \Psi^{0}_{2}+\sigma^{0}\lambda^{0}+\eth^{2}\widebar\sigma^{0}-\mathrm{c.c.}=0,
\end{equation}
as well as the evolution equation
\begin{equation}
  \label{eq:67}
  (\partial_{u}+6\gamma^{0})\Psi^{0}_{2}=\eth\Psi^{0}_{3}+\sigma^{0}\Psi^{0}_{4}.
\end{equation}
As a direct consequence of the generalization of the solution space, the
asymptotic symmetry group is enhanced to BMS-Weyl transformations, introduced in
the BMS set-up in~\cite{Barnich:2009se,Barnich:2010eb} and worked out in the context of the NP framework
in~\cite{Barnich:2016lyg,Barnich:2019vzx}.

We are interested in the combined coordinate transformation and frame rotation
that brings the RT solutions of Section~\ref{sec:rt-solutions} to the standard asymptotically flat
form. This transformation is obtained as a particular case of the results worked
out in Sections 6.3 and 6.4 of~\cite{Barnich:2016lyg}, that implements an asymptotic Weyl
transformation and brings the time-dependent conformal factor $P$ to the
conformal factor $P_{\circ}$ for the sphere. Details on these transformations are
provided in Appendix~\ref{sec:asympt-frame-rott}.

As discussed in Section 6.5 of~\cite{Barnich:2016lyg}, the relevant quantity in the
transformation of the free data is
\begin{equation}
  \label{eq:32}
  e^{{-E_{R0}}}\eth u'_{0}=P_{\circ}\widebar\partial u'_{0}=\eth_{\circ} u'_{0}.
\end{equation}
It then follows from equations (6.85) {--} (6.87) of~\cite{Barnich:2016lyg} and the NP
quantities for the RT waves in Section~\ref{sec:rt-solutions} that
\begin{equation}
  \label{eq:33}
  \begin{split}
    \sigma^{\prime 0}&=e^{-E_{R0}}[\eth(e^{-E_{R0}}\eth u'_{0} )-(e^{-E_{R0}}\eth u' )\partial_{u}(e^{-E_{R0}}\eth u'_{0} )
    ]\big\rvert=\widebar\partial(P^{2}_{\circ}\widebar\partial u')\big\rvert\\& =\eth^{2}_{\circ}u'_{0}\big\rvert,
  \end{split}
\end{equation}
where the vertical bar indicates that $\sigma^{\prime 0}$ depends on the ``Bondi'' time
$u'$, i.e., on the right hand side one has to replace $u=u_{0}(u',x)$
by using the inverse of
\begin{equation}
  \label{eq:105}
  u'_{0}(u,x)
  =\int_{0}^{u}dv\, V_\circ(v,x),
\end{equation}
cf.~\cite{Tafel2000} and~\cite{Adamo:2009vu}, equation (5.31).

It then follows directly from the solution space with standard
conformal factor that
\begin{equation}
  \label{eq:30}
     \lambda^{\prime 0}
     =  \partial_{u'} \widebar \sigma^{\prime 0}=V_\circ^{-1}\widebar \eth_{\circ}^{2}V_\circ\big\rvert =[-\widebar\eth_{\circ}^{2}\Phi_\circ+{(\widebar\eth_{\circ}\Phi_\circ)}^{2}]\big\rvert,
\end{equation}
which is also what equation (6.86) of~\cite{Barnich:2016lyg} yields, cf.~equation (4.101)
of~\cite{Singleton2020}. One also finds directly from the solution space with
standard conformal factor that
\begin{equation}
  \label{eq:35}
  \begin{split}
    \Psi^{\prime 0}_{4}
    &=-\partial_{u'}\lambda^{\prime 0}  =\big[V_\circ^{-1}
  \widebar\eth^{2}_{\circ}\partial_{u}\varphi+2V_\circ^{-2}\widebar\eth_{\circ}V_\circ
      \widebar\eth_{\circ}(\partial_{u}\varphi)\big]\big\rvert\\
    &=\big[V_\circ^{-3}\widebar\eth_\circ^2V_\circ\partial_u V_\circ-V_\circ^{-2}\widebar\eth_\circ^2\partial_u V_\circ\big]\big\rvert=\big[e^{\Phi_\circ}\big(\widebar\eth_{\circ}^{2}\partial_{u}\Phi_\circ
  -2\widebar\eth_{\circ}\partial_{u}\Phi_\circ\widebar\eth_{\circ}\Phi_\circ\big)\big]\big\rvert,
  \end{split}
\end{equation}
where the equation of motion~\eqref{eq:52} should be used to substitute
$\partial_u V_\circ,\partial_{u}\Phi_\circ$. This coincides with what one finds from the
transformation law (6.87) of~\cite{Barnich:2016lyg}. Both from the transformation law (6.87) of~\cite{Barnich:2016lyg}
and from the direct computation for solution space with standard conformal
factor, $\Psi^{\prime 0}_{3}=-\eth^{\prime}_{\circ}\lambda^{\prime 0}$, one
finds
\begin{equation}
  \label{eq:37}
  \begin{split} \Psi^{\prime 0}_{3}
    & =
      (V_\circ^{-2}\eth_\circ V_\circ\widebar\eth_\circ^2V_\circ-V_\circ^{-1}\eth_\circ\widebar\eth_\circ^2V_\circ)
      \big\rvert
      -\eth_{\circ}u'_0\big\rvert\Psi^{\prime 0}_{4}\\
     & =(\widebar\eth_{\circ}\eth_{\circ}\widebar\eth_{\circ}\Phi_\circ+\widebar\eth_{\circ}\Phi_\circ
      -2\widebar\eth_{\circ}\Phi_\circ\eth_{\circ} \widebar\eth_{\circ}\Phi_\circ)\big\rvert
      -\eth_{\circ}u'_0\big\rvert\Psi^{\prime 0}_{4}.
  \end{split}
\end{equation}
From the transformation law (6.87) of~\cite{Barnich:2016lyg},
we get
\begin{equation}
  \label{eq:39}
 \Psi^{\prime 0}_{2}=-\big[ V_\circ^{-3}\big\rvert +2\eth_{\circ}u'_{0}\big\rvert\Psi^{\prime 0}_{3}
    +{(\eth_{\circ}u'_{0} )}^{2}\big\rvert\Psi^{\prime 0}_{4}\big],
\end{equation}
and also
\begin{equation}
  \label{eq:40}
  \begin{split}
  & \Psi^{\prime 0}_{1}=-\big[3\eth_{\circ}u'_{0}\big\rvert\Psi^{\prime 0}_{2}+3{(\eth_{\circ}u'_{0})}^{2}\big\rvert\Psi^{\prime 0}_{3}
                       +{(\eth_{\circ}u'_{0})}^{3}\big\rvert\Psi^{\prime 0}_{4}\big],\\
  & \Psi^{\prime 0}_{0}=-\big[4\eth_{\circ}u'_{0}\big\rvert\Psi^{\prime 0}_{1}+6{(\eth_{\circ}u'_{0})}^{2}\big\rvert\Psi^{\prime 0}_{2}
                       +4{(\eth_{\circ}u'_{0})}^{3}\big\rvert\Psi^{\prime 0}_{3}+{(\eth_{\circ}u'_{0})}^{4}\big\rvert\Psi^{\prime 0}_{4}\big],
  \end{split}
\end{equation}
cf.~equation (6.15) of~\cite{Adamo:2009vu}.

\section{Memory effect for RT waves}\label{sec:memory-effect}

By construction, with the quantities~\eqref{eq:30},~\eqref{eq:35},~\eqref{eq:37}, equations~\eqref{eq:56} hold for RT
waves and so does the evolution equation~\eqref{eq:23}. The explicit verification to show
that the latter is indeed equivalent to the RT evolution equation in the form
of~\eqref{eq:52} is however somewhat tedious.

Similarly, the explicit verification of the reality condition~\eqref{eq:6},
which reduces to
\begin{equation}
  \label{eq:65}
  -2\eth_{\circ}u'_{0}\big\rvert \Psi^{\prime 0}_{3}-{(\eth_{\circ}u'_{0})}^{2}\big\rvert\Psi^{\prime 0}_{4}+\sigma^{\prime 0}\partial_{u'}\widebar\sigma^{\prime 0}+\eth_{\circ}^{\prime 2}\widebar\sigma^{\prime 0}
  -\mathrm{c.c.}=0,
\end{equation}
requires some algebra which involves
\begin{equation}
  \label{eq:57}
  \eth^\prime_{\circ}\widebar\sigma^{\prime 0}=\eth_{\circ}\widebar\eth^{2}_{\circ}u'_{0}\big\rvert -\eth_{\circ}u'_{0}\big\rvert
  \partial_{u'}\widebar\sigma^{\prime 0},
\end{equation}
\begin{multline}
  \label{eq:71}
  \eth^{\prime 2}_{\circ}\widebar\sigma^{\prime 0}=\Big[\eth^2_{\circ}\widebar\eth^2_{\circ}u'_{0}+ 2\eth_{\circ}u'_{0}\big(V_\circ^{-2}\eth_\circ V_\circ\widebar\eth_\circ^2V_\circ-V_\circ^{-1}\eth_\circ\widebar\eth_\circ^2V_\circ)\Big]\Big\rvert
  \\
  -\eth^{2}_{\circ}u'_{0}\big\rvert \partial_{u'}\widebar\sigma^{\prime 0}  -{(\eth_{\circ}u'_{0})}^{2}\big\rvert\Psi^{\prime 0}_{4}.
\end{multline}
We will also need
\begin{equation}
  \label{eq:78}
  \eth^\prime_{\circ}\sigma^{\prime 0}=\eth^{3}_{\circ}u'_{0}\big\rvert -\eth_{\circ}u'_{0}\big\rvert
  \partial_{u'}\sigma^{\prime 0}.
\end{equation}
Putting these results together, the generalized Bondi mass aspect~\eqref{eq:75} for RT
waves is given by
\begin{equation}
  \label{eq:72}
\Psi^{\prime} =\big[V_\circ^{-3}-\eth^2_{\circ}\widebar\eth^2_{\circ}u'_{0}\big]\big\rvert,
\end{equation}
cf.~equation (82) of~\cite{Tafel2000}, while the angular momentum aspect is
\begin{multline}\label{eq:77}
  \widebar \Psi^\prime_{J}=-\Big[3\eth_{\circ}u'_{0}V_\circ^{-3}
  +3{(\eth_{\circ}u'_{0})}^2
  (V_\circ^{-2}\eth_\circ V_\circ\widebar\eth_\circ^2V_\circ
  -V_\circ^{-1}\eth_\circ\widebar\eth_\circ^2V_\circ)\\+{(\eth_{\circ}u'_{0})}^3(V_\circ^{-2}\widebar\eth_\circ^2\partial_u V_\circ
  -V_\circ^{-3}\widebar\eth_\circ^2V_\circ\partial_u V_\circ)
  +\frac 32\eth_\circ^2 u^\prime_0(\eth_{\circ}\widebar\eth^{2}_{\circ}u'_{0} -\eth_{\circ}u'_{0}
  V_\circ^{-1}\widebar \eth_{\circ}^{2}V_\circ  )\\
  +\frac 12 \widebar\eth^2_\circ u^\prime_0(\eth_\circ^3u^\prime_0-\eth_\circ u^\prime_0V_\circ^{-1}\eth_\circ^2V_\circ)
  \Big]\Big\rvert.
\end{multline}

The displacement memory is thus given by
\begin{multline}
  \label{eq:252}
  \mathcal{D}_{[u^\prime_i,u^\prime_f]} \equiv \Delta\widebar\sigma^{\prime 0}= \int^{u'_f}_{u'_{i}}du'\,\lambda^{\prime 0}=
  \Big[\int^{u_f}_{u_{i}}du\, \widebar \eth_{\circ}^{2}V_\circ\Big]\Big\rvert
=\\\Big[\int^{u_f}_{u_{i}}du\,
e^{-\Phi_\circ}[-\widebar\eth_{\circ}^{2}\Phi_\circ
+{(\widebar\eth_{\circ}\Phi_\circ)}^{2}]\Big]\Big\rvert,
\end{multline}
while the energy that is radiated away is
\begin{multline}
  \label{eq:74}
  4\pi \mathcal F_{[u^\prime_i,u^\prime_f]} = \int^{u'}_{u'_{i}}dv'\, |\lambda^{\prime 0}|^{2}
  =  \Big[\int_{u_{i}}^{u}du\, V_\circ^{-1}
  \big|\eth^2_\circ V_\circ\big|^2 \Big]\Big\rvert
  \\=  \Big[\int_{u_{i}}^{u}du\, e^{-\Phi_\circ}
  \big|\eth_{\circ}^{2}\Phi_\circ-{(\eth_{\circ}\Phi_\circ)}^{2}\big|^{2}
  \Big]\Big\rvert,
\end{multline}
with the sum of~\eqref{eq:72} and~\eqref{eq:74} giving the total conserved mass
$4\pi \mathcal M'_{T}$ in~\eqref{eq:68}.

It is now possible to go back to natural RT time $u$. The expressions simplify
since it is now longer necessary to evaluate them in the implicitly defined
Bondi time,
\begin{equation}
  \label{eq:253}
  \boxed{\mathcal{D}_{[u_i,u_f]}= \int^{u_f}_{u_{i}}du\, \widebar \eth_{\circ}^{2}V_\circ}
=\int^{u_f}_{u_{i}}du\,
e^{-\Phi_\circ}[-\widebar\eth_{\circ}^{2}\Phi_\circ
+{(\widebar\eth_{\circ}\Phi_\circ)}^{2}],
\end{equation}
cf.~\cite{Mao_2019}, equation (3.4), while the generalized Bondi mass aspect
becomes
\begin{equation}
  \label{eq:189}
\Psi=V_\circ^{-3}-\eth^2_{\circ}\widebar\eth^2_{\circ}u'_{0},
\end{equation}
and finally, the local mass loss formula is
\begin{equation}
  \label{eq:73}
  \begin{split}
  & \boxed{\Delta\Psi =-4\pi \mathcal F_{[u_i,u_f]}},\\
  &  \boxed{4\pi\mathcal F_{[u_i,u_f]}
    =\int^{u_{f}}_{u_{i}}dv\, V_{\circ}^{-1}|\eth^{2}_{\circ}V_{\circ}|^{2}}
    =\int_{u_{i}}^{u_{f}}dv\, e^{-\Phi_\circ}
    {\big|\eth_{\circ}^{2}\Phi_\circ-{(\eth_{\circ}\Phi_\circ)}^{2}\big|}^{2}.
  \end{split}
\end{equation}
This expression for RT waves in natural RT time for the non-linear memory is the
first novel results of our analysis. In order to evaluate it concretely, one
needs the explicit dependence of $V_{\circ},\Phi_\circ$, on $u$. This will be discussed in
more details in Section~\ref{sec:settl-down-schw} below.

The equivalent version of equation~\eqref{eq:73} before integration,
\begin{equation}
  \label{eq:103}
\partial_{u}\Psi=-V_{\circ}^{-1}|\eth^{2}_{\circ}V_{\circ}|^{2} \iff
  \partial_u V^{-3}_\circ=\eth^2_\circ\widebar\eth^2_\circ V_\circ-V^{-1}_\circ|\eth^2_\circ V_\circ|^2,
\end{equation}
is the RT equation in the form of~\eqref{eq:302} with $n=-3$, while 
\begin{equation}
  \label{eq:112}
  3 e^{4\Phi_\circ}\partial_{u}\Phi_\circ-e^{\Phi_\circ}\eth^2_{\circ}\widebar\eth^2_{\circ}e^{-\Phi_\circ}=-
{\big|\eth_{\circ}^{2}\Phi_\circ-{(\eth_{\circ}\Phi_\circ)}^{2}\big|}^{2},
\end{equation}
is equivalent to the RT equation as written in~\eqref{eq:52}.

Independently of considerations on BMS transformations and transformations of
$u$ and $r$, it follows directly from the definition~\eqref{eq:47} and~\eqref{eq:148a} that $V_\circ$
transforms as
\begin{equation}
  \label{eq:356}
  \boxed{V^\prime_{\circ;g}(u,x^\prime)
  =[e^{E^g_R}(x)V_\circ(u,x)]|_{x=x_{g^{-1}}(x^\prime)}}. 
\end{equation}
and is thus a field of spin and boost weight~$(0,1)$.

Our analysis is somewhat tedious because one seems to be forced, in the
intermediate steps, to express $u$ in Bondi-time $u'$, which is only implicitly
defined. Note however that what is relevant for the geodesic deviation equation,
and also for the memory, is really the rotated frame, and not the change of
coordinates.

Both the change of coordinates and the change of frames are usually tied
together. In the metric approach, once suitable coordinates
$u',r',x^\prime$ are found such that the metric is asymptotically flat (in
the sense of Bondi-Metzner-Sachs), one constructs a Bondi frame which, after a
suitable rescaling of the radial coordinate, is in particular such that
$l=\partial_{r'}$ and $l=du'$. In the context of NU solution space, it has been shown
in~\cite{Barnich:2016lyg} that, when working out the combined frame rotation and coordinate
transformations that leave the solution space invariant, the frame rotation is
determined by the asymptotic change of coordinates. One may however proceed
differently and first work out the frame rotation that preserve the solution
space in terms of the old coordinates. If needed, one may then perform in a
second step the change of coordinates in which the Bondi frame takes the
standard form, but which is however only implicitly defined.

In Appendix~\ref{sec:stay-natur-robins}, we will explicitly construct the asymptotically flat frame, in
which one may compute the memory effect, by starting from the algebraically
special frame while staying in natural RT coordinates throughout.

\section{Manifestly positive mass aspect and supertranslations}\label{sec:supertr-robins-traut}

In order to work out the action of BMS transformations on RT waves, one may use
the standard results in asymptotically flat spacetimes to transform the
quantities computed in Section~\ref{sec:bms-data-rt}. Alternatively, one can also stay in natural
RT coordinates and apply, to the particular case of RT waves, the results
established in~\cite{Barnich:2016lyg}, Sections~6.4 and 6.5, on how BMS transformations act in
the generalized solution space.

In particular, in the context of the generalized solution space, pure supertranslations
originate from the ambiguity present in the definition of Bondi time: rather
than defining $u'_{0}(u,x)$ as in~\eqref{eq:105}, which is a manifestly
positive function that vanishes at $u=0$, $u'_{0}(0,x)=0$, one
may choose
\begin{equation}
  \label{eq:172}
  u'_{N}=\int^{u}_{\hat u}dv\, V_{\circ}=u'_{0}+a,\quad a=-\int^{\hat u}_{0}dv\, V_{\circ}.
\end{equation}
How $\hat u$ is related to the usual supertranslation ambiguity is discussed in
more detail in section 6.4 of~\cite{Barnich:2016lyg}.

Using this ambiguity, another natural definition is to define the Bondi time by
changing the origin and choosing $\hat u(x)=\infty$, $u'_{\infty}=-U_{\circ}$ where
\begin{equation}
  \label{eq:188}
  \boxed{U_{\circ}=\int^{\infty}_{u} dv\, V_{\circ}},
\end{equation}
so that $U_\circ$ is a manifestly positive function that vanishes at infinity,
$U_\circ(\infty,x)=0$. As a consequence, one may replace $u'_{0}$ by $-U_{\circ}$ in all
expressions in the previous section so that the generalized mass aspect
\begin{equation}
  \label{eq:190}
  \boxed{\Psi^{\infty}_{\circ}= V_\circ^{-3}+\eth^2_{\circ}\widebar\eth^2_{\circ}U_{\circ}},
\end{equation}
is also a manifestly positive function. Since its time evolution is given by
\begin{equation}
    \label{eq:191}
 \boxed{\partial_{u}\Psi^{\infty}_{\circ}=-V_{\circ}^{-1}|\eth^{2}_{\circ}V_{\circ}|^{2}},
\end{equation}
it can never increase but only decrease or stay constant. 

When considering the problem as a system of evolution equations for two
independent fields $(V_0>0,U_0>0)$, whose evolution is determined by
equation~\eqref{eq:27} together with
\begin{equation}
  \label{eq:118}
  \partial_{u} U_{\circ}=-V_\circ,
\end{equation}
a bona fide Lyapunov function is given by 
\begin{equation}
  \label{eq:240}
  L(V_\circ,U_\circ)=\Psi^{\infty}_{\circ},
\end{equation}
in addition to the Lyapunov functionals provided by the integral of the Gaussian
curvature~\cite{lukacs1984lyapunov}, its square or the integrated Bondi mass~\cite{Singleton2020,Chrusciel:1992cj}.
Constructing this Lyapunov function in terms of the generalized mass aspect is
another novel result for RT spacetimes.

One might be worried about late-time divergences in the definition of $U_{\circ}$
in~\eqref{eq:188}. Going back to~\eqref{eq:190}, one sees that only the harmonics with $j\geq 2$ of
$U_\circ$ are needed in the definition of the generalized mass aspect $\Psi^{\infty}$.
Denoting by
\begin{equation}
  \label{eq:173}
  \eta^{0,1}_{j\geq k}=\eta^{0,1}-\sum_{j<k}{}_{0,1}Y_{jm}\int_{\mathbb S^{2}}\eta^{0,1}\, {}_{0,1}\widebar Y_{jm},
\end{equation}
a field of spin and boost weights $(0,1)$ with its first $k-1$ (spin and boosted
weighted) spherical harmonics (cf.~Appendix~\ref{sec:spin weighted harmonics})
projected out, and using that the first four harmonics are in the kernel of
$\eth_{\circ}^{2}\widebar\eth^2_\circ$, one may write,
\begin{equation}
  \label{eq:174}
  \Psi^{\infty}_{\circ}= V_{\circ}^{-3}+\eth^2_{\circ}\widebar\eth^2_{\circ}
  U_{\circ j\geq 2},\ \partial_{u}\Psi^{\infty}=-V^{-1}_{\circ}|\eth_{\circ}V_{\circ j\geq 2}|^{2},\
  \partial_{u} U_{\circ j\geq 2}=-V_{\circ j\geq 2}.
\end{equation}
In this form, it is clear that one need only worry about the late-time
convergence of
\begin{equation}
  \label{eq:175}
  U_{\circ j\geq 2}=\int^{\infty}_{u}dv\, V_{\circ j\geq 2}.
\end{equation}

\section{Vacuum RT solutions}\label{sec:absence-news-1}

\subsection{Vacua as time-independent low harmonics}\label{sec:vacua-as-time}

The generalized Bondi mass aspect in natural time $u$, ${\Psi}^{\infty}_\circ$ is a
decreasing function in $u$ that remains constant if and only if there is no
news. We will call news-free solutions $P,\varphi,V_\circ,\Phi_\circ$ to the RT equation ``vacuum''
solutions and denote them by $P_v,\varphi_v,w_v,\phi_v$, with
\begin{equation}
  \label{eq:24a}
P_v=e^{-\varphi_v},\quad  w_v=\frac{P_v}{P_\circ}=e^{-\phi_v}. 
\end{equation}
When using that
\begin{equation}
  \label{eq:139}
  \eth^{2}\varphi+{(\eth\varphi)}^{2}=e^{-2\Phi_\circ}[\eth^{2}_{\circ}\Phi_{\circ}-{(\eth_{\circ}\Phi_\circ)}^{2}]
  =e^{-2\varphi}[\widebar\partial^{2}\varphi-{(\widebar\partial\varphi)}^{2}],
\end{equation}
absence of news, and thus zero mass loss together with a conserved generalized
Bondi mass aspect, is determined by the vanishing of any of the three
expressions,
\begin{equation}
  \label{eq:146}
  \boxed{\eth^{2}_v\varphi_v+{(\eth_v\varphi_v)}^{2}=0\iff \eth^{2}_{\circ}\phi_{v}-{(\eth_{\circ}\phi_{v})}^{2}=0\iff
  \widebar\partial^{2}\varphi_v-{(\widebar\partial\varphi_v)}^{2}=0},
\end{equation}
together with their complex conjugates. The middle of these equations is
equivalent to
\begin{equation}
  \label{eq:93}
  \boxed{\eth_{\circ}^{2}w_v=0},
\end{equation}
and its complex conjugate. Since $w_v$ is of spin and boost weight $(1,0)$,
this is the same condition that selects the $4$-dimensional Lorentz invariant
space of translations among the supertranslations (cf.~equation~\eqref{eq:374}).

When expanding in spherical harmonics,
\begin{equation}
  \label{eq:104}
  V_{\circ}=a^{jm}(u){}_{0,1}Y_{jm},\quad \widebar a^{jm}={(-)}^{m}a^{j,-m},
\end{equation}
and using~\eqref{eq:303} and~\eqref{eq:259}, the vacuum solutions $w_v$ thus only involve the 4 lowest
harmonics. Injecting into~\eqref{eq:27} then implies that $\partial_{u}w_v=0$, so that the four
harmonics in the decomposition of $w_v$ come with $u$-independent coefficients,
\begin{equation}
  \label{eq:119}
  w_v=c^{jm} {}_{0,1} Y_{jm}|_{{j\leq 1}}.
\end{equation}

\subsection{Vacua as a homogeneous space}\label{sec:vacua-as-homogeneous}

In order to clarify the physical interpretation of these vacuum
solutions, an alternative, equivalent, way to solve the no news condition is
instructive. The last equation of~\eqref{eq:146} and its complex conjugate imply that
vacuum solutions $\varphi_v$ are of the form
\begin{equation}
  \label{eq:151}
  \varphi_{v}=k-\ln X_{g},\quad
  \boxed{X_{g}(x)= |a\zeta+b|^{2}+|c\zeta +d|^{2}},
\end{equation}
with $a(u),b(u),c(u),d(u)\in \mathbb C$,
  $ad-bc=1$, $k(u)\in \mathbb R$.
Since
\begin{equation}
  \label{eq:156}
  \partial\widebar\partial \ln X_{g}= X^{-2}_{g},
\end{equation}
it follows from~\eqref{eq:152} that
\begin{equation}
  \label{eq:150}
\partial\widebar\partial\varphi_{v}=-e^{-2k} e^{2\varphi_{v}} \iff  \eth_{v}\widebar\eth_{v}\varphi_{v}=-e^{-2k}.
\end{equation}
Furthermore, since applying $\eth_{v}\widebar\eth_{v}$ a second time yields $0$, the
RT equation~\eqref{eq:36} requires $\partial_{u}\varphi_{v}=0$, or equivalently, $\partial_u P=0$, where
\begin{equation}
  \label{eq:121}
  P_{v}=e^{-k}X_g,
\end{equation}
and in particular, $P_{\circ}=P_{k_{\circ},e}$, $\varphi_{\circ}=\varphi_{k_{\circ},e}$ with $k_{\circ}=\frac 12\ln 2$, and $e$ the identity element in
${\rm SL}(2,\mathbb C)$,
$a=1=d$, $b=0=c$. Defining
\begin{equation}
  \label{eq:176}
  e^{-k}=\frac{1}{\sqrt 2 R}\iff k=\ln{R}+k_{\circ},\quad R_{\circ}=1,
\end{equation}
it follows that
\begin{equation}
  \label{eq:177}
  \boxed{P_{v}=\frac{X_{g}}{\sqrt 2 R},\quad \eth_{v}\widebar\eth_{v}\varphi_{v}=-\frac{1}{2R^{2}},\quad w_v=\frac{X_{g}}{R(|\zeta|^{2}+1)}}.
\end{equation}
In particular,
\begin{equation}
  \label{eq:197}
  P_{R,e}=\frac{|\zeta|^{2}+1}{R\sqrt 2},\quad P_{\circ}=P_{1,e},
\end{equation}
cf.~(4.15.116) of~\cite{Penrose:1984}, with $R$ taken here as a
dimensionless scale factor for the unit round $2$-sphere metric.

The vacuum sector of RT solutions coincides with the vacuum sector of Euclidean
Liouville theory,
\begin{equation}
  \label{eq:164}
  S=\int d\zeta d\widebar\zeta\, [\frac{1}{2}\partial\phi_{L}\widebar\partial \phi_{L}+\frac{\mu}{2\gamma^{2}} e^{\gamma\Phi_{L}}].
\end{equation}
Indeed, the Euler-Lagrange equations of motion agree with~\eqref{eq:150} while the
vanishing of the stress tensor
\begin{equation}
  \label{eq:142}
  T=-\frac 12{(\partial\phi_{L})}^{2}+\frac{1}{\gamma}\partial^{2}\phi_{L},
\end{equation}
and its complex conjugate agree with the last of the no news equations~\eqref{eq:146} if
\begin{equation}
  \label{eq:171}
  \gamma\phi_{L}=\frac 12\varphi_{R,g},\quad \mu=-4e^{-2k}=-\frac{2}{R^{2}}\iff k=-\ln\sqrt{\frac{-\mu}{4}},\, R=\sqrt{-\frac{\mu}{2}}.
\end{equation}

In terms of groups, vacuum solutions form a realization of the multiplicative
group of strictly positive real numbers $\mathbb R^*_+$, and those with
different values of the scale factor may be generated by acting on solutions
with unit scale factor $P_{1,g}$ through multiplication by $R^{-1}$,
\begin{equation}
  \label{eq:381}
  P_{1,g;R^{-1}}=P_{R,g}. 
\end{equation}
Furthermore, under a constant
$g\in \mathrm{SL}(2,\mathbb C)$ transformation that acts as $x^\prime_{\tilde g}(x)$ given explicitly in~\eqref{eq:147a}, it follows from~\eqref{eq:151} that
\begin{equation}
  \label{eq:166}
  X_{g}(x^\prime_{\tilde g}(x))=\frac{X_{g\circ \tilde g}(x)}{|\tilde c\zeta+\tilde d|^{2}}\iff
X_{g}(x_{\tilde g^{-1}}(x^\prime))=\frac{X_{g\circ \tilde g^{-1}}(x^\prime)}{|-\tilde c\zeta^\prime+\tilde a|^{2}},
\end{equation}
and in particular, when $g=e$,
\begin{equation}
  \label{eq:340}
  P_\circ(x^\prime_{\tilde g}(x))=\frac{P_{1,\tilde g}(x)}{|\tilde c\zeta+\tilde d|^{2}}\iff
  P_\circ (x_{\tilde g^{-1}}(x^\prime))=\frac{P_{1,\tilde g^{-1}}(x^\prime)}{|-\tilde c\zeta^\prime+\tilde a|^{2}}. 
\end{equation}
When combining~\eqref{eq:177},~\eqref{eq:148a},\eqref{eq:166}, the vacuum solutions transform simply as
\begin{equation}
  \label{eq:167}
P'_{R,g;\tilde g}(x^\prime)=P_{R,g\circ\tilde g^{-1}}(x^\prime)\iff   \varphi'_{R,g;\tilde g}(x^\prime)=\varphi_{R,g\circ \tilde g^{-1}}(x^\prime).
\end{equation}
This means that {\em vacuum solutions form a realization of}
$\mathrm{PSL}(2,\mathbb C)$.

As a consequence, the vacuum solutions can be obtained from $P_{R,e}$ in~\eqref{eq:197} by
acting with the $\mathrm{SL}(2,\mathbb C)$ transformation $g^{-1}$ described by
$x_{g}(x^\prime)$,
\begin{equation}
  \label{eq:82}
x_{g}(x^\prime):\  \zeta=\frac{a\zeta'+b}{c\zeta'+d},\ \widebar \zeta=\frac{\widebar a\widebar \zeta'+\widebar b}{\widebar c\widebar \zeta'+\widebar d},
\end{equation}
so that
\begin{equation}
  \label{eq:83}
  P'_{R,e;g^{-1}}(x^\prime)=P_{R,g}(x^\prime)\iff \varphi'_{R,e;g^{-1}}(x^\prime)=\varphi_{R,g}(x^\prime).
\end{equation}
The stabilizer is determined by group elements such that
\begin{equation}
  \label{eq:86}
  X_g(
  x^\prime)=X_e(x^\prime)=|\zeta'|^{2}+1,
\end{equation}
which requires 
\begin{equation}
  \label{eq:87}
  |a|^{2}+|c|^{2}=1,\ |b|^{2}+|d|^{2}=1,\
  a\widebar b+c\widebar d=0,\ b\widebar a+d\widebar c=0.
\end{equation}
It is given by $\mathrm{PSU}(2)$, where the associated $\mathrm{SU}(2)$ elements
may be parametrized by one complex parameter $B$ and one real parameter $E_I$ as
\begin{equation}
  \label{eq:84}
  \mathrm{SU}(2)\ni\begin{pmatrix} a & b\\ c& d
  \end{pmatrix}=\begin{pmatrix} \frac{e^{-\imath E_I/2}}{\sqrt{1+|B|^2}}
    & \frac{\widebar Be^{\imath E_I/2}}{\sqrt{1+|B|^2}} \\
    -\frac{Be^{-\imath E_I/2}}{\sqrt{1+|B|^2}} & \frac{e^{\imath E_I/2}}{\sqrt{1+|B|^2}}
  \end{pmatrix}.
\end{equation}
This can be obtained by starting with the parametrization of
$\mathrm{SL}(2,\mathbb C)$ as in~\eqref{eq:45}. The constraints~\eqref{eq:87} then require
\begin{equation}
  e^{E_R}=1+|B|^2,\quad A=-\frac{B}{1+|B|^2}.
\end{equation}
As a consequence, vacuum solutions correspond to a 4 parameter family of
solutions that may be parametrized by $R$ and the parameters contained in the
boosts of $\mathrm{PSU}(2)\backslash \mathrm{PSL}(2,\mathbb C)$. A convenient way to see
this is to use the polar decomposition $g=uh$ for elements of
$\mathrm{SL}(2,\mathbb C)$, with $u$ in $\mathrm{SU}(2)$ and $h$ hermitian,
positive definite and of determinant $1$. (For instance, in an exponential
parametrization using the Pauli matrices $\vec\sigma$,
$u=e^{\frac{-\imath}{2}\vec\theta\cdot\vec\sigma}$, $h=e^{\frac 12 \vec\omega\cdot\vec\sigma}$). When using
$\partial_u\frac{P_v}{P_\circ}=0$ and the independence of the basis elements, it follows
that $R$ and the three boost parameters in $h$ are $u$-independent. Note that
$w_{R,g}$ in equation~\eqref{eq:177} reduces to $R^{-1}$ if $g\in \mathrm{SU}(2)$ and thus
only depends on $h$.

The vacuum sector is thus described by the homogeneous space
$\mathbb R^*_+\times \mathrm{PSU}(2)\backslash \mathrm{PSL}(2,\mathbb C)$ of strictly positive
real numbers times boosts. The explicit relation between $R$ and the boost
parameters on the one hand, and the coefficient of the expansion in spherical
harmonics on the other is given in Appendix~\ref{sec:boosts-conf-fact}.

In the following, a group element will be denoted by 
\begin{equation}
  \label{eq:260}
  v=R,g.
\end{equation}

\subsection{Vacua as boosted Schwarzschild black holes}\label{sec:vacua-as-boosted}

In terms of spacetime solutions, the frame, co-frame and line-element given in~\eqref{eq:9}
and~\eqref{eq:20} with $P=P_{\circ}$ and $U=U_{\circ}=(-\frac 12+r^{-1})$ is the Schwarzschild
solution with $M=1$ in ingoing Eddington-Finkelstein coordinates. Explicitly,
\begin{equation}
  \label{eq:215}
  \begin{split}
    & l=\partial_r=du,\quad n=\partial_u+U_{\circ} \partial_r=-U_{\circ} du+dr,\quad m=r^{-1}P_\circ \widebar\partial=-\frac{r}{P_\circ}d\zeta,\\
    & ds^2=(1-\frac{2}{r})du^2+2dudr-r^2\frac{4d\zeta d\widebar\zeta}{{(|\zeta|^2+1)}^2},
  \end{split}
\end{equation}
with non-vanishing Weyl scalar $\Psi_2=-r^{-3}$ and spin-coefficients
\begin{equation}
  \label{eq:232}
  \begin{split}
    & \rho=-r^{-1},\quad \gamma=\frac 12 r^{-2},\quad \mu=r^{-1}\eth_\circ\widebar\eth_{\circ}\varphi_\circ+r^{-2}=-\frac 12 r^{-1}+r^{-2},\\
    & \alpha=-\frac 12 r^{-1}\widebar\eth_\circ\varphi_\circ=\frac{1}{2\sqrt 2}r^{-1}\widebar\zeta,\quad \beta=\frac 12 r^{-1}\eth_\circ\varphi_\circ=-\frac{1}{2\sqrt 2}r^{-1}\zeta.
  \end{split}
\end{equation}
For later use, note that the Schwarzschild solution with mass parameter $M$
corresponds to $U=U_M=(-\frac 12+r^{-1}M)$ with
\begin{equation}
  \label{eq:215a}
  \begin{split}
    & l=\partial_r=du,\quad n=\partial_u+U_{M} \partial_r=-U_{M} du+dr,\quad m=r^{-1}P_\circ \widebar\partial=-\frac{r}{P_\circ}d\zeta,\\
    & ds^2=(1-\frac{2M}{r})du^2+2dudr-r^2\frac{4d\zeta d\widebar\zeta}{{(|\zeta|^2+1)}^2},
  \end{split}
\end{equation}
so that $\Psi_2=-r^{-3}M$, $\gamma=\frac 12 r^{-2}M$,
$\mu=-\frac 12 r^{-1}+r^{-2}M$ and all other spin coefficients unchanged.

For general vacuum solutions one uses $P=P_v$ and
$U=U_{v}=(-\frac{1}{2R^2}+r^{-1})$ instead, so that
\begin{equation}
  \label{eq:223}
  \begin{split}
    & l=\partial_r=du,\quad n=\partial_u+U_{v} \partial_r=-U_{v} du+dr,\quad m=r^{-1}P_v \widebar\partial=-\frac{r}{P_v}d\zeta,\\
    & ds^2=(\frac{1}{R^2}-\frac{2}{r})du^2+2dudr
      -r^2R^2\frac{4d\zeta d\widebar\zeta}{{(|a\zeta+b|^2+|c\zeta+d|^2)}^2},
  \end{split}
\end{equation}
where the non-vanishing Weyl scalars $\Psi_2=-r^{-3}$ and spin coefficients
$\rho=-r^{-1},\gamma=\frac 12 r^{-2}$ are as before, while
\begin{equation}
  \label{eq:233}
  \begin{split}
    & \alpha=-\frac 12 r^{-1}\widebar\eth_v\varphi_v=\frac 12 r^{-1}\partial P_v
      =\frac{1}{2\sqrt 2}r^{-1}\frac{(|a|^2+|c|^2)\widebar\zeta+a\widebar b+c\widebar d}{R},\\
    &\beta=\frac 12 r^{-1}\eth_v\varphi_v=-\frac 12 r^{-1}\widebar \partial P_v
      =-\frac{1}{2\sqrt 2}r^{-1}\frac{(|a|^2+|c|^2)\zeta+b \widebar a+d\widebar c}{R},\\ & \mu=r^{-1}\eth_v\widebar\eth_v\varphi_v+r^{-2}=-\frac{1}{2R^2}r^{-1}+r^{-2}.
  \end{split}
\end{equation}

General vacuum solutions may be generated from the Schwarzschild solution with
mass parameter
\begin{equation}
  \label{eq:148}
  \boxed{M=R^3},
\end{equation}
through the coordinate transformations $x_g(x^\prime)$ associated to $g^{-1}$ and the
rescaling $u=R^{-1}u^\prime$, $r=Rr^\prime$ together with the class
$\mathrm{III}$ Lorentz rotation
\begin{equation}
  \label{eq:234}
  l'=e^{-E_R}l,\ n'=e^{E_R}n ,\ m'=e^{\imath E_I}m, \ e^{-E_R}=R,\quad
  m'=e^{\imath E_I}m, \  e^{\imath E_I}=\frac{-\widebar c\widebar\zeta+\widebar a}{-c\zeta+a},
\end{equation}
after dropping the primes. In the associated frame, there is no outgoing
radiation $(\Psi_3=0=\Psi_4)$ no shear $(\sigma=0)$, no news ($\lambda=0$), and no incoming
radiation ($\Psi_0=0=\Psi_1$). It follows that {\em the mass parameter of the
  Schwarzschild solution corresponds to the dimensionless radius of the
  celestial sphere to the power $3$}.

Furthermore, the vacuum solutions are locally asymptotically flat at $\scrip$.
Indeed, when performing the reverse of the above transformations on the vacuum
solutions, they are mapped to locally asymptotically flat Schwarzschild black
holes. One can also make general vacuum solutions locally asymptotically flat
at $\scrip$ by a particular case of the BMS-Weyl transformation discussed in
Section~\ref{sec:bms-data-rt} and Appendix~\ref{sec:asympt-frame-rott} which brings the conformal factor $P_v$ to $P_\circ$.
In this case, we have to replace at the beginning of the analysis $P\to P_v$,
$\varphi\to \varphi_v$, $V_\circ \to w_v$, $\Phi_\circ\to \phi_v$, in all formulas. One uses in addition that
$P_v,\varphi_v,w_v,\phi_v$ are time independent and that, by definition, $\eth^2_\circ w_v=0$,
$\eth^2_\circ \phi_v-{(\eth_\circ \phi_v)}^2=0$. Note furthermore that
\begin{equation}
  \label{eq:263}
  \eth_\circ\widebar\eth_\circ w_{vj=1}=-w_{vj=1},\quad {w}_{vj=0}=\frac{1}{2R}
  (|a|^2+|c|^2+|b|^2+|d|^2),
\end{equation}
and also,
\begin{equation}
  \label{eq:262}
\eth_\circ\widebar\eth_\circ w_v=-w_v(\eth_\circ\widebar\eth_\circ\phi_v
  -\eth_\circ\phi_v\widebar\eth_\circ\phi_v),\quad  \eth_\circ\widebar\eth_\circ\phi_v=\frac 12 -\frac{1}{2R^2}{w_v}^{-2}.
\end{equation}
where the last relation follows from equation~\eqref{eq:177}. As a consequence,
\begin{equation}
  \label{eq:265}
  E_0=E_{R0}=-\phi_v,\ u'_0=w_v u,\ B_0=-w^{-1}_v\eth_\circ w_v u'=\eth_\circ\phi_v u',\ A_0=-\eth_\circ \phi_v,
\end{equation}
 and to lower orders,
\begin{equation}
  \label{eq:264}
  \begin{split}
    E&=-\phi_v+2[-\frac 12\eth_\circ\phi_v\widebar\eth_\circ\varphi_\circ+\frac 12 \widebar\eth_\circ \phi_v\eth_\circ\varphi_\circ+\eth_\circ\phi_v\widebar\eth_\circ\phi_v] u' r^{\prime -1}+\mathcal O(r^{\prime -2}),\\
    A&=-\eth_\circ \phi_v+[\eth_\circ\phi_v\eth_\circ \widebar\eth_\circ\phi_v-\frac 12 \eth_\circ \phi_v-{(\eth_\circ\phi_v)}^2\widebar\eth_\circ\phi_v] u' r^{\prime -1}+\mathcal O(r^{\prime -2}),
  \end{split}
\end{equation}
and also
$\chi_1=-w_v^{-1}(\eth_\circ\widebar\eth_\circ w_v) u'$, so that
\begin{equation}
  \label{eq:266}
  \begin{split}
    u&={w_v}^{-1}u' -[\eth_\circ\phi_v\widebar\eth_\circ\phi_v]{w_v}^{-1}{(u')}^2r^{\prime -1}+\mathcal O(r^{\prime -2}), \\
    r&=e^{E_R}[r'+(\eth_\circ\widebar\eth_\circ \phi_v-\eth_\circ\phi_v
       \widebar\eth_\circ\phi_v) u']
       +\mathcal O(r^{\prime -1}),\\
    \zeta&=\zeta' -\eth_\circ\phi_v u' P_\circ r^{\prime -1}+\mathcal O(r^{\prime -2}).
  \end{split}
\end{equation}
The asymptotically flat data corresponding to the $4$-parameter family of vacuum
solutions is
\begin{equation}
  \label{eq:267}
  \begin{split}
    & \sigma^{\prime 0}=0,\quad \lambda^{\prime 0}=0,\quad \Psi^{\prime 0}_4=0=\Psi^{\prime 0}_3,\\
    & \Psi^{\prime 0}_2=-w_v^{-3},\quad \Psi^{\prime 0}_1=3w_v^{-4}u' \eth^\prime_\circ w_v,\quad
      \Psi^{\prime 0}_0=-6 w_v^{-5}{(\eth^\prime_\circ w_v u')}^2,
  \end{split}
\end{equation}
so that, to leading order, there is no outgoing radiation or shear, and no news
but there is incoming radiation in the asymptotically flat frame.

As a consistency check, let us first set $g=e$. In this case, $w_{R,e}=R^{-1}$,
$\phi_v=\ln R$, all subleading terms of the transformations and the frame
rotations vanish. They are explicity given by
\begin{equation}
  \label{eq:147}
  e^E=e^{E_R}=R^{-1},\quad A=0=B, \quad u=Ru^\prime,\quad r=R^{-1}r^\prime,
  \quad \zeta=\zeta^\prime,
\end{equation}
so that the coordinate transformations reduces to a simple rescaling of the time
and the radial coordinate, while the frame rotation reduces to a real class
$\mathrm{III}$ rotation $l^\prime=R l,n^\prime=R^{-1}n$. After dropping the primes, this
transforms~\eqref{eq:223} into~\eqref{eq:215a}, the Schwarzschild solution with mass parameter $M=R^3$,
as it should since this corresponds to the inverse of the rescalings previoulsy
discussed to generate the vacuum solutions from the Schwarzschild black holes.

Once this is done, the transformations associated to $w_{1,g}$ correspond to
applying a boost that belongs to the $\mathrm{BMS}_4$ group at $\scrip$ to a
general Schwarzschild solution with mass parameter $M$ (see also~\cite{Madler:2017qlu} for a
related discussion).

In summary, RT vacuum solutions correspond to boosted Schwarzschild back holes.

\subsection{Vacuum charges}\label{sec:charg-vacc-solut}

In the time-independent vacuum sector, both the improved and the standard
generalized mass aspect reduce to the standard Bondi mass aspect and are given
by
\begin{equation}
  \label{eq:212}
4\pi \mathcal M_B = \Psi= w_v^{-3}=M w_{1,g}^{-3}. 
\end{equation}

\paragraph*{Vacuum BMS supermomentum} BMS supermomentum is the charge associated to an
infinitesimal supertranslation, parametrized by a real smooth ``function'' $T$
on the sphere. For the vacuum sector, it is given by
\begin{equation}
  P= \frac{1}{4\pi }\int_{\mathbb S^2} d^2\Omega\,\Psi\, T =
  M \int_{\mathbb S^2} \frac{d^2\Omega}{4\pi}\,w_{1,g}^{-3}\, T.
\end{equation}
As discussed above, only Bondi four momentum associated to the first four
spherical harmonics $T_{j\leq 1}$ is supertranslation invariant, whereas BMS
supermomentum associated to $T_{j\geq 2}$ may be set to zero by an appropriate
supertranslation.

In order to explicitly perform the integrals, spinorial methods are presumably
the most convenient way to proceed. Alternatively, one may use the
parametrization~\eqref{eq:216} in terms of the radius and a future directed normalized
time-like vector, $w_{1,g}=u_0-u_i n^i$, where $n^i$ denote the components of
the unit radial vector and decompose $T$ into symmetric trace-free (STF)
harmonics rather than spherical harmonics. The STF harmonics are given by
$\hat n^{L}\equiv n^{\langle i_1}\cdots n^{i_\ell\rangle}$, where angle brackets refer to the STF part of
the tensor and we use the multi-index notation $X^L\equiv X^{\langle i_1 \cdots i_\ell\rangle}$ for a
tensor of rank $\ell$. In this basis, the supermomenta are given by
\begin{equation}
	P_L = M\int_{\mathbb S^2} \frac{d^2\Omega}{4\pi}\, w_{1,g}^{-3}{(-\hat{n})}_L.
\end{equation}
In Appendix~\ref{app:STF Integrals}, we explain how to compute such integrals. The integral
at hand corresponds to $k=3$ case in~\eqref{eq:242}, which yields
\begin{equation}
  P_L 
  = M\,\mathcal{I}_{3L}
  = \,\frac{M}{2}\,\frac{\hat{u}_L}{|\vec u|^3}\,Q''_\ell(z),
  \quad z=\frac{u_0}{|\vec u|}.
\end{equation}
In particular, Bondi four-momentum splits into energy $E=P_0$ at $\ell=0$ and
linear momentum $P_i$ at $\ell=1$,
	\begin{equation}
		\boxed{E=Mu_0, \quad P_i=M u_i}. 
	\end{equation}

Higher order supermomenta can be computed in a similar way. Moreover, we can set
$k=3$ in~\eqref{eq:246} to multipole expand the Bondi mass aspect $\mathcal M_B$ as
\begin{equation}
  \label{eq:247}
  \mathcal M_B=M\sum_{\ell=0}^{\infty}
	\,\frac{(2\ell+1)!!}{2\ell!|\vec u|^3}\,
	Q''_\ell(u_0/|\vec u|)\,\hat u_{L}\,	\hat n^{L}=
    M\sum_{\ell=0}^{\infty}
	\,\frac{2\ell+1}{2|\vec u|^{3}}\,
	Q''_\ell(u_0/|\vec u|) P_\ell (u_i n^i).
\end{equation}

\paragraph*{Rest frame}\label{sec:rest-frame}

By definition, the ``rest frame'' for the vacuum solutions is defined by
coordinates $u_r,r_r,x_r$ such that
\begin{equation}
  \label{eq:342}
  P_0=M,\quad P_i=0.
\end{equation}
For a general vacuum solution parametrized by $R$ and $u_i$, it may be obtained
by performing the inverse of the fractional linear transformations and the
associated complex dyad rotation that generates the general vacuum solution from
the general Schwarzschild solution. In these coordinates and in this frame,
$w_v(x_r)=w_{R,e}=R^{-1}$, $u_0=1, u_i=0$, so that in addition to~\eqref{eq:342}, all higher
BMS supermomenta vanish.

\paragraph*{Angular momentum aspect}

The total angular momentum~\eqref{eq:34} of the vacuum solutions vanishes,
\begin{multline}
  \label{eq:97}
  Q_{\mathcal Y,\widebar{\mathcal Y}}=\frac{u' }{8\pi G}\int_{\mathbb S^2}
  \frac{d\zeta\wedge d\widebar\zeta}{\imath P_\circ^2}\big[\mathcal Y (-3w_v^{-4}\eth_\circ w_v)
  + \widebar{\mathcal Y} (-3w_v^{-4}\widebar \eth_\circ w_v)\\
+(\eth_\circ\mathcal Y+\widebar\eth_\circ\widebar{\mathcal Y})w_v^{-3}
  \big]=0,
\end{multline}
in line with the discussion in Appendix C of~\cite{Bonga:2018gzr}. Moreover, the
time-dependent angular momentum aspect~\eqref{eq:312} vanishes
\begin{equation}
  \label{eq:313}
  \widebar \Psi^\prime_{J}(u^\prime)=-3u^\prime w_v^{-4}\eth^\prime_\circ w_v-u^\prime \eth^\prime_\circ w_v^{-3}=0.
\end{equation}

\section{Symmetry actions on RT solutions}\label{sec:expand-around-vacu}

\subsection{Deviation expansion around any vacuum}

Any of the equivalent vacuum solutions of the previous section may be used as an
alternative background instead of $P_{\circ},\varphi_{\circ}$ for the deviation expansion. A
treatment that keeps all of them simultaneously on the same footing proceeds as
follows. Let us denote by
\begin{equation}
  \label{eq:88}
  \varphi=\varphi_{v}+\Phi_{v} \iff V_{v}=\frac{P}{P_{v}}=e^{-\Phi_{v}},
\end{equation}
which implies in particular that
\begin{equation}
  \label{eq:196}
  \Phi_{v}=\Phi_{\circ}-\phi_{v},\quad V_{v}=w_v^{-1}V_{\circ}.
\end{equation}
Since the analog of the two equations~\eqref{eq:62} used to expand the RT equations are
now
\begin{equation}
  \label{eq:198}
  \eth^{2}_{v}\varphi_{v}+{(\eth\varphi_{v})}^{2}=0,\quad \eth_{v}\widebar\eth_{v}\varphi_{v}=-\frac{1}{2R^{2}},
\end{equation}
the relevant equations of
Sections~\ref{sec:rt-solutions},\ref{sec:bms-data-rt},\ref{sec:memory-effect},\ref{sec:supertr-robins-traut} may also be written with the replacements
\begin{equation}
  \label{eq:194}
  P_{\circ},\varphi_{\circ},\eth_{\circ},V_{\circ},\Phi_{\circ}\quad \longrightarrow  P_{v},\varphi_{v},\eth_{v},V_{v},\Phi_{v}
\end{equation}
and suitable factors of $w_v$ and/or $R$. For instance,
\begin{equation}
  \label{eq:199}
  \eth\widebar\eth\varphi=\eth_v V_v\widebar\eth_v V_v -V_v \eth_v\widebar\eth_v V_v-\frac{1}{2R^2} V_v^2
  =e^{-2\Phi_{v}}(\eth_{v}\widebar\eth_{v}\Phi_{v}-\frac{1}{2R^{2}}),
\end{equation}
and
\begin{equation}
  \label{eq:218}
  \quad \eth^{2}_{v}\widebar\eth^{2}_{v}\eta^0=\big[{(\eth_{v}\widebar\eth_{v})}^2+\frac{1}{R^2}\eth_{v}\widebar\eth_{v}\big]\eta^0,
\end{equation}
while equation~\eqref{eq:302} and~\eqref{eq:52} become
  \begin{equation}
	\label{eq:133}
    \frac{3}{n} \partial_{u}V^n_{v}=-V_v^{3+n}\eth^{2}_{v}\widebar\eth^{2}_{v}V_{v}+V_{v}^{2+n}|\eth^{2}_{v} V_{v}|^{2},
  \end{equation}
\begin{multline}
  \label{eq:200}
  3e^{4\Phi_v}\, \partial_{u}\Phi_v=-\eth^2_{v}\widebar\eth^2_{v}\Phi_v
+\frac{2}{R^{2}}\eth_{v}\Phi_v\widebar\eth_{v}\Phi_v   +2\widebar\eth_{v}\Phi_v \eth_{v}\widebar\eth_{v}\eth_{v} \Phi_v
  \\ +
  2\eth_{v}\Phi_v\widebar\eth_{v}\eth_{v}
  \widebar\eth_{v}\Phi_v+2\eth_{v}\widebar\eth_{v}\Phi_v \eth_{v}\widebar\eth_{v}\Phi_v
  -4\eth_{v}\Phi_v\widebar\eth_{v}\Phi_v\eth_{v}\widebar\eth_{v}\Phi_v.
\end{multline}

In order to relate previous computations around the ``round'' vacuum to the
current expansion around a general one, the following relations are useful:
\begin{equation}
  \label{eq:99}
  \eth_\circ \eta^s=w_v^{-1}(\eth_v -sw_v^{-1}\eth_v w_v)\eta^s,\quad \widebar\eth_\circ \eta^s
  =w^{-1}_v(\widebar\eth_v +sw_v^{-1}\widebar\eth_v w_v)\eta^s,
\end{equation}
In particular, the defining equation~\eqref{eq:93} for vacuum solutions translates into
\begin{equation}
  \label{eq:8}
  w_v^{-1}\eth^2_v w_v=2{(w_v^{-1}\eth_v w_v)}^2 \iff \eth^2_v w_v^{-1}=0. 
\end{equation}
which implies in turn that 
\begin{equation}
  \label{eq:202}
  \begin{split}
& \eth^{2}_{\circ} V_{\circ}=w_v^{-1}\eth^{2}_{v}V_{v},\quad \widebar\eth^{2}_{\circ} V_{\circ}=w_v^{-1}\widebar\eth^{2}_{v}V_{v},\\
& \eth^{2}_{\circ}\widebar\eth^{2}_{\circ}V_{\circ}=w_v^{-3}\eth^{2}_{v}\widebar\eth^{2}_{v}V_{v},\quad V_{\circ}^{-1}|\eth^{2}_{\circ}V_{\circ}|^{2}
  =w_v^{-3}V_{v}^{-1}|\eth^{2}_{v}V_{v}|^{2}.
  \end{split}
\end{equation}
If 
\begin{equation}
  \label{eq:205}
  U_{v}=\int^{\infty}_{u}dv\, V_{v}=w_v^{-1}U_{\circ},\quad \Psi^\infty_v=V_v^{-3}+\eth^{2}_{v}\widebar\eth^{2}_{v}U_{v},
\end{equation}
it follows that
\begin{equation}
  \label{eq:204}
  \Psi^{\infty}_{\circ}=w_v^{-3}\Psi^\infty_v,\quad \partial_{u}\Psi^{\infty}_{v}=-V_{v}^{-1}|\eth^{2}_{v}V_{v}|^{2}.
\end{equation}

\subsection{Action of $\mathrm{SL}(2,\mathbb C)$ on RT solutions}\label{sec:acti-resc-boost}

Under $\mathrm{SL}(2,\mathbb C)$, solutions to the RT equation~\eqref{eq:302} are mapped to
solutions. More explicitly, when taking into account that $V_\circ$ is of spin and
boost weight $(0,1)$ (in the form adapted to the inverse transformation as in
equation~\eqref{eq:347}), it follows that
\begin{equation}
  \label{eq:357}
  V^\prime_{\circ;g^{-1}}(u,x^\prime)
  =e^{-E^g_R}(x^\prime)V_{\circ}(u,x_g(x^\prime)).
\end{equation}
Using~\eqref{eq:380} (with $x^\prime\to x$, $\tilde x\to x^\prime$,
$g\to g^{-1}$ and using also~\eqref{eq:132}), we get
\begin{equation}
  \label{eq:382}
  \begin{split}
    \eth^{\prime 2}_\circ V^\prime_{\circ;g^{-1}}(u,x^\prime)
    &=e^{E^g_R-2\imath E^I_g}(x^\prime)
      \big(\eth^2_\circ V_\circ \big)(u,x_g(x^\prime)),\\
    \widebar\eth^{\prime 2}_\circ V^\prime_{\circ;g^{-1}}(u,x^\prime)
    &=e^{E^g_R+2\imath E^I_g}(x^\prime)\big(\widebar\eth^2_\circ V_\circ \big)
      (u,x_g(x^\prime)),\\
    \eth^{\prime 2}_\circ \widebar\eth^{\prime 2}_\circ
    V^\prime_{\circ;g^{-1}}(u,x^\prime)
    &=e^{3E^g_R}(x^\prime)\big(\eth^2_\circ\widebar\eth^2_\circ V_\circ \big)(u,x_g(x^\prime)).
  \end{split}
\end{equation}
It follows that
\begin{multline}
  \label{eq:390}
  \frac{3}{n}\partial_{u}V^{\prime n}_{\circ;g^{-1}}+V^{\prime 3+n}_{\circ;g^{-1}}\eth^{\prime 2}_\circ\widebar\eth^{\prime 2}_\circ V^\prime_{\circ;g^{-1}}-V^{\prime 2+n}_{\circ;g^{-1}}|\eth^{\prime 2}_\circ V^\prime_{\circ;g^{-1}}|^2\\=
  e^{-nE^g_R}(x^\prime)\Big[\frac{3}{n} \partial_u V^n_\circ+ V_\circ^{3+n}\eth^{2}_\circ\widebar\eth^{2}_\circ V_\circ-V_\circ^{2+n}|\eth^{ 2}_\circ  V_\circ|^2\big](u,x_g(x^\prime)),
\end{multline}
which proves the result.

In an expansion in terms of spin and boost weighted spherical harmonics, 
\begin{equation}
  \label{eq:358}
  V_\circ(u,x)=a^{jm}(u){}_{0,1}Y_{jm}(x),
\end{equation}
is a solution to the RT equation iff
\begin{equation}
  \label{eq:385}
  V^\prime_{\circ;g^{-1}}(u,x^\prime)=a^{jm}(u)w_{1,g}(x^\prime){}_{0,1}Y_{jm}(x_g(x^\prime)),
\end{equation}
is a solution to RT equation in terms of primed variables. Note that only
electric multipoles appear here because $V_\circ$ is real.

Solutions to the RT equation may thus be classified in terms of equivalence
classes of solutions related by this $\mathrm{SL}(2,\mathbb C)$ transformations.

\subsection{Action of scaling symmetry}\label{sec:acti-scal-symm}

Defining $u=R^{-4}u^\prime$, the scaling transformation
\begin{equation}
  \label{eq:138}
  V^\prime_{\circ; R^{-1}}(u^\prime,x)=R^{-1}V_{\circ}(R^{-4}u^\prime,x), 
\end{equation}
maps solutions to solutions. Indeed, the RT equations~\eqref{eq:302} are covariant in the sense that 
\begin{multline}
  \label{eq:227}
  \frac{3}{n}\partial_{u^\prime}V^{\prime n}_{\circ; R^{-1}}
  +V^{\prime 3+n}_{\circ;R^{-1}}\eth^{2}_\circ\widebar\eth^{ 2}_\circ V^\prime_{\circ;R^{-1}}-V^{\prime 2+n}_{\circ;R^{-1}}|\eth^{2}_\circ V^\prime_{\circ;R^{-1}}|^2\\=
  R^{-4-n}\Big[\frac{3}{n} \partial_u V^n_\circ+ V_\circ^{3+n}\eth^{2}_\circ\widebar\eth^{2}_\circ V_\circ-V_\circ^{2+n}|\eth^{ 2}_\circ  V_\circ|^2\big](
  R^{-4}u^\prime,x),
\end{multline}
so that now a solution~\eqref{eq:358} corresponds to a solution 
\begin{equation}
  \label{eq:82a}
  V^\prime_{\circ;R^{-1}}(u^\prime,x)=a^{jm}(R^{-4}u^\prime)R^{-1}{}_{0,1}Y_{jm}(x).
\end{equation}

The combination of the two transformations will be denoted by
\begin{equation}
  \label{eq:268}
  V^\prime_{\circ; v^{-1}}(u^\prime,x^\prime)=w_v(x^\prime)V_{\circ}(R^{-4}u^\prime,x_g(x^\prime)), 
\end{equation}
and is a solution to~\eqref{eq:302} in terms of the primed coordinates.

\subsection{Generating solutions around an arbitrary
  vacuum}\label{sec:gener-solut-around}

As a corollary, one may generate solutions around an arbitray vacuum from
solutions around the round one by a suitable scaling and boost. More precisely,
a solution $V_v(u^\prime,x^\prime)$ to the RT equation~\eqref{eq:133} (in terms of primed
coordinates) may be obtained from a solution $V_\circ(u,x)$ to the RT equation~\eqref{eq:302}
by a scaling by $R^{-1}$ of $V_\circ(u,x)$ together with a change of the time
coordinate, $u=R^{-4}u^\prime$, and a boost associated to $g^{-1}$ that acts as
$x_g(x^\prime)$ in~\eqref{eq:82}.

Equation~\eqref{eq:268} may be re-expressed as
\begin{equation}
  \label{eq:344}
  V^\prime_{\circ;v^{-1}}(u^\prime,x^\prime)=w_v(x^\prime) V_v(u^\prime,x^\prime),\quad  V_v(u^\prime,x^\prime)=V_{\circ}(R^{-4}u^\prime,x_g(x^\prime)).
\end{equation}
Furthermore, using~\eqref{eq:202} (in terms of primed variables) on the left hand side, yields
after simplification by a factor of ${w_v(x^\prime)}^{-1}=Re^{E^g_R}(x^\prime)$ for the
first two equations and of $w_v(x^\prime)^{-3}$ for the last equation,
\begin{equation}
  \label{eq:383}
  \begin{split}
    \eth^{\prime 2}_v V_v(x^\prime)&=R^{-2}e^{-2\imath E^I_g}(x^\prime)\big(\eth^2_\circ V_\circ \big)(R^{-4}u^\prime,x_g(x^\prime)),\\
    \widebar\eth^{\prime 2}_v V_v(x^\prime)&=R^{-2}e^{-2\imath E^I_g}(x^\prime)\big(\widebar\eth^2_\circ V_\circ \big)(R^{-4}u^\prime,x_g(x^\prime)),\\
    \eth^{\prime 2}_v \widebar\eth^{\prime 2}_v V_v(x^\prime)&=R^{-4}\big(\eth^2_\circ\widebar\eth^2_\circ V_\circ \big)(R^{-4}u^\prime,x_g(x^\prime)).
  \end{split}
\end{equation}
As a consequence, the RT equation~\eqref{eq:133} (in terms of primed coordinates) becomes
\begin{multline}
  \label{eq:384}
  \frac{3}{n}\partial_{u^\prime}V^n_v+V_v^{3+n}\eth^{\prime 2}_v\widebar\eth^{\prime 2}_v V_v-V_v^{2+n}|\eth^{\prime 2}_v V_v|^2=\\
  R^{-4}\Big[\frac{3}{n} \partial_u V_\circ
  + V_\circ^{3+n}\eth^{2}_\circ\widebar\eth^{2}_\circ V_\circ-V_\circ^{2+n}|\eth^{ 2}_\circ  V_\circ|^2\big](R^{-4}u^\prime,x_g(x^\prime)),
\end{multline}
which proves the result.

In terms of spherical harmonics and their eigenvalues, one may focus on
what happens under a scaling. In this case, $w_{R,e}=R^{-1}$,
$P_{R,e}=R^{-1}P_\circ$. The spin and boost weighted spherical harmonics are
scaled by a factor of $R^{-1}$, ${}_{sw} Y^{R,e}_{jm}=R^{-1}{}_{sw}Y_{jm}$ so
that the orthonormality relations~\eqref{eq:379} hold in terms of
${}_{sw} Y^{R,e}_{jm}$ and the measure
$\frac{2d\zeta d\widebar\zeta}{\imath P^2_{R,e}}$. Furthermore, the actions
$\eth_{R,e}\, {}_s Y^{R,w}_{jm},\widebar\eth_{R,e}\, {}_s Y^{R,e}_{jm}$ are
given by $R^{-1}$ times the right hand sides of~\eqref{eq:304} with
${}_s Y_{jm}$ is replaced by ${}_{sw} Y^{R,e}_{jm}$, cf.~equation (4.15.106)
of~\cite{Penrose:1984}. Since each of the two terms on the right hand side of
the RT equations involves $4$ spatial derivatives, this leads to a power of
$R^{-4}$ in the eigenvalues.

To a solution $V_\circ(u,x)$ of the form~\eqref{eq:358} to the RT equation~\eqref{eq:302} corresponds a
solution
\begin{equation}
  \label{eq:371}
  V_v(u^\prime,x^\prime)=a^{jm}(R^{-4}u^\prime){}_{0,1}Y_{jm}(x_g(x^\prime)). 
\end{equation}
to the RT equation~\eqref{eq:133} in terms of primed coordinates.

In other words, even if there is no reason to privilege the round vacuum for a
perturbation expansion over any other vacuum solution at the outset, results are
simply related by applying the appropriate transformation.

\subsection{Initial conditions and rest frame}
\label{sec:init-cond-inst}

We want to show that one may assume that there are no low harmonics with
$j\leq 1$ in the initial conditions by performing a suitable fractional linear
transformation and a scaling of $u$. Indeed, let
\begin{equation}
  \label{eq:337}
  V_\circ= V_{\circ j\leq 1}+ V_{\circ j\geq 2}.
\end{equation}
According to~\eqref{eq:225},~\eqref{eq:351}, there exist a unique $R$ and
$3$ boost parameters contained in $g$ such that
\begin{equation}
  \label{eq:338}
  V_{\circ j\leq 1}(0,x)\equiv a^{jm}(0)\,{}_{0,1}Y_{jm}\big\rvert_{j\leq 1}
  =w_{v}(x).
\end{equation}
When performing a scaling associated to $R$ and a fractional linear
transformation associated to $g$, that acts as $x_{g^{-1}}(x^\prime)$ and taking into
account~\eqref{eq:132}, one gets from~\eqref{eq:344},
\begin{equation}
  \label{eq:386}
  V^\prime_{\circ;v}(0,x^\prime)
  =1+w_{v^{-1}}(x^\prime)V_{\circ j\geq 2}(0,x_{g^{-1}}(x^\prime)).
\end{equation}
Hence, when taking into account the discussion in
Section~\ref{sec:acti-resc-boost},~\ref{sec:acti-scal-symm},
\begin{equation}
  \label{eq:214}
  V^\prime_{\circ;v}(u^\prime,x^\prime)=a^{jm}(R^4u^\prime)w_{v^{-1}}(x^\prime){}_{0,1}Y_{jm}(x_{g^{-1}}(x^\prime)),
\end{equation}
is a solution to RT equation in terms of primed variables that does not contain
low harmonics in the initial conditions in the sense that
$a^{jm}(0)w_{v^{-1}}(x^\prime)Y_{jm}(x_{g^{-1}}(x^\prime))|_{j\leq 1}=1$.

By covariance and duality, the transformation that turns off the low harmonics
in $V_\circ$ in the above sense also turns off the low harmonics in $V_\circ^{-3}$ and
puts the system in its rest frame. Indeed, if
\begin{equation}
  \label{eq:392}
  \begin{split}
    & 4\pi P_0(u)=\int_{\mathbb S^2} d^2\Omega\, \Psi^\infty_\circ=\int_{\mathbb S^2} {d^2\Omega}\, V_\circ^{-3},\\
    & 4\pi P_i(u)=\int_{\mathbb S^2}d^2\Omega\, \Psi^\infty_\circ (-n_i)=\int_{\mathbb S^2}d^2\Omega\, V_\circ^{-3} (-n_i),
  \end{split}
\end{equation}
we may write
\begin{equation}
  \label{eq:393}
  4\pi P_a(u)u^a= \int_{\mathbb S^2} d^2\Omega\, V_\circ^{-3}w_{1,g},
\end{equation}
by using the parametrization of $w_v$ as in~\eqref{eq:216} and the
$u^a$ explicitly given in~\eqref{eq:243}. When evaluating at $u=R^4u^\prime$, performing the
change of variables $x=x_{g^{-1}}(x^\prime)$ and
using~\eqref{eq:268},~\eqref{eq:391} with $R\to R^{-1},g\to g^{-1}$ as well
as~\eqref{eq:132}, it follows that
\begin{equation}
  \label{eq:396}
  4\pi P_a(R^4u^\prime)u^a= R^3\int d^2\Omega^\prime V^{\prime -3}_{\circ;v}(u^\prime,x^\prime)= 4\pi P^\prime_a(u^\prime) u^{\prime a}_{;v},
\end{equation}
with $u^{\prime a}_{;v}=\delta^a_0$. In other words, the primed coordinates define the
initial rest frame since $P^\prime_i(0)=0$. The normalization $4\pi P^\prime_0(0)=M^\prime$
is achieved by fixing $R$ so that
\begin{equation}
  \label{eq:395}
  M^\prime =R^3 \int d^2\Omega^\prime V^{\prime -3}_{\circ;v}.
\end{equation}
The system does not stay in its rest frame on account of the news. Indeed, when
using the RT equation for $V^{\prime -3}_{\circ;v}$, i.e., the LHS of~\eqref{eq:390}, with $n=-3$
and $\tilde v^{-1}\to v$, equals $0$, one finds
\begin{equation}
  \label{eq:397}
  4\pi \partial_{u^\prime} P^\prime_a(u^\prime)u^a=-\int d^2\Omega^\prime\, V^{\prime -1}_{\circ;v}|\eth^{\prime 2}_\circ V^\prime_{\circ;v}|^2. 
\end{equation}

\section{Collective coordinates and instantaneous rest frame}\label{sec:coll-coord}

As proposed in~\cite{Singleton_1990,Singleton2020,Chrusciel:1992cj,Chrusciel:1992rv,Chrusciel:1992tj}, the low harmonics may be
controlled by allowing the parameters characterizing the vacuum solutions to
depend on time and by fixing their time-dependence in a suitable way.

The way this plays out from the current perspective is that the dynamics of the
group elements may be fixed by the requirement that the system should remain in
its instantaneous rest frame with a fixed normalization condition set by the
mass.

\paragraph*{Scalings and zero mode}

Let us start by dealing with the zero mode. This can be done through the change
of time defined by
\begin{equation}
  \label{eq:193}
  u'(u)=\int^u_0 dv\, \tilde R^4(v), 
\end{equation}
which implies in particular that
$ \partial_{u^\prime}f(u(u^\prime))=(\tilde R^{-4}\partial_u f)(u(u^\prime))$. 
If one defines
\begin{equation}
  \label{eq:256}
  V^\prime_{\circ;\tilde R^{-1}(u)}(u^\prime,x)=\frac{P^\prime_{\circ;\tilde R^{-1}}}{P_\circ}
  =\big(\tilde R^{-1}V_\circ\big)(u(u^\prime),x), 
\end{equation}
the modified RT equations equivalent to~\eqref{eq:302} are
\begin{multline}
  \label{eq:238}
  \frac{3}{n} \partial_{u^\prime} V^{\prime n}_{\circ;\tilde R^{-1}(u)}
  =-V^{\prime 3+n}_{\circ;\tilde R^{-1}(u)}\eth_\circ^2\widebar\eth_\circ^2V^\prime_{\circ;\tilde R^{-1}(u)}
  +V^{\prime 2+n}_\circ|\eth_\circ^{2}V^\prime _{\circ;\tilde R^{-1}(u)}|^2
                      \\-3 V^{\prime n}_{\circ;\tilde R^{-1}(u)}  (\tilde R^{-1}\partial_{u}\tilde R)(u(u^\prime)).
\end{multline}

\paragraph*{Boosts and $j=1$ harmonics}

In order to control the harmonics at $j=1$, one considers time-dependent
fractional linear transformation $\tilde g^{-1}(u)$,
\begin{equation}
  \label{eq:183}
  \zeta^\prime =\frac{\tilde d(u)\zeta -\tilde b(u)}{-\tilde c(u)\zeta+a(u)} \iff
  \zeta=\frac{\tilde a(u)\zeta^\prime+\tilde b(u)}{\tilde c(u)\zeta^\prime+\tilde d(u)}, 
 \end{equation}
under which
\begin{multline}
  \label{eq:273}
  P^\prime_{;\tilde g^{-1}(u)}(u,x^\prime)=P(u,x_{\tilde g(u)}(x^\prime))
  |\frac{\partial\zeta^\prime}{\partial\zeta}|\\ \iff \varphi^\prime_{;\tilde g^{-1}(u)}(u,x^\prime)=\varphi(u,x_{\tilde g(u)}(x^\prime))-\ln |\frac{\partial\zeta^\prime}{\partial\zeta}|. 
\end{multline}
Starting from
\begin{equation}
  \label{eq:286}
  \partial_u\varphi^\prime_{;\tilde g^{-1}(u)}(u,x^\prime)=(\partial_u \varphi)|+
  \big[(\partial \varphi)|\partial_u\zeta(\zeta^\prime)-\partial_u\ln{(\tilde c\zeta^\prime +\tilde d)}
  +\mathrm{c.c.}\big],
\end{equation}
and using the same reasoning that has lead to~\eqref{eq:134} for the first term
on the right hand side, as well as the result~\eqref{eq:181} for the second term, this
yields

{\em $\varphi$ satisfies the RT equation~\eqref{eq:36} if and only if
    $\varphi^\prime_{;\tilde g^{-1}(u)}$ satisfies the modified RT equation}, 
\begin{equation}
  \label{eq:187}
  3 \partial_u\varphi^\prime_{;\tilde g^{-1}(u)} =-{(\eth^\prime\widebar\eth^\prime)}^2\varphi^\prime_{;\tilde g^{-1}(u)}
  +\frac{3}{2} (\eth^\prime\mathcal Y^\prime+\widebar\eth^\prime \widebar{\mathcal{Y}}^\prime).  
\end{equation}
If one defines 
\begin{equation}
  \label{eq:256a}
  V^\prime_{\circ;\tilde g^{-1}(u)}(u,x^\prime)=\frac{P^\prime_{;\tilde g^{-1}(u)}(u,x^\prime) }{P_\circ(x^\prime)}
\end{equation}
it follows that
\begin{equation}
  \label{eq:154}
  V^\prime_{\circ;\tilde g^{-1}(u)}(u,x^\prime)=e^{-E^{\tilde g(u)}_R}V_\circ(u,x_{\tilde g(u)}(x^\prime))=
  w_{1,\tilde g(u)}(x^\prime)
  V_\circ(u,x_{\tilde g(u)}(x^\prime)).
\end{equation}
When using~\eqref{eq:282}, the modified RT equations equivalent to~\eqref{eq:302} are now
\begin{multline}
  \label{eq:182}
  \frac{3}{n} \partial_{u} V^{\prime n}_{\circ;\tilde g^{-1}(u)}
  =-V^{\prime 3+n}_{\circ;\tilde g^{-1}(u)}\eth_\circ^{\prime 2}\widebar\eth_\circ^{\prime 2}V^\prime_{\circ;\tilde g^{-1}(u)}+V^{\prime 2+n}_{\circ;\tilde g^{-1}(u)}|\eth_\circ^{\prime 2}V^\prime _{\circ;\tilde g^{-1}(u)}|^2
  \\
  -3 V^{\prime n}_{\circ;\tilde g^{-1}(u)} \big[
  \frac 12(
  \eth^\prime_\circ\mathcal Y^\prime_{\tilde g^{-1}(u)\circ}+\widebar\eth^\prime_\circ \widebar{\mathcal{Y}}^\prime_{\tilde g^{-1}(u)\circ})
  +\mathcal Y^\prime_{\tilde g^{-1}(u)\circ}
  \eth^\prime_\circ \Phi^\prime_{\circ;\tilde g^{-1}(u)}\\+\widebar{\mathcal{Y}}^\prime_{\tilde g^{-1}(u)\circ}\widebar\eth^\prime_\circ\Phi^\prime_{\circ;\tilde g^{-1}(u)}\big]. 
\end{multline}
For $n=-2$, this is the modified flow associated to the Calabi flow~\eqref{eq:125} in
conformally flat coordinates, which can be re-written as
\begin{multline}
  \label{eq:290}
  -\frac 32  \partial_{u} V^{\prime -2}_{\circ;\tilde g^{-1}(u)}= \frac 12
  \eth^\prime_\circ\big(\eth^\prime_\circ V^\prime_{\circ;\tilde g^{-1}(u)}\widebar\eth^{\prime 2}_\circ V^\prime_{\circ;\tilde g^{-1}(u)}
  -V^\prime_{\circ;\tilde g^{-1}(u)} \eth^\prime_\circ \widebar\eth^{\prime 2}_\circ V^\prime_{\circ;\tilde g^{-1}(u)} -3 V^{\prime -2}_{\circ;\tilde g^{-1}(u)}\mathcal Y^\prime_{\tilde g^{-1}(u)\circ}   \big)\\
  +\frac 12 \widebar\eth^\prime_\circ\big(\widebar\eth^\prime_\circ V^\prime_{\circ;\tilde g^{-1}(u)}\eth^{\prime 2}_\circ V^\prime_{\circ;\tilde g^{-1}(u)}
  -V^\prime_{\circ;\tilde g^{-1}(u)}\widebar\eth^\prime_\circ \eth^{\prime 2}_\circ V^\prime_{\circ;\tilde g^{-1}(u)}  -3 V^{\prime -2}_{\circ;\tilde g^{-1}(u)} \widebar{\mathcal{Y}}^\prime_{\tilde g^{-1}(u)\circ}  \big).
\end{multline}
As a consequence, the normalization condition on the volume is also compatible
with this modified flow: if the zero mode of $V^{\prime -2}_\circ$ is $1$ initially, it
remains $1$.

The modified flow corresponding to~\eqref{eq:249} is then
\begin{multline}
  \label{eq:299}
  3\partial_u f^\prime_{;\tilde g^{-1}(u)} +\eth^{\prime 2}_\circ \widebar\eth^{\prime 2}_\circ f^\prime_{;\tilde g^{-1}(u)}=
  3\eth_\circ^\prime [(1+f^\prime_{;\tilde g^{-1}(u)})\mathcal Y^\prime_{\tilde g^{-1}(u)\circ}]\\+3\widebar\eth_\circ^\prime [(1+f^\prime)\widebar{\mathcal Y}^\prime_{\tilde g^{-1}(u)\circ}]+3g^\prime(f^\prime_{;\tilde g^{-1}(u)}).
\end{multline}
By suitably adjusting the
parameters in $\mathcal Y^\prime_\circ$, this has been used to prove results on the
non-linear convergence towards a Schwarzschild black hole. 

Finally, if one combines scalings with boosts 
it follows that
\begin{equation}
  \label{eq:399}
  V^\prime_{\circ;\tilde v^{-1}(u)}(u^\prime,x^\prime)=
  \big[w_{\tilde v(u)}(x^\prime)
  V_\circ(u,x_{\tilde g(u)}(x^\prime))\big]\big\rvert_{u=u(u^\prime)},
\end{equation}
and the modified equations are
\begin{multline}
  \label{eq:394}
  \frac{3}{n} \partial_{u^\prime} V^{\prime n}_{\circ;\tilde v^{-1}(u)}
  =-V^{\prime 3+n}_{\circ;\tilde v^{-1}(u)}\eth_\circ^{\prime 2}\widebar\eth_\circ^{\prime 2}V^\prime_{\circ;\tilde v^{-1}(u)}+V^{\prime 2+n}_{\circ;\tilde v^{-1}(u)}|\eth_\circ^{\prime 2}V^\prime _{\circ;\tilde v^{-1}(u)}|^2
  \\
  -3 V^{\prime n}_{\circ;\tilde v^{-1}(u)} \big[\tilde R^{-1}\partial_{u}\tilde R+
  \frac 12(
  \eth^\prime_\circ\mathcal Y^\prime_{\tilde g^{-1}(u)\circ}+\widebar\eth^\prime_\circ \widebar{\mathcal{Y}}^\prime_{\tilde g^{-1}(u)\circ})
  +\mathcal Y^\prime_{\tilde g^{-1}(u)\circ}
  \eth^\prime_\circ \Phi^\prime_{\circ;\tilde v^{-1}(u)}\\
  +\widebar{\mathcal{Y}}^\prime_{\tilde g^{-1}(u)\circ}\widebar\eth^\prime_\circ\Phi^\prime_{\circ;\tilde v^{-1}(u)}\big](u(u^\prime)). 
\end{multline}

\paragraph{Instantaneous rest frame}

The instantaneous rest frame is defined by following Section~\ref{sec:init-cond-inst}, except that
the low harmonics at each $u$ are parametrized by $w_{v(u)}$ through $u$-dependent
parameters $v(u)=R(u),g(u)$, respectively through $v^a(u)$. The system is put in
its instantaneous rest frame by performing the appropriate transformations with $u$
dependent parameters, so that equation~\eqref{eq:396} becomes
\begin{multline}
  \label{eq:157}
  4\pi P_a(u(u^\prime))u^a(u(u^\prime))=R^3(u(u^\prime))\int d^2\Omega^\prime V^{\prime -3}_{\circ;v(u)}(u^\prime,x^\prime)\\
  = 4\pi P^\prime_a(u^\prime) u^{\prime a}(u(u^\prime)),\quad u^{\prime a}_{;v(u)}(u(u^\prime))=\delta^a_0,
\end{multline}
in terms of $V^{\prime -3}_{\circ;v(u)}(u^\prime,x^\prime)$ that satisfies the modified RT
equation~\eqref{eq:394} at $n=-3$ (with $\tilde v^{-1}(u)$ replaced by $v(u)$). We now have
\begin{equation}
  \label{eq:163}
  P^\prime_i(u^\prime)=0,
\end{equation}
at each $u^\prime$ while the radius is fixed through the normalization condition
\begin{equation}
  \label{eq:161}
M^\prime  =R^{-3}(u(u^\prime))\int d^2\Omega^\prime V^{\prime -3}_{\circ;v(u)}(u^\prime,x^\prime). 
\end{equation}
These four equations defining the normalized instantaneous rest frame fix the
dynamics of the four collective coordinates $v(u)$. At the same time, time evolution of
$V^{\prime -3}_{\circ;v(u)}(u^\prime,x^\prime)$ is governed by the modified RT
equation.

In the next section, we will explicitly implement these conditions in
perturbation theory.

\section{Evolution}\label{sec:settl-down-schw}

\subsection{General structure}\label{sec:general-structure}

The RT equations for $V_\circ,\Phi_\circ,f$, equation~\eqref{eq:302} at $n=1$, equation~\eqref{eq:52}, equation~\eqref{eq:249} may be written as
\begin{equation}
  \label{eq:336}
  3\partial_u X+\eth_\circ^2\widebar\eth_\circ^2 X= S_X,
\end{equation}
where
\begin{equation}
  \label{eq:367}
  S_{V_\circ}=-(V_{\circ}^4-1)\eth_\circ^2\widebar\eth_\circ^2 V_\circ+V_\circ^3|\eth_\circ^2 V_\circ|^2,
\end{equation}
\begin{multline}
  \label{eq:368}
  S_{\Phi_0}=-(e^{-4\Phi_\circ}-1)\eth_\circ^2\widebar\eth_\circ^2\Phi_\circ+
  e^{-4\Phi_\circ}\big[2\eth_{\circ}\Phi_\circ\widebar\eth_{\circ}\Phi_\circ   +2\widebar\eth_{\circ}\Phi_\circ \eth_{\circ}\widebar\eth_{\circ}\eth_{\circ} \Phi_\circ
  \\ +
  2\eth_{\circ}\Phi_\circ\widebar\eth_{\circ}\eth_{\circ}
  \widebar\eth_{\circ}\Phi_\circ+2\eth_{\circ}\widebar\eth_{\circ}\Phi_\circ \eth_{\circ}\widebar\eth_{\circ}\Phi_\circ
  -4\eth_{\circ}\Phi_\circ\widebar\eth_{\circ}\Phi_\circ\eth_{\circ}\widebar\eth_{\circ}\Phi_\circ\big],
\end{multline}
and $S_f=3g(f)$. Similar results hold when using as unknown $h$ defined as
$V_\circ^{-3}=1+h$. Again, one may write the RT equation~\eqref{eq:302} at $n=3$
in the form of~\eqref{eq:336} with $X=h$ and a suitable non-linear source term
$S_h$ on the right hand side.

In order to understand generic features of
perturbation theory, it is useful to consider the equation where $S_X$ is
replaced by an arbitrary $u$-dependent source $S$,
\begin{equation}
  \label{eq:361}
  3\partial_u X+\eth_\circ^2\widebar\eth_\circ^2 X= S,
\end{equation}
When expanding in spherical
harmonics using a ``variations of constants'' ansatz,
\begin{equation}
  \label{eq:364}
  X=X^{jm}(u)e^{-\omega_j u}\,{}_0Y_{jm},\quad \omega_j=\frac{s_j}{3}, \quad S=S^{jm}(u){}_0 Y_{jm},
\end{equation}
with $s_j$ given in~\eqref{eq:259}, and in particular $s_0=0=s_1$, the general
solution to~\eqref{eq:361} is
\begin{equation}
  \label{eq:365}
  X^{jm}=X^{jm}(0)+\int^u_0dv\, e^{\omega_j v}S^{jm}(v). 
\end{equation}

\subsection{Linearized theory}\label{sec:line-appr}

\paragraph*{Algebraically special quasi-normal modes}

As originally discussed in~\cite{Foster:1967}, one may set up perturbation theory around the
Schwarzschild solution with $M=1$ by introducing a small parameter $\epsilon$, and
considering
\begin{equation}
  \label{eq:301}
  V_\circ=1+\sum_{n=1}\epsilon^n V_{\circ n},\quad \Phi_\circ=\sum_{m=1}\epsilon^m\Phi_{\circ m},
\end{equation}
with the two series related by $\Phi_\circ=-\ln V_\circ$ and
$-\ln (1+x)=\sum_{l=1}{(-)}^l\frac{x^l}{l}$, so that $V_{\circ 1}=-\Phi_{\circ 1}$,
and $V_{\circ 2}=-\Phi_{\circ 2}+\frac 12 \Phi^2_{\circ 1}$.

The linearized RT equation becomes
\begin{equation}
  \label{eq:114}
  3\partial_{u} V_{\circ 1}+\eth_\circ^2\widebar\eth_\circ^2 V_{\circ 1}=0,
\end{equation}
without a source term. When the perturbation is expanded in spherical
harmonics,
\begin{equation}
  \label{eq:113}
   V_{\circ 1}=a^{jm}_1(u)\,{}_{0}Y_{jm}(\zeta,\widebar\zeta),\quad j\in \mathbb N,\ |m|\leq j,
 \end{equation}
one gets
\begin{equation}
  \label{eq:116}
3\, \partial_{u}  a^{jm}_1(u)=-s_j a^{jm}_1(u),
\end{equation}
with $s_j$ given in~\eqref{eq:259}. The first four harmonics with $j=0,1$ are
constant, while the higher harmonics with $j>1$ are exponentially decreasing,
\begin{equation}
  \label{eq:117}
  a^{jm}_1(u)=c^{jm}_1,\ j\leq 1,\quad a^{jm}_1(u)=c^{jm}_1e^{-\omega_j u},\ j>1,
\end{equation}
with the slowest exponential decrease for $j=2$ determined by $s_2=6$, $\omega_2=2$
and given by $e^{-2u}$. In the spectrum of quasi-normal modes of the
Schwarzschild back hole (see e.g.~\cite{Berti:2009kk} equation~(83) and Figure 5 for a
review), they correspond to the algebraically special ones~\cite{Couch:1973zc,Chandrasekhar:1984mgh,Qi:1993ey}.

\paragraph*{Displacement memory}\label{sec:displacement-memory}

It now follows directly from~\eqref{eq:253} that, in the linearized approximation, the
displacement memory is given by
\begin{equation}
  \label{eq:362}
  \boxed{\mathcal D_{[u_i,u_f]1}= \frac{3}{\sqrt{s_j}}c^{jm}_1[e^{-\omega_j u_i}-e^{-\omega_j u_f}]
  \,{}_{-2}Y_{jm}\big\rvert_{j\geq 2}}. 
\end{equation}

In the linearized approximation, the generalized Bondi mass aspect is conserved
and explicitly given by
\begin{equation}
  \label{eq:141}
  \boxed{\Psi^\infty_\circ=1 -\epsilon\, 3  c^{jm}_1\, {}_0Y_{jm}\big\rvert_{j\leq 1}+O(\epsilon^{2})},
\end{equation}
in-line with the fact that there is no non-linear memory effect at first order.

\paragraph*{Radiated energy}\label{sec:radiated-energy}

The first order solution $V_{\circ 1}$ determines the energy that is radiated
away to second order in perturbation theory,
\begin{equation}
  \label{eq:261}
  4\pi \mathcal F_{2[u_i,u_f]}=\int^{u_f}_{u_i}du\, |\eth_\circ^2 V_{\circ 1}|^2.
\end{equation}
When using~\eqref{eq:291},
\begin{equation}
  \label{eq:269}
  |\eth_\circ^2 V_{\circ 1}|^2=\sqrt{s_{j_1}s_{j_2}}c^{j_1m_1}_1c^{j_1m_1}_1
  e^{-(\omega_{j_1}+\omega_{j_2})u}
  \mathcal A^{j}_{-2,j_1m_1;2,j_2m_2}\delta^m_{m_1+m_2}\big\rvert_{j_1,j_2\geq 2}\,{}_0 Y_{jm}, 
\end{equation}
so that
\begin{multline}
  \label{eq:285}
  \boxed{ 4\pi \mathcal F_{2[u_i,u_f]}=}\\\boxed{3\sqrt{s_{j_1}s_{j_2}}
    c^{j_1m_1}_1c^{j_1m_1}_1
  \frac{1-e^{-(\omega_{j_1}+\omega_{j_2})(u_f-u_i)}}{s_{j_1}+s_{j_2}}
  \mathcal A^{j}_{-2,j_1m_1;2,j_2m_2}\delta^m_{m_1+m_2}\big\rvert_{j_1,j_2\geq 2}\,{}_0 Y_{jm}}.
\end{multline}

\paragraph*{Absorption of the constant low harmonics}\label{sec:absorpt-const-harm}

For reasons detailed below, it is useful to get rid of the constant harmonics by
an appropriate scaling and boost before going to second order. Note that, to
first order in perturbation theory, absorbing the first four harmonics in either
$V_{\circ 1},\Phi_{\circ 1},f_{1},h_{1}$ is equivalent, so that this puts the system into
its rest frame to first order in perturbation theory.

We start from~\eqref{eq:344}, and with $v$ replaced by $v^{-1}_1$, i.e., a
scaling parametrized by multiplication through $R_1$ and a boost parametrized
by $g_1$,
where
\begin{equation}
  \label{eq:317a}
  R_1=1+\epsilon r_1,\quad  a_1=1+\epsilon\alpha_1,\quad  b_1=\epsilon\beta_1,
  \quad  c_1=\epsilon\gamma_1,\quad  d_1=1-\epsilon\alpha_1,
\end{equation}
with $r_1,\alpha_1,\beta_1,\gamma_1$ to be determined presently. This gives
instead of~\eqref{eq:386}
\begin{multline}
  \label{eq:257}
  V^{(1)}_{\circ;v_1}(u_1,x_1)=w_{v^{-1}_1}(x_1)\big(1+\epsilon c^{jm}_1\,{}_{0,1}Y_{jm}|_{j\leq 1}
  \\+\epsilon c_1^{jm}e^{-\omega_j u}{}_{0,1}Y_{jm}|_{j\geq 2}\big)(R^4_1u_1,x_{g^{-1}_1}(x_1)).
\end{multline}
The low constant harmonics are then absorbed by choosing $v_1$ such that
\begin{equation}
  \label{eq:258}
  w_{v^{-1}_1}(x_1)=1-\epsilon c^{jm}_1\,{}_{0,1}Y_{jm}|_{j\leq 1}(x_{g^{-1}_1}(x_1)),
\end{equation}
At first order, one may replace the argument $(R^4_1u_1,x_{g^{-1}_1}(x_1))$ on the
right hand side by $u_1,x_1$ and
\begin{equation}
  \label{eq:315}
  V^{(1)}_{\circ;v_1}(u_1,x_1)=1+\epsilon V^{(1)}_{\circ 1}+\mathcal O(\epsilon^2),\ V^{(1)}_{\circ 1}
  =c_1^{jm}e^{-\omega_j u_1}{}_{0,1}Y_{jm}(x_1)\big\rvert_{j\geq 2}=\mathcal O(e^{-2u_1}),
\end{equation}
with $V^{(1)}_{\circ;v_1}(u_1,x_1)$ a solution to the RT equation in terms of the
variables $u_1,x_1$ without low harmonics in the linearized approximation.

To find the appropriate parameters, one uses~\eqref{eq:319} to get
\begin{equation}
  \label{eq:314}
  \begin{split}
  r_1=-\sqrt{\frac{1}{4\pi}}c^{00}_1,\quad \beta_1+\widebar\gamma_1=\sqrt{\frac{3}{4\pi}}c^{1-1}_1,\\
  \alpha_1+\widebar\alpha_1=\sqrt{\frac{3}{4\pi}}c^{10}_1,\quad
    \widebar\beta_1+\gamma_1=-
    \sqrt{\frac{3}{4\pi}}c^{11}_1.
  \end{split}
\end{equation}

\subsection{Second order perturbation theory}\label{sec:second-order-pert}

\paragraph*{Multipole expansion}

To second order in perturbation theory, one now gets
\begin{equation}
  \label{eq:136}
  3\partial_{u_1}V^{(1)}_{\circ 2}+\eth^{(1)2}_\circ\widebar\eth^{(1)2}_\circ V_{\circ 2}^{(1)}=
  -4V_{\circ 1}^{(1)}\eth^{(1)2}_\circ\widebar\eth^{(1)2}_\circ V_{\circ 1}^{(1)}+|\eth^{(1)2}_\circ V_{\circ 1}^{(1)}|^2. 
\end{equation}
The left hand side involves the same operator acting on the unknown second order
perturbation that acts on the first order perturbation in the linearized case in
equation~\eqref{eq:114}, while the right hand side is a source term that is entirely
determined by the known $V^{(1)}_{\circ 1}$ in~\eqref{eq:315}. When
using~\eqref{eq:115} and~\eqref{eq:303}, the ansatz
\begin{equation}
  \label{eq:221}
  V^{(1)}_{\circ 2}=a^{jm}_2(u_1)e^{-\omega_j u_1}\,{}_0Y_{jm}(x_1),
\end{equation}
yields
\begin{multline}
  \label{eq:224}
  3\partial_{u_1} a^{jm}_2e^{-\omega_j u_1}\,{}_0Y_{jm}
  =c^{j_1m_1}_1c^{j_2m_2}_1
  e^{-(\omega_{j_1}+\omega_{j_2})u_1}\\\big[-4s_{j_2}\,{}_0Y_{j_1m_1}
  \,{}_0Y_{j_2m_2}+\sqrt{s_{j_1}s_{j_2}}\,{}_{-2}Y_{j_2m_2}
  \,{}_2Y_{j_1m_1}\big]\big\rvert_{j_1,j_2\geq 2}.
\end{multline}
{\em At this stage, it appears that the elimination of the low harmonics at first order guarantees
  that the right hand side is $\mathcal O(e^{-4u_1})$ so that the integration of the
  equation for $a^{jm}_2$ does not produce divergent terms at late times.}

For the multipole expansion of the source, one uses~\eqref{eq:291}, which allows one to
write the right hand side as $S^{jm}\, {}_0Y_{jm}$ with
\begin{equation}
  \label{eq:328}
  S^{jm}=c^{j_1m_1}_1c^{j_2m_2}_1
  \mathcal B^{j}_{j_1m_1;j_2m_2}\big\rvert_{j_1,j_2\geq 2}\delta^m_{m_1+m_2}\
  e^{-(\omega_{j_1}+\omega_{j_2})u_1},
\end{equation}
and where
\begin{equation}
  \label{eq:318}
  \mathcal B^{j}_{j_1m_1;j_2m_2}=-2(s_{j_1}+s_{j_2})
  \mathcal A^{j}_{0,j_1m_1;0,j_2m_2}+\sqrt{s_{j_1}s_{j_2}}
  \mathcal A^{j}_{-2,j_1m_1;2,j_2m_2},
\end{equation}
the sum over $j$ being restricted to
$\max(0,|j_1-j_2|,|m_1+m_2|),\dots, j_1+j_2$. Equation~\eqref{eq:224}
thus becomes
\begin{equation}
  \label{eq:320}
  3\partial_{u_1} a^{jm}_2=c^{j_1m_1}_1c^{j_2m_2}_1
  \mathcal B^{j}_{j_1m_1;j_2m_2}\big\rvert_{j_1,j_2\geq 2}\delta^m_{m_1+m_2}\
  e^{-(\omega_{j_1}+\omega_{j_2}-\omega_j)u_1},
\end{equation}
with the understanding that $\mathcal B^{j}_{j_1m_1;j_2m_2}$ vanishes when
$j\notin\max(0,|j_1-j_2|,|m_1+m_2|),\dots, j_1+j_2$.

\paragraph*{Resonances}

The time dependence on the right hand side of~\eqref{eq:320} implies that the behaviour of
the solution changes, from a decreasing exponential to a linearly growing
function when
\begin{equation}
  \label{eq:321}
  \omega_j= \omega_{j_1}+\omega_{j_2} \iff f_{(j,j_1,j_2)}=0
\end{equation}
where 
\begin{equation}
  \label{eq:330}
  f(j,j_1,j_2)=\frac{(j+2)!}{(j-2)!}- \frac{(j_1+2)!}{(j_1-2)!}-\frac{(j_2+2)!}{(j_2-2)!}.
\end{equation}
More explicitly, the coefficients $a^{jm}_2(u_1)$ are given by
\begin{multline}
  \label{eq:329}
  a^{jm}_2=c^{jm}_2+ c^{j_1m_1}_1c^{j_2m_2}_1\mathcal B^{j}_{j_1m_1;j_2m_2}\big\rvert_{j_1,j_2\geq 2}\delta^m_{m_1+m_2}(1-\delta^0_{f_{(j,j_1,j_2)}})
  \frac{\big(e^{-(\omega_{j_1}+\omega_{j_2}-\omega_j)u_1}-1\big)}{s_j-s_{j_1}-s_{j_2}}
  \\+ c^{j_1m_1}_1c^{j_2m_2}_1\mathcal B^{j}_{j_1m_1;j_2m_2}\big\rvert_{j_1,j_2\geq 2}\delta^m_{m_1+m_2} 
  \delta^0_{f_{(j,j_1,j_2)}}\frac{u_1}{3}, 
\end{multline}
What remains to be discussed is the values of $(j,j_1,j_2)$ for which resonances
might occur and whether the coefficient at these values vanishes or not. The
first solutions to~\eqref{eq:321} are
\begin{equation}
	(j,j_1,j_2) = (6,5,5) \,,\quad (198, 188, 130)\,,\quad \dots . 
\end{equation}
The question is then whether the resonances actually occur for generic initial
conditions, i.e., whether the coefficients vanish or not. The first one is given
by,
\begin{equation}
  \label{eq:333}
  \mathcal B^{6}_{5,m_1;5,m_2}=-s_5(4\mathcal A^{6}_{0,5,m_1;0,5,m_2}-
  \mathcal A^{6}_{-2,5,m_1;2,5,m_2}),
\end{equation}
where $s_5=210$ and the exact expression for the non-vanishing parenthesis on
the right hand side is given in~\eqref{eq:334}.

\paragraph*{Memory at second order}

To second order, equations~\eqref{eq:221} and~\ref{eq:329} imply that the
solution is given by
\begin{multline}
  \label{eq:331}
  V^{(1)}_{\circ 2} =\,{}_0 Y_{jm}\Big[c^{jm}_2 e^{-\omega_j u_1}
  \\ +  c^{j_1m_1}_1c^{j_2m_2}_1\mathcal B^{j}_{j_1m_1;j_2m_2}\big\rvert_{j_1,j_2\geq 2}\delta^m_{m_1+m_2} (1-\delta^0_{f_{(j,j_1,j_2)}})\
  \frac{e^{-(\omega_{j_1}+\omega_{j_2})u_1}-e^{-\omega_j u_1}}{s_j-s_{j_1}-s_{j_2}}
  \\
  + c^{j_1m_1}_1c^{j_2m_2}_1\mathcal B^{j}_{j_1m_1;j_2m_2}\big\rvert_{j_1,j_2\geq 2}
  \delta^m_{m_1+m_2}\delta^0_{f_{(j,j_1,j_2)}}\ \frac{{u_1}}{3}e^{-\omega_j u_1}
  \Big]. 
\end{multline}
Note that the second and third lines are zero at $u=0$ and when $u\to \infty$. To that
order, the displacement memory is given by
\begin{multline}
  \label{eq:122}
  \mathcal D_{[{u_1}_i,{u_1}_f]2}=\sqrt{s_j}\,{}_{-2}Y_{jm}\big\rvert_{j\geq 2}\Big[3c^{jm}_2\frac{e^{-\omega_j {u_1}_i}-e^{-\omega_j {u_1}_f}}{s_j}\\
+  c^{j_1m_1}_1c^{j_2m_2}_1\mathcal B^{j}_{j_1m_1;j_2m_2}\big\rvert_{j_1,j_2\geq 2}\delta^m_{m_1+m_2} (1-\delta^0_{f_{(j,j_1,j_2)}})
\int^{{u_1}_f}_{{u_1}_i}du_1\, \frac{e^{-(\omega_{j_1}+\omega_{j_2})u_1}-e^{-\omega_j u_1}}{s_j-s_{j_1}-s_{j_2}}\\
+c^{j_1m_1}_1c^{j_2m_2}_1\mathcal B^{j}_{j_1m_1;j_2m_2}\big\rvert_{j_1,j_2\geq 2}
\delta^m_{m_1+m_2}\delta^0_{f_{(j,j_1,j_2)}}\int^{{u_1}_f}_{{u_1}_i}du_1\,
\frac{{u_1}}{3}e^{-\omega_j u_1}
  \Big]
,
\end{multline}
with
\begin{equation}
  \label{eq:135}
  \int^{{u_1}_f}_{{u_1}_i}du_1\, \frac{e^{-(\omega_{j_1}+\omega_{j_2})u_1}-e^{-\omega_j u_1}}{s_j-s_{j_1}-s_{j_2}}=3 \frac{\frac{e^{-(\omega_{j_1}+\omega_{j_2})u_i}
      -e^{-(\omega_{j_1}+\omega_{j_2})u_f}}{s_{j_1}+s_{j_2}}
  -\frac{e^{-\omega_j u_i}-e^{-\omega_j u_f}}{s_j}}{s_j-s_{j_1}-s_{j_2}},
\end{equation}
and
\begin{equation}
  \label{eq:284}
  \int^{{u_1}_f}_{{u_1}_i}du_1\,\frac{{u_1}}{3}e^{-\omega_j u_1}
  =\frac{{u_1}_ie^{-\omega_j {u_1}_i}-{u_1}_f e^{-\omega_j {u_1}_f}}{s_j}
  +3\frac{e^{-\omega_j {u_1}_i}-e^{-\omega_j{u_1}_f}}{s_j^2},
\end{equation}
while the generalized mass aspect is
\begin{equation}
  \label{eq:295}
  \Psi^{\infty (1)}_\circ=1+\epsilon^2\Big[-3V_{\circ 2}^{(1)}+\eth^{(1) 2}_\circ\widebar\eth^{(1)2}_\circ\int^\infty_{u_1}\,dv_1 V_{\circ 2}^{(1)}+6(V^{(1)}_{\circ 1})^2\Big]+\mathcal O(\epsilon^3).
\end{equation}
The term at second order turns out to be 
\begin{multline}
  \label{eq:130}
  \Psi^{\infty(1)}_{\circ 2}=3\ {}_0 Y_{jm}\Big[-c^{jm}_2\big\rvert_{j\leq 1}
  -c^{j_1m_1}_1c^{j_2m_2}_1\mathcal B^{j}_{j_1m_1;j_2m_2}\big\rvert_{j_1,j_2\geq 2}\delta^m_{m_1+m_2} (1-\delta^0_{f_{(j,j_1,j_2)}})\\
  \frac{e^{-(\omega_{j_1}+\omega_{j_2})u_1}
-s_j\frac{e^{-(\omega_{j_1}+\omega_{j_2})u_1}}{s_{j_1}+s_{j_2}}}{s_j-s_{j_1}-s_{j_2}}\\
+c^{j_1m_1}_1c^{j_2m_2}_1\mathcal B^{j}_{j_1m_1;j_2m_2}\big\rvert_{j_1,j_2\geq 2}
\delta^m_{m_1+m_2}\delta^0_{f_{(j,j_1,j_2)}} \frac{e^{-\omega_j u_1}}{s_j}\\
+2c^{j_1m_1}_1c^{j_2m_2}_1\mathcal A^{j}_{0,j_1m_1;0,j_2m_2}\big\rvert_{j_1,j_2\geq 2}\delta^m_{m_1+m_2}
e^{-(\omega_{j_1}+\omega_{j_2})u_1}\Big].
\end{multline}
According to~\eqref{eq:191},
\begin{equation}
  \label{eq:311}
  \partial_{u_1}\Psi^{\infty(1)}_{\circ 2}=-|\eth^{(1)2}_\circ V^{(1)}_{\circ 1}|^2, 
\end{equation}
where the right hand side may be read off~\eqref{eq:269} which, at second order, may be
evaluated at $u_1,x_1$. The explicit verification is
instructive.

\paragraph*{Absorption of the time-dependent low harmonics}

Besides similar constant terms than in $V_{\circ 1}$, the other contributions to
the low harmonics in $V^{(1)}_{\circ 2}$ from the second and third lines
in~\eqref{eq:331} are in general time-dependent. Their absorption thus requires
time-dependent parameters $r_2,\alpha_2,\beta_2,\gamma_2$, so that the relevant
RT equation at second order will be the modified one.

\subsection{Higher order perturbations and late-time behavior}\label{sec:high-order-pert}

Rather than computing the detailed perturbative coefficients at higher orders,
one may instead analyze the possible resonances. At order $\epsilon^n$, equation~\eqref{eq:336}
for $V_{\circ n}$ involves a source term with sums of products of $V_{\circ m}$ such
that the sums of the $m$'s add up to $n$,
\begin{equation}
  \label{eq:254}
  (3\partial_u+\eth_\circ^2\widebar\eth_\circ^2)V_{\circ n}=S_n[V_{\circ 1},\dots V_{\circ n-1}]. 
\end{equation}

In the generic case when the source terms involve only simple exponential
decays, there will be new resonances whenever
\begin{equation}\label{resonance condition generalized}
	s_{j}=s_{j_1}+s_{j_2}+\cdots +s_{j_n},
\end{equation}
for some integers $j,j_1,\dots, j_n$. From the definition~\eqref{eq:259}, we note that 
\begin{equation}
	s_j=\frac{j+2}{j-2}s_{j-1}=(1+\frac{4}{j-2})s_{j-1}. 
\end{equation}
This provides some, but not all, solutions to~\eqref{resonance condition generalized}.

As shown in Section~\ref{sec:antip-symm-solut}, the RT equation is equivariant
under the antipodal map. Upon expanding the solution in spherical harmonics,
which satisfy 
\begin{equation}
  \label{eq:326}
  {}_{0}Y_{jm}(A(x))={(-)}^j{}_{0}Y_{jm}(x),
\end{equation}
the geometric symmetry becomes the statement that parity is preserved by the
evolution. In turn, this implies that the source term projects trivially onto
the target harmonic unless the parity selection rule
\begin{equation}
  \label{eq:255}
  (-1)^j=(-1)^{\sum_i j_i},
\end{equation}
is satisfied. 

The lowest solutions of the resonance condition at $n$-th order in perturbation
theory are listed below together with their parity classification.
\begin{equation}
	\begin{matrix}{}
		n=2: &&j=6&s_6=2 s_5&&\omega_j=140 && \mathrm{parity\ allowed}\\
		n=3: &&j=4&s_4=3 s_3&& \omega_j=30 && \mathrm{parity\ forbidden}\\
    n=3: &&j=5&s_5= 2s_4+s_3&& \omega_j=70 && \mathrm{parity\ allowed}\\
		n=4: &&j=6&s_6=s_5+2s_4 +s_3&& \omega_j=140u && \mathrm{parity\ allowed}\\
		n=5: &&j=3&s_3=5 s_2&& \omega_j=10 && \mathrm{parity\ forbidden}\\
    n=5: &&j=5&s_5=s_4+4s_3 && \omega_j=70 && \mathrm{parity\ forbidden}\\
    n=5: &&j=8&s_8=2s_6+4s_5&& \omega_j=420 && \mathrm{parity\ allowed}\\
    n=6: &&j=6&s_6=4s_4+2s_3&& \omega_j=140 && \mathrm{parity\ allowed}\\
    n=7: &&j=4&s_4=2s_3+5s_2&& \omega_j=30 && \mathrm{parity\ allowed}
	\end{matrix}
\end{equation}
Since $j=2,3$ cannot satisfy both the parity and the resonance condition, the
first parity allowed resonance is $j=4$ with $\omega_j=30$ at seventh order in
perturbation theory.

The same mechanism has appeared originally in the non-perturbative asymptotic
analysis, where parity preservation delays the onset of logarithmic terms in the
late time expansion. By using the modified flow discussed in Section~\ref{sec:coll-coord},
Chru\'sciel has shown that the late-time behavior of $\Phi_\circ$ satisfying~\eqref{eq:52} is of
the form (equation (4.21) of~\cite{Chrusciel:1992rv}),
\begin{equation}
  \label{eq:300}
  \Phi_\circ=\sum_{k=1}^{4} f_{k,0}e^{-2kum^{-1}}+f_{5,1}um^{-1}e^{-10um^{-1}}+f_{5,0}e^{-10 um^{-1}} +\dots,
\end{equation}
with the onset of the $u$-dependence corresponding to the above resonance at
order 5 in perturbation theory. Using parity arguments, it is then shown
(equation (4.22) and Proposition 4.2 of~\cite{Chrusciel:1992rv}) that the actual onset is
\begin{equation}
  \label{eq:217}
  \Phi_\circ=\sum_{k=1}^{14} f_{k,0}e^{-2kum^{-1}}+f_{15,1}um^{-1}e^{-30um^{-1}}+f_{15,0}
  e^{-30um^{-1}}+\dots,
\end{equation}
The
associated late-time behavior of $V_\circ$ satisfying~\eqref{eq:27} is of the
form (equation (2.8) of~\cite{Chrusciel:1992tj}),
\begin{equation}
  \label{eq:300a}
  V_\circ=1+\sum_{k=1}^{14} V_{k,0}e^{-2kum^{-1}}
  +V_{15,1}um^{-1}e^{-30um^{-1}}+V_{15,0}e^{-30 um^{-1}} +\dots,
\end{equation}
and it has been shown in~Proposition 2.1 of~\cite{Chrusciel:1992tj} that the coefficient
$V_{15,1}$ generically does not vanish. This corresponds precisely to the
resonance at order $7$ in perturbation theory above (when $m=1$).

More generally, the non-perturbative late time behavior is of the form
$u^i e^{-\sum_j\omega_j u }$. It should be instructive to compare in more details the
proofs of this behavior and the results based on it to (a completed) small
amplitude expansion as set-up here.

\section{Discussion}\label{sec:discussion}

In this paper, we have explicitly worked out the displacement and the non-linear
memory for Robinson-Trautman waves. Using these results, it should not be too
difficult to also work out their spin memory. Note that it is straightforward to
translate the results of this paper to the, maybe more familiar, Bondi-Sachs
framework: the change of coordinates given explicitly in~\eqref{eq:107}-~\eqref{eq:111} allows one to
write the RT metrics as asymptotically flat metrics of Bondi-Sachs form.

It would be of interest to try to
generalize some of the results presented in this paper here to the case of
algebraically special twisting spacetimes and to include magnetic multipoles.

Another interesting question is to re-formulate considerations on memory effects
in locally asymptotically flat spacetimes at null infinity in terms of
asymptotically time-like geodesic congruences and the associated geodesic
deviation equation. Preliminary results indicate that the generalized mass
aspect that has played an important role here as a Lyapunov function, naturally
arises in terms of the expansion of these geodescis. We plan to address this
question elsewhere.

\section*{Acknowledgments}\label{sec:acknowledgements}

\addcontentsline{toc}{section}{Acknowledgments}

This work is supported by the F.R.S.-FNRS Belgium through convention IISN
4.4514.08. G.B.~is grateful to C.~Troessaert for past collaboration on this
subject. The authors acknowledge useful discussions with L.~Blanchet,
M.~Campiglia, T.~Damour, F.~Diaz and S.~Speziale. A.S.~is grateful to the
Institut des Hautes Études Scientifiques (IHES) for supporting a research stay
during the final stages of this project.

\appendix

\section{Newman-Unti and asymptotically flat spacetimes}\label{sec:locally-asympt-flat}

Newman-Unti solution space~\cite{Newman:1962cia} (see also~\cite{newman:1980xx,newman_spin-coefficient_2009} for reviews), is
defined by
\begin{equation}
  \label{eq:79}
  \kappa=\epsilon=\pi=0,
\end{equation}
which are equivalent to $Dl=0=Dn=Dm$, i.e., $l$ is an
affinely parametrized null geodesic, and the Newman-Penrose null tetrad is
parallely transported along $l$.
Furthermore,
\begin{equation}
  \label{eq:129}
  \rho=\widebar\rho,\quad \tau=\widebar\alpha+\beta,
\end{equation}
which imply that the null geodesic generator is hypersurface orthonormal and a
gradient.

One chooses coordinates $x^{\mu}=(u,r,\zeta,\widebar\zeta)$ such that $u={\mathrm cte}$
are null with normal covector $l=du$, while $r$ is the affine parameter,
\begin{equation}
  \label{eq:1}
  l=\partial_{r},\quad n=\partial_{u}+U\partial_{r}+X^{\zeta}\partial+X^{\widebar\zeta}\widebar\partial,
  \quad m=\omega\partial_{r}+\xi^{\zeta}\partial+\xi^{\widebar\zeta}\widebar\partial,
\end{equation}
with $\partial=\partial_{\zeta}$. Furthermore, one requires
\begin{equation}
  \label{eq:89}
  \Psi_0=r^{-5}\Psi^0_0+O(r^{-6}),\quad
  \rho=-r^{-1}+O(r^{-3}),\quad \tau=O(r^{-1}),
\end{equation}
as well as
\begin{equation}
  \label{eq:126}
  \xi^{\widebar\zeta}=r^{-1}P(u,\zeta,\widebar\zeta)+O(r^{-2}),\quad
  \xi^\zeta=O(r^{-2}).
\end{equation}

If one imposes Einstein's equations in vacuum, this solution space contains both
RT waves and also asymptotically flat spacetimes at $\scrip$ in the sense of
Newman and Penrose. The latter is obtained from Newman-Unti solution space when
one requires in addition that
\begin{equation}
  \label{eq:127}
  P(u,\zeta,\widebar\zeta)=P_\circ=\frac{1}{\sqrt 2}(|\zeta|^2+1).
\end{equation}
As a consequence, the metric on a cut $u=\mathrm{cte}$ of $\scrip$ is that for
the round sphere, and all spin-coefficients vanish when $r\to\infty$.

More details on the solution space can be found in~\cite{Newman:1962cia}, and in
the conventions used here in~\cite{Barnich:2016lyg}.
  
\section{Asymptotic coordinate transformation and frame
  rotation}\label{sec:asympt-frame-rott}

The combined coordinate transformation and local frame rotation that preserves
Newman-Unti solution space is of the form
\begin{equation}
  \label{eq:48}
  u=u(u',r',\zeta',\widebar\zeta'),\quad r=r(u',r',\zeta',\widebar\zeta'),\quad \zeta=\zeta(u',r',\zeta',\widebar\zeta'),
\end{equation}
and
\begin{equation}
  \label{eq:43}
  e\indices{_{a}^{\prime\mu}}\frac{\partial x^{\nu}}{\partial x^{\prime\mu}}
  =\Lambda\indices{_{a}^{b}} e\indices{_{b}^{\nu}},\quad
  e\indices{_{a}^{\mu}}=(l^{\mu},n^{\mu},m^{\mu},\widebar m^{\mu}).
\end{equation}
The Lorentz rotation $\Lambda\indices{^{a}_{b}}$ is determined by an
$SL(2,\mathbb C)$ group element according to
\begin{equation}
  \label{eq:46}
  g\check \jmath_{a}\widebar g^{-1}=\check \jmath_{b}\Lambda\indices{^{b}_{a}},
\end{equation}
where\footnote{There is a typo in the first of equation~(6.17) of~\cite{Barnich:2016lyg} where the bar on $g^{-1}$ is missing.}
$\check\jmath_{1}=\Delta\indices{_{1}^{2}},\check\jmath_{2}=-\Delta\indices{_{2}^{1}}, \check\jmath_{3}=-\Delta\indices{_{1}^{1}}$,
$\check\jmath_{4}=\Delta\indices{_{2}^{2}}$, and
$(\Delta\indices{_{i}^{j}})\indices{^{k}_{l}}=\delta_{i}^{k}\delta^{j}_{l}$. In turn, the
$\mathrm{SL}(2,\mathbb C)$ group element is parametrized as
\begin{equation}
  \label{eq:45}
  g=\begin{pmatrix} a& b\\ c& d
  \end{pmatrix}
  =\begin{pmatrix} e^{-E/2} & -\widebar A e^{E/2}\\ -Be^{-E/2} & (1+\widebar AB)e^{E/2}
  \end{pmatrix},
\end{equation}
where $E=E,A,B\in \mathbb{C}$ so as to implement a combined class null rotation
of type $\mathrm{II\circ I\circ III}$. For later use, note that in the absence of a coordinate
transformation, $e\indices{^{\prime}_{a}^{\mu}}=\Lambda\indices{_{a}^{b}}e\indices{_{b}^{\mu}}$
is given in terms of $g$ by
\begin{equation}
  \label{eq:96}
  \begin{pmatrix}l' \\ n' \\ m' \\ \widebar m'
  \end{pmatrix}=\begin{pmatrix} d\widebar d & c\widebar c & -d\widebar c
    & -c\widebar d\\
    b\widebar b &  a\widebar a & -b\widebar a & -a \widebar b\\
    -d\widebar b & -c\widebar a & d\widebar a & c\widebar b \\
    -b\widebar d & -a\widebar c & b\widebar c & a\widebar d
  \end{pmatrix}\begin{pmatrix}l \\ n \\ m \\ \widebar m
  \end{pmatrix}.
\end{equation}

The transformation that brings the RT waves of Section~\ref{sec:rt-solutions} to the standard
asymptotically flat form can be obtained as a particular case of the results
worked out in Sections 6.3 and 6.4 of~\cite{Barnich:2016lyg}, that implements, as in equation
(6.77) of that reference, a change of the time-dependent conformal factor by a
finite amount in Newman-Unti solution space,
\begin{equation}
  \label{eq:29}
  P'(u',\zeta',\widebar\zeta)=P(u_{0},\zeta_{0},\widebar\zeta_{0})
  e^{{-E_{R{0}}(u',\zeta',\widebar\zeta')}},\quad E_{I0}=0.
\end{equation}
In other words, it amounts to performing an asymptotic real Weyl transformation,
without any supertranslation, (super) rotation or purely imaginary asymptotic
Weyl transformation (which corresponds to an asymptotic rotation of the spatial
dyads).

The asymptotic expansions of the various quantities, adapted to this case of an
asymptotic (real) Weyl transformation, can be read off from the equation after
(6.47) of~\cite{Barnich:2016lyg}, where $P=\widebar P$ and only the non-vanishing spin
coefficients for RT waves need to be taken into account:
\begin{equation}
  \label{eq:42}
  \begin{split}
    u&=u_{0}-B_{0}\widebar B_{0}e^{-E_{R0}}r^{\prime -1}+O(r^{{\prime -2}}),\\
    \chi& \equiv r e^{-E_{R}} =r'+\chi_{1}+O(r^{\prime -1}),\\
    \zeta&=\zeta_{0}-B_{0}e^{-E_{0}}P r^{\prime -1}+O(r^{\prime -2}),\\
    B&=B_{0}r^{\prime -1}+O(r^{\prime -2}),\\
    A&=A_{0}+\big[e^{-2E_{R0}+\imath E_{I0}}B_{0}\widebar B_{0}\widebar \nu^{0}
    +e^{{-2E_{R0}}}B_{0}\widebar \mu^{0}+A^{2}_{0}\widebar B_{0}\big]r^{\prime -1}+O(r^{\prime -2}),\\
    E&=E_{{0}}+2[B_{0}\widebar B_{0}e^{-E_{R0}}\gamma^{0}+B_0e^{-E_0}\alpha^0-\widebar B_0e^{-\widebar E_0}\widebar \alpha^0
           -\widebar A_{0}B_{0}\big]r^{\prime -1}+O(r^{\prime -2}).
  \end{split}
\end{equation}
The coefficients $u_{0},\zeta_{0},\widebar\zeta_{0},B_{0},A_{0},\chi_{1},E_{R0}$,
are functions of $(u',\zeta',\widebar\zeta')$. In particular, the asymptotic
change of coordinates $u_{0}=u_{0}(u',\zeta,\widebar\zeta')$,
$\zeta_0=\zeta_{0}(u',\zeta,\widebar\zeta')$,
$\widebar\zeta_{0}=\widebar\zeta_{0}(u',\zeta,\widebar\zeta')$ is invertible
with inverse denoted for later convenience by
$u'_{0}=u'_{0}(u,\zeta,\widebar\zeta)$,
$\zeta'_{0}=\zeta'_{0}(u,\zeta,\widebar\zeta)$,
$\widebar\zeta'_{0}=\widebar\zeta' _{0}(u,\zeta,\widebar\zeta)$.

In these terms, we have according to Section 6.3 of~\cite{Barnich:2016lyg},
adapted to the case at hand, that all quantities are determined by the
asymptotic change of coordinates where only the retarded time undergoes a
non-trivial change, $u_{0}=u_{0}(u' ,\zeta' ,\widebar\zeta') $, $\zeta_{0}=\zeta'$,
$\widebar\zeta_{0}=\widebar\zeta'$:
\begin{equation}
  \label{eq:49}
  e^{-{E_{R0}}}=\frac{\partial u_{0}}{\partial u'},\quad B_{0}=-e^{-E_{R0}}\eth u'_{0},\quad A_{0}=-\partial_{u'}B_{0},
\end{equation}
\begin{equation}
  \label{eq:53}
  \chi_{1}=e^{-3E_{R0}}\eth u'_{0}\widebar\eth u' _{0}[\partial_{u}E_{R0}+\gamma^{0}+\widebar\gamma^{0}]
  -e^{-2E_{R0}}\eth\widebar\eth u'_{0},
\end{equation}
and also
\begin{equation}
  \label{eq:50}
  \frac{\partial u'_{0}}{\partial u}=e^{E_{R0}},\quad \frac{\partial u'_{0}}{\partial \zeta}=e^{E_{R0}}\frac{\partial u_{0}}{\partial\zeta'}
\end{equation}
where $E_{R0}$, when considered as a function of $u,\zeta,\widebar\zeta$, is explicitly given by  $E_{R0}(u,\zeta,\widebar\zeta)=E_{R0}(u'_{0}(u,\zeta,\widebar\zeta),\zeta,\widebar\zeta)$.
In particular,
\begin{equation}
  \label{eq:51}
  \frac{\partial \zeta_{0}}{\partial u'}=0,\quad \frac{\partial \zeta_{0}}{\partial \zeta'}=1,\quad \frac{\partial \zeta_{0}}{\partial \widebar \zeta'}=0,\quad \frac{\partial \zeta'_{0}}{\partial u}=0,\quad \frac{\partial \zeta'_{0}}{\partial \zeta}=1,\quad
  \frac{\partial \zeta'_{0}}{\partial \widebar\zeta}=0.
\end{equation}
Finally\footnote{In equation (6.63) of~\cite{Barnich:2016lyg}, the arguments of $P$ should be $u_{0},\zeta_{0},\widebar\zeta_{0}$ rather than $u,\zeta,\widebar\zeta $.},
\begin{equation}
  \label{eq:54}
  e^{E_{R0}}=\frac{P(u_{0}(u',\zeta',\widebar\zeta'),\zeta',\widebar\zeta')}{P'(u',\zeta',\widebar\zeta')}.
\end{equation}

Contrary to the presentation in~\cite{Barnich:2016lyg}, we will not simplify the
notation and keep the subscript $0$ on the asymptotic change of coordinates and
on the Weyl parameter.

As in equation (6.77) we want to implement a pure (real) Weyl transformations.
In addition, we require that, after the transformation, we remain with the
standard factor
$P'=P_{\circ}(\zeta',\widebar\zeta')=e^{-\varphi_{\circ}(\zeta',\widebar\zeta')}$ for
the round metric on the sphere. Let
$\Phi_\circ=\varphi(u,\zeta,\widebar\zeta)-\varphi_{\circ}(\zeta,\widebar\zeta)$. The Weyl
parameter relevant for us here is
\begin{equation}
  \label{eq:31}
  E_{R0}(u',\zeta',\widebar\zeta')=\varphi_{\circ}(\zeta',\widebar\zeta')-\varphi(u_{0}(u',\zeta',\widebar\zeta'),\zeta',\widebar\zeta')=-\Phi_\circ\big\rvert,\quad E_{I0}=0,
\end{equation}
where the bar means the substitution $u=u_{0}(u',\zeta,\widebar\zeta)$,
and all quantities are determined from (the inverse of)
\begin{equation}
  \label{eq:55}
u'_{0}(u,\zeta,\widebar\zeta)=\int_{0}^{u}dv\ e^{E_{R0}(v,\zeta,\widebar\zeta)}
=\int_{0}^{u}dv\frac{P(v,\zeta,\widebar\zeta)}{P_{\circ}(\zeta,\widebar\zeta)}
=\int_{0}^{u}dv\,e^{-\Phi_\circ(v,\zeta,\widebar\zeta)}.
\end{equation}
In particular, for the Lorentz rotation, we have in addition to~\eqref{eq:31},
\begin{equation}
  \label{eq:94}
  B_{0}=-(\eth_{\circ}u'_{0})|,\quad A_{0}=-(\eth_{\circ}\Phi_\circ)|,
\end{equation}
\begin{equation}
  \label{eq:95}
  \begin{split}
  a&=e^{\frac{\Phi_\circ}{2}}+\mathcal O(r^{\prime -1}),\ b=e^{-\frac{\Phi_\circ}{2}}\widebar\eth_{\circ}\Phi_\circ+\mathcal O(r^{\prime -1}),\\ c&=r^{\prime -1}(\eth_{\circ}u'_{0})|e^{\frac{\Phi_\circ}{2}}+\mathcal O(r^{\prime -2}),\ d=e^{-\frac{\Phi_\circ}{2}}
                                                                                                        +\mathcal O(r^{\prime -1}),
  \end{split}
\end{equation}
while
\begin{equation}
  \label{eq:3}
  \chi_{1}=-\eth_{\circ}\widebar\eth_{\circ}u'_{0}|.
\end{equation}
Starting from fields of the form $\eta^{s} (u,\zeta,\widebar\zeta)$, and considering composed fields of the form
\begin{equation}
  \label{eq:64}
  \eta^{\prime s}(u',\zeta,\widebar\zeta)=\eta^{s}(u_{0}(u',\zeta,\widebar\zeta),\zeta,\widebar\zeta)
  =\eta^{s}\rvert,
\end{equation}
by taking into account the additional dependence on $\zeta$ through $u=u_{0} (u',\zeta,\widebar\zeta)$,
and the relation (6.58), which reduces here to
\begin{equation}
  \label{eq:38}
  \partial u'_{0} =-\frac{P}{P_{\circ}}\partial u_{0}\iff \eth_{\circ} u_{0}=-\frac{P_{\circ}}{P}\eth_{\circ}u'_{0},
\end{equation}
it follows that
\begin{equation}
  \label{eq:69}
  \eth_{\circ}\eta^{\prime s}(u',\zeta,\widebar\zeta)=\big[\eth_{\circ}\eta^{s}-e^{\Phi}\eth_{\circ}u'_{0}\partial_{u}\eta^{s}\big]\big\rvert,
\end{equation}
and furthermore, if $u_{f0}=u_{0}(u'_{f},\zeta,\widebar\zeta)$,
$u_{i0}=u_{0}(u'_{i},\zeta,\widebar\zeta)$,
\begin{equation}
  \label{eq:70}
  \Delta'\eta^{\prime s}=\Delta \eta^{s},\quad \Delta \eta^{s}=\eta^{s}(u_{f0},\zeta,\widebar\zeta)-\eta^{s}(u_{i0},\zeta,\widebar\zeta).
\end{equation}

Collecting the results, we have for the Lorentz rotation,
\begin{multline}
  \label{eq:106}
    E=-\Phi_\circ|+2\big[\frac 12 \eth_{\circ}u'_{0}\widebar\eth_{\circ}u'_{0}e^{\Phi_\circ}\partial_{u}\Phi_\circ-\eth_{\circ}u'_{0} e^{\Phi_\circ}\alpha^0+\widebar\eth_{\circ}u'_{0} e^{\Phi_\circ}\widebar \alpha^0
    \\-\widebar\eth_{\circ}\Phi_\circ\eth_{\circ}u'_{0}\big]\big\rvert r^{\prime -1}+\mathcal O(r^{\prime -2}),
\end{multline}
\begin{multline}
    \label{eq:108}
    A=-(\eth_{\circ}\Phi_\circ)|\\+\big[\eth_{\circ}u'_{0}
       \widebar\eth_{\circ}u'_{0}e^{\Phi_\circ}\eth_{\circ}\partial_{u}\Phi_\circ
       -\eth_{\circ}u'_{0}\eth_{\circ}\widebar\eth_{\circ}\Phi_\circ+\frac 12 \eth_{\circ}u'_{0}-{(\eth_{\circ}\Phi_\circ)}^{2}\widebar\eth_{\circ}u'_{0}\big]\big\rvert r^{\prime -1}
       +\mathcal O(r^{\prime -2}),
\end{multline}
\begin{equation}
  \label{eq:109}
  B=-(\eth_{\circ}u'_{0})|r^{\prime -1}+\mathcal O(r^{\prime -2}),
\end{equation}
while the coordinate change is given by
\begin{equation}
  \label{eq:107}
  u=u_{0}(u',\zeta,\widebar\zeta)-\big[\eth_{\circ}u'_{0}\widebar\eth_{\circ}u'_{0}e^{\Phi_\circ}\big]\rvert
    r^{\prime -1}+\mathcal O(r^{\prime -2}),
  \end{equation}
\begin{multline}
    \label{eq:110}
    r=e^{E_{R}}\big(r'-(\eth_{\circ}\widebar\eth_{\circ}u'_{0})\big)\big\rvert+\mathcal O(r^{\prime -1})
    =e^{-\Phi_\circ|}r'\\+\big[e^{-\Phi_\circ}\big(-\eth_{\circ}\widebar\eth_{\circ}u'_{0}+e^{\Phi_\circ}\eth_{\circ}u'_{0}\widebar\eth_{\circ}
    u'_{0}\partial_{u}\Phi_\circ-\eth_{\circ}\Phi_\circ\widebar\eth_{\circ}u'_{0}
    -\widebar\eth_{\circ}\Phi_\circ\eth_{\circ}u'_{0}\big)\big]\big\rvert
    +\mathcal O(r^{\prime -1}),
\end{multline}
\begin{equation}
    \label{eq:111}
    \zeta=\zeta'+(\eth_{\circ}u'_{0})|P_{\circ}(\zeta',\widebar\zeta')r^{\prime -1}+\mathcal O(r^{\prime -2}).
  \end{equation}

\section{Staying in natural Robinson-Trautman time}\label{sec:stay-natur-robins}

In~Section~\ref{sec:bms-data-rt} and Appendix~\ref{sec:asympt-frame-rott}, we have derived the asymptotically flat data
for RT waves by a combined coordinate transformation and frame rotation. An
alternative approach is to stay in the natural RT coordinate system and perform
a pure frame rotation, from the algebraically special frame, to the
asymptotically flat frame associated with the Newman-Unti gauge, and read off
the asymptotically flat data from the transformed spin coefficients.

The starting point is the $\mathrm{SL}(2,\mathbb{C})$ transformation~\eqref{eq:45}, under
which the NP tetrad transforms as~\eqref{eq:46}. By grouping the spin coefficients in the
form
\begin{equation}
\label{eq:128}
\begin{split} & \check \omega_{1}=\begin{pmatrix} \epsilon & \pi \\ -\kappa & -\epsilon
\end{pmatrix},\ \check \omega_{2}=\begin{pmatrix} \gamma & \nu \\ -\tau & -\gamma
\end{pmatrix},\ \check \omega_{3}=\begin{pmatrix} \beta & \mu \\ -\sigma & -\beta
\end{pmatrix},\ \check \omega_{4}=\begin{pmatrix} \alpha & \lambda \\ -\rho & -\alpha
\end{pmatrix},
\end{split}
\end{equation}
the transformation of the spin coefficients is derived from 
\begin{equation}
\label{eq:179}
g^{-1}\check \omega_{a}'g=\Lambda\indices{_{a}^{c}}[\check\omega_{c}-g^{-1}e_{c}(g)].
\end{equation}

For RT waves in the algebraically special frame, we have in particular
\begin{equation}
  \label{eq:128RT}
  \begin{split}
  & \check \omega_{1}=\begin{pmatrix} 0 & 0
\\ 0 & 0
  \end{pmatrix},\\
  & \check \omega_{2}=\begin{pmatrix} \frac 12 \partial_{u}\Phi_\circ -\frac 12 r^{-2}\Psi^{0}_{2} & e^{-\Phi_\circ}\widebar\eth_{\circ}\partial_{u}\Phi_\circ-r^{-1}\Psi^{0}_{3}
\\ 0 & -\frac 12 \partial_{u}\Phi_\circ+\frac 12 r^{-2}\Psi^{0}_{2}
  \end{pmatrix},\\
  & \check \omega_{3}=\begin{pmatrix} \frac 12 r^{-1} e^{-\Phi_\circ}(\eth_{\circ}\varphi_{\circ}+\eth_{\circ}\Phi_\circ) & r^{-1}
                                                                                                          e^{-2\Phi_\circ}(\eth_{\circ}\widebar\eth_{\circ}\Phi_\circ-\frac 12)-r^{-2}\Psi^{0}_{2}
\\ 0 & -\frac 12 r^{-1} e^{-\Phi}(\eth_{\circ}\varphi_{\circ}+\eth_{\circ}\Phi_\circ)
  \end{pmatrix},\\
  & \check \omega_{4}=\begin{pmatrix} -\frac 12 r^{-1} e^{-\Phi_\circ}(\widebar\eth_{\circ}\varphi_{\circ}+\widebar\eth_{\circ}\Phi_\circ) & 0
\\ r^{-1} & \frac 12 r^{-1} e^{-\Phi}(\widebar\eth_{\circ}\varphi_{\circ}+\widebar\eth_{\circ}\Phi_\circ)
  \end{pmatrix}.
  \end{split}
\end{equation}
Apart from the conditions~\eqref{eq:79} and~\eqref{eq:129} which are satisfied
for RT waves, in asymptotically flat spacetimes at $\scrip$ in the sense of
Newman and Penrose, all spin coefficients vanish when $r\to \infty$. This is not
the case for RT waves since $\gamma,\nu$ do not vanish asymptotically. We thus first
perform a rotation that turns of the leading order terms in these
spin-coeffcients. As a consequence, the gauge conditions are violated, and we
have to perform additional compensating frame rotations that restore the gauge
conditions. In practice, it is enough to impose the gauge condition
asymptotically. Consider a frame rotation~\eqref{eq:45} of the asymptotic form
\begin{equation}
  A=A_0+\frac{A_1 e^{-\Phi_\circ}}{r}+\mathcal{O}(r^{-2}),\ E=E_0+\frac{E_1e^{-\Phi_\circ}}{r}
  +\mathcal{O}(r^{-2}),\ B=\frac{B_1e^{-\Phi_\circ}}{r}+\mathcal{O}(r^{-2}).
\end{equation}
Note that the we have made explicit the exponential factor $e^{-\Phi_\circ}$ in the
asymptotic expansion. Although it can be absorbed into the accompanying
coefficients, it simplifies the presentation of the results.

Let us start by enforcing the boundary conditions. Demanding
$\gamma'=\mathcal{O}(r^{-1}),\nu'=\mathcal{O}(r^{-1})$ and also the gauge condition
$\tau'-(\bar{\alpha}'+\beta')=\mathcal{O}(r^{-2})$ implies respectively
\begin{equation}
E_0=-\Phi_\circ,\quad A_0= -\eth_\circ\Phi_\circ,\quad B_1=-\eth_\circ u'_0,
\end{equation}
where $u'_0= \int^u_0 dv \,e^{-\Phi_\circ(v,\zeta,\widebar\zeta)}$. Next we impose the
gauge conditions $\epsilon=\mathcal{O}(r^{-3}),\pi=\mathcal{O}(r^{-3})$ which imply
respectively
\begin{equation}
E_1= 2\imath \Im(\eth_\circ u^\prime_0 \,\widebar{\eth}_\circ \phi_\circ),\quad A_1= -E_1\eth_\circ\Phi_\circ +\frac12{\eth}_\circ u^\prime_0 +\widebar{\eth}_\circ u^\prime_0 \big( {\eth}_{\circ}^{2}\Phi_\circ-{({\eth}_{\circ}\Phi_\circ)}^{2}\big).
\end{equation}
In fact $\rho$ turns out to be real to $\mathcal{O}(r^{-3})$ and does not impose
additional constraints. Moreover, $\kappa=\mathcal{O}(r^{-3})$ without imposing further conditions. 

As a result of these frame rotations, the new  spin coefficients turn out to be
\begin{equation}
  \sigma'=r^{-2}\sigma^{\prime 0}+\mathcal O(r^{-3}),\quad \lambda'=r^{-1}\lambda^{\prime 0}+\mathcal O(r^{-2}),\quad \Psi^\prime_n=r^{n-5} \Psi_n^{\prime 0}
  +\mathcal O(r^{n-6}),
\end{equation}
where
\begin{equation}
\sigma^{\prime 0}=e^{-2\Phi_\circ}\eth^{2}_{\circ}u^\prime_0,\quad \lambda^{\prime 0}=e^{-\Phi_\circ}[-{\widebar\eth}_{\circ}^{2}\Phi_\circ+{(\widebar{\eth}_{\circ}\Phi_\circ)}^{2}],\quad \Psi_4^{\prime 0}=\widebar\eth_{\circ}^{2}\partial_{u}\Phi_\circ
-2\widebar\eth_{\circ}\partial_{u}\Phi_\circ\widebar\eth_{\circ}\Phi_\circ.
\end{equation}
Comparing this result with~\eqref{eq:33},~\eqref{eq:30},~\eqref{eq:35}, we find perfect agreement after
realizing that $r'=r e^{\Phi_\circ}+\mathcal{O}(r^0)$ and, since we are not
performing a change of coordinate, that the $\vert$ symbol denoting substitution
of old by new coordinates is now not needed.

Previous relations in asymptotically flat spacetimes in Bondi coordinates, such
as $\lambda^{\prime 0}=\partial_{u'}\widebar\sigma^{\prime 0}$ and $\Psi_4^{\prime 0}=-\partial_{u'}\widebar \lambda^{\prime 0}$
(cf.~\eqref{eq:30} and~\eqref{eq:35}) are not applicable in the present context because $u'$ as a
coordinate is not defined. Instead,
\begin{equation}
  \lambda^{\prime 0}=e^{2\Phi_\circ}[\partial_u+2\partial_u\Phi_\circ]
  \widebar \sigma^{\prime 0},\quad \Psi_4^{\prime 0}=-e^{\Phi_\circ}[\partial_u+\partial_u\Phi_\circ]\lambda^{\prime 0}.
\end{equation}
  
\section{Boosts, spherical harmonics and conformal factor}\label{sec:boosts-conf-fact}

\paragraph*{Boosts}\label{sec:boosts}

In the polar decomposition $g=uh$ of $\mathrm{SL}(2,\mathbb C)$, a
parametrization for $h$ and thus for the boosts in terms of null rotations is
\begin{equation}
  \label{eq:90}
  h=\begin{pmatrix} a & b\\ c & d
  \end{pmatrix}=\begin{pmatrix} e^{-\mathcal E_{R}/2}
    & - \widebar{\mathcal{B}} e^{\mathcal E_{R}/2}\\ -\mathcal B e^{-\mathcal E_R/2} & (1+|\mathcal B|^{2})e^{\mathcal E_R/2}
  \end{pmatrix},
\end{equation}
which is obtained from~\eqref{eq:45} by setting
\begin{equation}
  \label{eq:92}
  E=\mathcal E_{R},\quad B=A=\mathcal B.
\end{equation}
Other useful parametrizations are in terms of boost or rapidity vectors, $\vec \beta,\vec \omega$,
\begin{multline}
  \label{eq:228}
 h=\begin{pmatrix} a & b\\ c & d
  \end{pmatrix}  =\frac{1}{\sqrt 2}\begin{pmatrix} {(1+\gamma)}^{\frac 12}+\gamma\beta^3{(1+\gamma)}^{-\frac 12}& \gamma {(\beta^{1}+\imath\beta^{2})}{(1+\gamma)}^{-\frac 12}\\ \gamma (\beta^{1}-\imath\beta^{2}){(1+\gamma)}^{-\frac 12} & {(1+\gamma)}^{\frac 12}-\gamma\beta^3{(1+\gamma)}^{-\frac 12}
  \end{pmatrix}
  \\=\begin{pmatrix} \cosh{\frac{\omega}{2}}+\hat\omega^{3}\sinh{\frac{\omega}{2}}& (\hat\omega^{1}+\imath \hat\omega^{2})\sinh{\frac{\omega}{2}}\\ (\hat \omega^{1}-\imath \hat \omega^{2})\sinh{\frac{\omega}{2}} & \cosh{\frac{\omega}{2}}-\hat\omega^{3}\sinh{\frac{\omega}{2}}
  \end{pmatrix},
\end{multline}
where $\gamma={(1-|\vec\beta|^2)}^{-\frac 12}$ and the latter is the exponential parametrization, $h=e^{\frac 12 {\vec\omega\cdot\vec\sigma}}$,
with $\hat\omega^{i}=\frac{\omega^{i}}{\omega}$. These parametrizations are related by
\begin{equation}
  \label{eq:236}
  |\vec \beta|=\tanh \omega \iff
  \omega=\frac 12\ln{{(\frac{1+|\vec \beta|}{1-|\vec \beta|})}},
\end{equation}
so that
\begin{equation}
  \label{eq:222}
  \cosh{\frac{\omega}{2}}=\sqrt{\frac{1+\gamma}{2}},\quad
  \sinh
  \frac{\omega}{2}=\frac{\gamma|\vec\beta|}{\sqrt{2(1+\gamma)}},\quad \cosh\omega=\gamma,\quad \sinh \omega=\gamma|\vec \beta|.
\end{equation}
In particular, for a boost in the $z$ direction, $\mathcal B=0$, $\beta^1=0=\beta^2,\beta^3=\beta$, respectively $\omega^1=0=\omega^2,\omega^3=\omega$, so that  $b=0=c$ and
\begin{equation}
  \label{eq:229}
  a=e^{-\mathcal E_{R}/2}={(\frac{1+\beta}{1-\beta})}^{\frac 14}=\cosh{\frac{\omega}{2}}+\sinh{\frac{\omega}{2}},\quad
  d=\frac{1}{a}=e^{\mathcal E_{R}/2}={(\frac{1-\beta}{1+\beta})}^{\frac 14}=\cosh{\frac{\omega}{2}}-\sinh{\frac{\omega}{2}},
\end{equation}
and also
\begin{equation}
  \label{eq:241}
  \frac 12(a^2+\frac{1}{a^2})=\gamma,
  \quad \frac 12(a^2-\frac{1}{a^2})=\gamma\beta.
\end{equation}

\paragraph*{Relation to the first four harmonics}

When using that $\zeta=\cot{\frac{\theta}{2}e^{\imath\phi}}$, the first four spherical harmonics are ${}_{0}Y_{00}=\frac{1}{\sqrt{4\pi}}$,
\begin{equation}
  \label{eq:162}
  \begin{split}
    & {}_{0}Y_{1,-1}=\sqrt{\frac{3}{8\pi}}\sin\theta e^{-\imath\phi}=\sqrt{\frac{3}{2\pi}}
      \frac{\widebar\zeta}{|\zeta|^{2}+1}, \\
     & {}_{0}Y_{10}=\sqrt{\frac{3}{4\pi}}\cos\theta
      =\sqrt{\frac{3}{4\pi}}\frac{|\zeta|^{2}-1}{|\zeta|^{2}+1},
 \\
         & {}_{0}Y_{11}=-\sqrt{\frac{3}{8\pi}}\sin\theta e^{\imath\phi}=-\sqrt{\frac{3}{2\pi}}
      \frac{\zeta}{|\zeta|^{2}+1}.
  \end{split}
\end{equation}
In the current context, it is sometimes convenient to work with real spherical
harmonics. They may be defined by
\begin{equation}
  \label{eq:231}
  \begin{split}
    & {}^R Y_{jm}=\frac{\imath}{\sqrt 2}({}_0Y_{jm}-{(-)}^m{}_0Y_{j,-m}),\  m<0,\\
    & {}^{R}Y_{j0}= {}_0Y_{j0},\\
    & {}^{R}Y_{jm}=\frac{1}{\sqrt 2}({}_0Y_{j,-m}+{(-)}^m{}_0Y_{jm}),
      \ m>0,    
  \end{split}
\end{equation}
so that 
\begin{equation}
  \label{eq:235}
  \begin{split} ({}^{R}Y_{1,-1},{}^{R}Y_{1,0},{}^{R}Y_{1,1})&
=\sqrt{\frac{3}{4\pi}}\frac{1}{|\zeta|^2+1}(\imath(\widebar\zeta-\zeta),|\zeta^2|-1,\zeta+\widebar\zeta)\\ &=\sqrt{\frac{3}{4\pi}}(\sin\theta\sin\phi,\cos\theta,\sin\theta\cos\phi)         =\sqrt{\frac{3}{4\pi}}(n^2,n^3,n^1),
\end{split}
\end{equation}
where $n^i$ denote the components of the radial unit vector in $\mathbb R^3$. 
Since
\begin{multline}
  \label{eq:239}
  X_g=|a\zeta+b|^2+|c\zeta+d|^2=(|a|^2+|c|^2)|\zeta|^2+(a\widebar b+c\widebar d)\zeta+(b\widebar a+d\widebar c)\widebar \zeta+|b|^2+|d|^2\\=
  \frac 12 (|a|^2+|c|^2+|b|^2+|d|^2)(|\zeta|^2+1)+(a\widebar b+c\widebar d)\zeta+(b\widebar a+d\widebar c)\widebar \zeta\\+\frac 12(|a|^2+|c|^2-|b|^2-|d|^2)(|\zeta|^2-1),
\end{multline}
it follows that $R$ and the three boost parameters contained in $h$ of~\eqref{eq:90} are
directly related to the $4$ real paramaters that appear in the expansion of
$w_v=c^{jm}_R {}^{R}Y_{jm}|_{j\leq 1}$ in equation~\eqref{eq:177} in terms of the first four
real spherical harmonics, 
\begin{equation}
  \label{eq:192}
  \begin{split}
    & \sqrt{\frac{1}{4\pi}}c^{00}_R=\frac{|a|^{2} + |c|^2+ |b|^{2}+|d|^{2}}{2R},\\
    & \sqrt{\frac{3}{4\pi}}c^{1-1}_R=\frac{\imath(a\widebar b+c\widebar d-b\widebar a-d\widebar c)}{2R},\\
    &\sqrt{\frac{3}{4\pi}}c^{10}_R
      =\frac{|a|^{2}+|c|^{2}-|b|^{2}-|d|^{2}}{2R},\\
    & \sqrt{\frac{3}{4\pi}}c^{11}_R=\frac{a\widebar b+c\widebar d
      +b\widebar a+d\widebar c}{2R}.
  \end{split}
\end{equation}
We will continue to work here with complex harmonics and the expansion in~\eqref{eq:119},
so that $R$ and the three boost parameters contained in $h$ of~\eqref{eq:90} are directly
related to the $4$ real paramaters in $c^{00}=\widebar c^{00}$,
$c^{10}=\widebar c^{10}$, $c^{11}=-\widebar c^{1,-1}$ through the invertible
relations
\begin{equation}
  \label{eq:225}
  \begin{split}
    & \sqrt{\frac{1}{4\pi}}c^{00}=\frac{|a|^{2} + |c|^2+ |b|^{2}+|d|^{2}}{2R},\\
    & \sqrt{\frac{3}{2\pi}}c^{1-1}=\frac{b\widebar a+d\widebar c}{R},\\
    &\sqrt{\frac{3}{4\pi}}c^{10}
      =\frac{|a|^{2}+|c|^{2}-|b|^{2}-|d|^{2}}{2R},\\
    & \sqrt{\frac{3}{2\pi}}c^{11}=-\frac{a\widebar b+c\widebar d}{R}.
  \end{split}
\end{equation}
Note that for ``small'' parameters,
\begin{equation}
  \label{eq:317}
  R=1+\epsilon r,\quad a=1+\epsilon\alpha,\quad b=\epsilon\beta,\quad c=\epsilon\gamma,\quad d=1-\epsilon\alpha,
\end{equation}
\begin{equation}
  \label{eq:319}
  \begin{split}
    & \sqrt{\frac{1}{4\pi}}c^{00}=1-\epsilon r+O(\epsilon^2),\\
    & \sqrt{\frac{3}{4\pi}}c^{1-1}=\epsilon(\beta+\widebar\gamma)+O(\epsilon^2),\\
    &\sqrt{\frac{3}{4\pi}}c^{10}
      =\epsilon(\alpha+\widebar\alpha)+O(\epsilon^2),\\
    & \sqrt{\frac{3}{4\pi}}c^{11}=-\epsilon(\widebar\beta+\gamma)+O(\epsilon^2).
  \end{split}
\end{equation}

The factor $w_v$ may also be parametrized in terms of rapidity and boost vectors
 \begin{equation}
   \label{eq:195}
   w_v=\frac{\cosh \omega (|\zeta|^{2}+1)+(\hat\omega^{1}
     -\imath\hat\omega^{2})\sinh\omega\zeta
     +(\hat\omega^{1}+\imath\hat\omega^{2})\sinh\omega\widebar\zeta+\hat\omega^{3}\sinh\omega(|\zeta|^{2}-1)}{R(|\zeta|^{2}+1)},
 \end{equation}
 \begin{equation}
   \label{eq:158} w_v=\gamma \frac{(|\zeta|^{2}+1)+( \beta^{1}-\imath \beta^{2})\zeta
     +(\beta^{1}+\imath\beta^{2})\widebar\zeta+\beta^{3}(|\zeta|^{2}-1)}{R(|\zeta|^{2}+1)}.
 \end{equation}
 In other words, one may parametrize $w_v$ in terms of a future directed
 time-like vector $v^a$, respectively in terms of $R$ and a normalized future
 directed time-like vector $u^a$,
 \begin{equation}
   \label{eq:216}
   w_v=v_0-v_i n^i=R^{-1}w_{1,g},\quad w_{1,g}=u_0-u_i n^i,
 \end{equation}
where one may assume that $u_0=u^0>0$ and, as a consequence of the determinant
condition,
\begin{equation}
  \label{eq:237}
  u^a u_a\equiv(u^0)^2-|\vec u|^2=1,\quad z\equiv \frac{u_0}{|\vec u|}=\sqrt{1+|\vec u|^{-2}}.  
\end{equation}
More explicitly,
 \begin{equation}
   \label{eq:243}
   \begin{split}
     u^0&=\frac{|a|^{2} + |c|^2+ |b|^{2}+|d|^{2}}{2}=\gamma=\cosh\omega,\\
     u^2&=\frac{\imath(a\widebar b+c\widebar d-b\widebar a-d\widebar c)}{2}=\gamma\beta^2=\sinh\omega\hat\omega^2,\\
     u^3&=\frac{|a|^{2}+|c|^{2}-|b|^{2}-|d|^{2}}{2}=\gamma\beta^3=
          \sinh\omega\hat\omega^3,\\
     u^1&=\frac{a\widebar b+c\widebar d
          +b\widebar a+d\widebar c}{2}=\gamma\beta^1=\sinh\omega\hat\omega^1.
   \end{split}
 \end{equation}
 For the inverse relations that determine the Lorentz boosts
 from the coefficients of the spherical harmonics, one may use
 \begin{equation}
   \label{eq:351}
   \begin{split}
    & R={\big[\frac{1}{4\pi} {(c^{00})}^2-\frac{3}{4\pi}{(c^{10})}^2-\frac{3}{2\pi}|c^{1-1}|^2\big]}^{-\frac 12},\quad 
     \omega=\cosh^{-1}{\big(R\sqrt{\frac{1}{4\pi}}c^{00}\big)},\\
    & \hat\omega^{1}+\imath\hat\omega^{2}=\frac{R\sqrt{\frac{3}{2\pi}}c^{1-1}}{\sinh\omega},\quad
  \hat\omega^3= \frac{R\sqrt{\frac{3}{4\pi}}c^{10}}{\sinh\omega}.
   \end{split}
 \end{equation}

\paragraph*{Conformal factors}\label{sec:conformal-factors}

Under the fractional linear transformations associated to $g^{-1}$ which acts as
$x_g(x^\prime)$ in~\eqref{eq:82}, the line element of the unit sphere in stereographic
coordinates $\zeta=\cot\frac{\theta}{2}e^{\phi}$,
\begin{equation}
  \label{eq:25}
  d\widebar s^2=d\theta^2+\sin^2\theta d\phi^2=\frac{4d\zeta d\widebar\zeta}{{(|\zeta|^2+1)}^2},
\end{equation}
transforms as
\begin{equation}
  \label{eq:230}
  d\widebar s^2=\frac{4d\zeta^\prime d\widebar\zeta^\prime}{{(X_g(x^\prime))}^2}
  =K^2_g(x^\prime)\frac{4d\zeta^\prime d\widebar\zeta^\prime}{{(|\zeta^\prime|^2+1)}^2},\quad
  K_g(x)=\frac{|\zeta|^{2}+1}{X_{g}(x)},
\end{equation}
cf.~\cite{Sachs1962},~\cite{Held1970}. As a consequence, the square root of the
conformal factor $K_g$ and $w_{1,g}$ are related as
 \begin{equation}
   \label{eq:140}
   K_{g}={(Rw_v)}^{-1}={(w_{1,g})}^{-1}.
 \end{equation}

 Rather than giving the new as functions of the old coordinates as
 in~\eqref{eq:147a}, one may also give the old as functions of the new
 coordinates. More explicitly, under fractional linear transformations
 $\tilde g^{-1}$ that acts as $x_{\tilde g}(x^\prime)$, we have
\begin{equation}
   \label{eq:143}
   w_{1,g}(x_{\tilde g}(x^\prime))
   =\frac{X_{g\circ\tilde g}}{X_{\tilde g}}(x^\prime)
   =\big[\frac{w_{1,g\circ\tilde g}}{w_{1,\tilde g}}\big](x^\prime).
 \end{equation}
 Defining
\begin{equation}
  \label{eq:153}
  e^{E^g_R}(x)=(w_{1,g}(x))^{-1},\quad e^{\imath E_I^g}(x)=\frac{\widebar c\widebar\zeta+\widebar d}{c\zeta+d},
\end{equation}
we also have
\begin{equation}
  \label{eq:346}
  e^{E^g_R}(x_{\tilde g}(x^\prime))
  =[e^{-E_R^{\tilde g}}e^{E_R^{g\circ\tilde g}}](x^\prime),\quad
  e^{\imath E^g_I}(x_{\tilde g}(x^\prime))
  =[e^{-\imath E_I^{\tilde g}}e^{\imath E_I^{g\circ\tilde g}}](x^\prime).
\end{equation}

In particular, when $\tilde g=g^{-1}$, $\zeta=\frac{d\zeta^\prime-b}{-c\zeta^\prime+a}$
this becomes, 
 \begin{equation}
   \label{eq:132}
   \begin{split}
   & w_{1,g}(x_{g^{-1}}(x^\prime))
     ={\big(w_{1,g^{-1}}(x^\prime)\big)}^{-1},\\ & {E^g_R}(x_{g^{-1}}(x^\prime))
    ={-E^{g^{-1}}_R}(x^\prime),\quad
{\imath E^g_I}(x_{g^{-1}}(x^\prime))={-\imath E^{g^{-1}}_I}(x^\prime),
   \end{split}                                 
 \end{equation}
since $w_{1,e}=1=e^{E^e_R}=e^{\imath E^e_I}$.

\paragraph*{Spin and boost weighted functions}\label{sec:spin-bost-weighted}

Under the fractional linear transformations $g^{-1}$ as in~\eqref{eq:82}, spin
and boost weighted ``functions'' $\eta^{s,w}(x)$ of weights $(s,w)$ transform as
\begin{equation}
  \label{eq:347}
  \eta^{\prime s,w}_{;g^{-1}}(x^\prime)=e^{-w E^g_R-s\imath E^g_I}(x^\prime)\eta^{s,w}(x_{g}(x^\prime)),
\end{equation}
and from a realization of $\mathrm{PSL}(2,\mathbb C)$,
\begin{multline}
  \label{eq:359}
  {(\eta^{\prime s,w}_{;g^{-1}})}^\prime_{;\tilde g^{-1}}(x^{\prime\prime})=
  e^{-w E^{\tilde g}_R-s\imath E^{\tilde g}_I}(x^{\prime\prime})\eta^{\prime s,w}_{;g^{-1}}(x^\prime_{\tilde g}(x^{\prime\prime}))=e^{-w E^{g\circ \tilde g}_R-s\imath E^{g\circ \tilde g}_I}(x^{\prime\prime})\eta^{s,w}(x_{g\circ\tilde g}(x^{\prime\prime}))\\=
  \eta^{\prime s,w}_{;{(g\circ \tilde g)}^{-1}}(x^{\prime\prime}),
\end{multline}
on account of~\eqref{eq:346}. The relation to the standard classification of
representations of the Lorentz group in the form presented
in~\cite{gelfand1966generalized} has been worked out in~\cite{Held1970} (see also~\cite{Pasterski:2016qvg,Pasterski:2017kqt}
for recent related discussions). For our purposes here, one may show by direct
computation that
\begin{multline}
  \label{eq:348}
  [\eth_\circ \eta^{s,w}](x_g(x^\prime))=e^{w E^g_R+(s+1)\imath E^g_I}(x^\prime)  \big[\eth^\prime_{1,g}  +w\eth^\prime_{1,g}E^g_R(x^\prime)\big]\eta^{\prime s,w}_{;g^{-1}}(x^\prime)\iff\\
  \big[\eth^\prime_{1,g}  +w\eth^\prime_{1,g}E^g_R(x^\prime)\big]\eta^{\prime s,w}_{;g^{-1}}(x^\prime)=e^{-w E^g_R-(s+1)\imath E^g_I}(x^\prime)[\eth_\circ \eta^{s,w}](x_g(x^\prime)),
\end{multline}
with the understanding that $\eth^\prime_{1,g}$ corresponds to the same operator than
$\eth_{1,g}$ but in terms of the primed variables, and that the spin weight of
$E^g_{R}(x^\prime)$ is $0$. When using~\eqref{eq:99} in the form
\begin{equation}
  \label{eq:349}
  \eth_{1,g}\eta^{s}=e^{-E^g_R}(\eth_\circ \eta^s-s\eth_\circ E^g_R\eta^s),
\end{equation}
in terms of the primed variables, this yields
\begin{equation}
  \label{eq:350}
  \eth^\prime_\circ\eta^{\prime s,w}_{;g^{-1}}(x^\prime)+(w-s)\eth_\circ^\prime E^g_R(x^\prime)\eta^{\prime s,w}_{;g^{-1}}(x^\prime)=e^{-(w-1) E^g_R-(s+1)\imath E^g_I}(x^\prime)[\eth_\circ \eta^{s,w}](x_g(x^\prime)),
\end{equation}
which is compatible with~\eqref{eq:7} when using~\eqref{eq:132} and 
\begin{equation}
  \label{eq:372}
  e^{E^{g^{-1}}_R-\imath E^{g^{-1}}_I}(x_g(x^\prime))\frac{P_\circ(x^\prime)}{P_\circ(x_g(x^\prime))}
  \frac{\partial\widebar\zeta}{\partial\widebar\zeta^\prime}=1,
\end{equation}
cf.~equation (2.19) of~\cite{Held1970} (up to the change of conventions $s\to -s$ so as to
agree with the convention of~\cite{Penrose:1984} for the eth operator).

Similarly,
\begin{equation}
  \label{eq:168}
  \big[\widebar \eth^\prime_{1,g} +w\widebar\eth^\prime_{1,g}E^g_R(x^\prime)\big]\eta^{\prime s,w}_{;g^{-1}}(x^\prime)=
  e^{-w E^g_R-(s-1)\imath E^g_I}(x^\prime)[\widebar\eth_\circ \eta^{s,w}](x_g(x^\prime)),
\end{equation}
and
\begin{equation}
  \label{eq:169}
  \widebar\eth^\prime_\circ\eta^{\prime s,w}_{;g^{-1}}(x^\prime)+(w+s)\eth_\circ^\prime E^g_R(x^\prime)\eta^{\prime s,w}_{;g^{-1}}(x^\prime)=e^{-(w-1) E^g_R-(s-1)\imath E^g_I}(x^\prime)[\eth_\circ \eta^{s,w}](x_g(x^\prime)).
\end{equation}
It follows that
\begin{multline}
  \label{eq:389}
  \big[\eth^\prime_{1,g}\widebar \eth^\prime_{1,g}+w\eth^\prime_{1,g}\widebar\eth^\prime_{1,g}E^g_R+
  w\widebar\eth^\prime_{1,g}E^g_R\eth^\prime_{1,g}+
  w\eth^\prime_{1,g}E^g_R\widebar\eth^\prime_{1,g}+w^2\eth^\prime_{1,g}E^g_R
  \widebar\eth^\prime_{1,g}E^g_R
  \big]\eta^{\prime s,w}_{;g^{-1}}(x^\prime)\\=
  e^{w E^g_R-s\imath E^g_I}(x^\prime)[\eth_\circ\widebar\eth_\circ \eta^{s,w}](x_g(x^\prime)). 
\end{multline}

Note that equations~\eqref{eq:348},~\eqref{eq:168},~\eqref{eq:389} may be re-written as
\begin{equation}
  \label{eq:360}
  \begin{split}
  \eth^\prime_{1,g}(e^{wE^g_R}\eta^{\prime s,w}_{;g^{-1}})(x^\prime)&=e^{-(s+1)\imath E^g_I}(x^\prime)
                                                 [\eth_\circ \eta^{s,w}](x_g(x^\prime)),\\
    \widebar\eth^\prime_{1,g}(e^{wE^g_R}\eta^{\prime s,w}_{;g^{-1}})(x^\prime)&=e^{-(s-1)\imath E^g_I}(x^\prime)                                                                              [\eth_\circ \eta^{s,w}](x_g(x^\prime)),\\
   e^{-wE^g_R}\eth^\prime_{1,g}\Big(e^{wE^g_R}\big(\widebar\eth^\prime_{1,g}(e^{wE^g_R}\eta^{\prime s,w}_{;g^{-1}})\big)\Big)(x^\prime)&= e^{-s\imath E^g_I}(x^\prime)                                                                              [\eth_\circ \widebar\eth_\circ \eta^{s,w}](x_g(x^\prime)).
  \end{split}
\end{equation}
In particular, 
\begin{equation}
  \label{eq:342a}
  \frac{d\zeta d\widebar\zeta}{\imath P^2_\circ(x)}[\eth_\circ \eta^{-1,-2}](x)=\frac{d\zeta^\prime d\widebar\zeta^\prime}{\imath P^2_{1,g}(x^\prime)}[\eth^\prime_{1,g} (e^{-2E^g_R}\eta^{-1,-2}_{;g^{-1}})](x^\prime),
\end{equation}
so that integrations by parts with respect to $\eth_{1,g}$ using the measure
determined by $P_{1,g}$ follow from those for $\eth_\circ$ with the standard
integration measure determined by $P_\circ$.

For later purposes, in the $u$-independent case, $\eta^{\prime s,w}=\eta^{\prime s,w}(x^\prime)$,
equation~\eqref{eq:366} becomes
\begin{equation}
  \label{eq:352}
  \tilde\eta^{s,w}_{;g}(\tilde x)=\big(e^{w E_R^g+s\imath E_I^g}\eta^{\prime s,w}\big)\big\rvert,
\end{equation}
while equation~\eqref{eq:7} reduces to
\begin{equation}
  \label{eq:373}
  \tilde\eth_\circ \tilde\eta^{s,w}_{;g}=\big( e^{(w-1)E^g_R+(s+1)\imath E^g_I}[\eth_\circ^\prime+(w-s)\eth_\circ^\prime E^g_R]\eta^{\prime s,w}\big)\big\rvert.
\end{equation}
Furthermore,
\begin{equation}
  \tilde{\widebar\eth}_\circ \tilde\eta^{s,w}_{;g}=\big( e^{(w-1)E^g_R+(s-1)\imath E^g_I}[\widebar\eth_\circ^\prime+(w+s)\widebar\eth_\circ^\prime E^g_R]\eta^{\prime s,w}\big)\big\rvert,
\end{equation}
When using $\eth_\circ^{\prime 2} E^g_R={(\eth_\circ^\prime E^g_R)}^2$ and its complex conjugate, this
yields
\begin{multline}
  \label{eq:377}
  \tilde\eth^2_\circ \tilde\eta^{s,w}_{;g}=\big( e^{(w-2)E^g_R+(s+2)\imath E^g_I}\\
  [\eth_\circ^{ \prime 2}+(w-s)(w-s-1){(\eth_\circ^\prime E^g_R)}^2+2(w-s-1)\eth_\circ^\prime E^g_R\eth_\circ^\prime]\eta^{\prime s,w}\big)\big\rvert.
\end{multline}
\begin{multline}
  \label{eq:377a}
  \tilde{\widebar\eth}^2_\circ \tilde\eta^{s,w}_{;g}=\big( e^{(w-2)E^g_R+(s-2)\imath E^g_I}\\
  [\widebar\eth_\circ^{ \prime 2}+(w+s)(w+s-1){(\widebar\eth_\circ^\prime E^g_R)}^2+2(w+s-1)\widebar\eth_\circ^\prime E^g_R\widebar
  \eth_\circ^\prime]\eta^{\prime s,w}\big)\big\rvert.
\end{multline}
\begin{multline}
  \label{eq:345}
  \tilde\eth_\circ \tilde{\widebar\eth}_\circ \tilde\eta^{s,w}_{;g}=\big( e^{(w-2)E^g_R+s\imath E^g_I}
  [\eth_\circ^{ \prime}\widebar\eth_\circ^{ \prime }+(w+s)\eth_\circ^\prime\widebar\eth_\circ^{ \prime } E^g_R+(w^2-s^2)\eth_\circ^\prime E^g_R\widebar\eth_\circ^{ \prime } E^g_R
  \\+(w+s)\widebar\eth_\circ^\prime E^g_R\eth_\circ^\prime+(w-s)\eth_\circ^\prime E^g_R\widebar\eth_\circ^\prime]\eta^{\prime s,w}\big)\big\rvert,
\end{multline}
\begin{multline}
  \label{eq:387}
  \tilde{\widebar\eth}_\circ\tilde\eth_\circ  \tilde\eta^{s,w}_{;g}=\big( e^{(w-2)E^g_R+s\imath E^g_I}
  [\widebar\eth_\circ^{ \prime }\eth_\circ^{ \prime}+(w-s)\widebar\eth_\circ^{ \prime } \eth_\circ^\prime E^g_R+(w^2-s^2)\eth_\circ^\prime E^g_R\widebar\eth_\circ^{ \prime } E^g_R
  \\+(w+s)\widebar\eth_\circ^\prime E^g_R\eth_\circ^\prime+(w-s)\eth_\circ^\prime E^g_R\widebar\eth_\circ^\prime]\eta^{\prime s,w}\big)\big\rvert,
\end{multline}
with
\begin{equation}
  \label{eq:388}
  \eth_\circ^\prime\widebar\eth_\circ^{ \prime } E^g_R(x^\prime)=\frac 12(1-e^{2E^g_R}(x^\prime)), 
\end{equation}
cf.~\eqref{eq:262}. 

In particular, when acting on a field of spin and boost weight $(0,1)$, the
operators
$\eth_\circ^{ \prime 2},\widebar\eth_\circ^{ \prime 2},\eth_\circ^{ \prime 2}\widebar\eth_\circ^{ \prime 2}$
transform covariantly,
\begin{equation}
  \label{eq:380}
  \begin{split}
  \tilde\eth^2_\circ \tilde\eta^{0,1}_{;g}&=\big( e^{-E^g_R+2\imath E^g_I}\eth_\circ^{\prime 2}
                                  \eta^{\prime 0,1}\big)\big\rvert,\\
  \tilde{\widebar\eth}^2_\circ \tilde\eta^{0,1}_{;g}&=\big( e^{-E^g_R-2\imath E^g_I}\widebar\eth_\circ^{\prime 2}
                                                      \eta^{\prime 0,1}\big)\big\rvert,\\
    \tilde\eth^2_\circ  \tilde{\widebar\eth}^2_\circ\tilde\eta^{0,1}_{;g}&=\big( e^{-3E^g_R}\eth_\circ^{\prime 2}\widebar\eth_\circ^{\prime 2} \eta^{\prime 0,1}\big)\big\rvert,
  \end{split}
\end{equation}
while
\begin{equation}
  \label{eq:355}
  \tilde\eth_\circ \tilde{\widebar\eth}_\circ \tilde\eta^{0,1}_{;g}=\big( e^{-E^g_R}
  [\eth_\circ^{ \prime}\widebar\eth_\circ^{ \prime }+\eth_\circ^\prime\widebar\eth_\circ^{ \prime } E^g_R+\eth_\circ^\prime E^g_R\widebar\eth_\circ^{ \prime } E^g_R
  +\widebar\eth_\circ^\prime E^g_R\eth_\circ^\prime+\eth_\circ^\prime E^g_R\widebar\eth_\circ^\prime]\eta^{\prime 0,1}\big)\big\rvert.
\end{equation}

\paragraph*{Time-dependent transformations}\label{sec:time-depend-transf}

Consider now a time-dependent fractional linear transformations
$\tilde g^{-1}(u)$ as in~\eqref{eq:183}, for which
\begin{equation}
   \label{eq:272}
   \frac{\partial \zeta^\prime}{\partial \zeta}=\frac{1}{{(-\tilde c\zeta+\tilde a)}^2}
   ={(\tilde c\zeta^\prime +\tilde d)}^2.
\end{equation}
and let
\begin{equation}
   \label{eq:276}
   \partial_u \zeta(\zeta^\prime)=\frac{Y^\prime_{\tilde g^{-1}(u)}(\zeta^\prime)}{{{(\tilde c\zeta^\prime+\tilde d)}^2}},\quad
   \partial_u \zeta(\zeta^\prime)|_{\zeta^\prime(\zeta)}=Y_{\tilde g^{-1}(u)}(\zeta), 
\end{equation}
where
\begin{multline}
  \label{eq:277}
  Y^\prime_{\tilde g^{-1}(u)}(\zeta^\prime)=(\dot{\tilde a}\zeta^\prime +\dot{\tilde b})(\tilde c\zeta^\prime +\tilde d)
  -(\dot{\tilde c}\zeta^\prime +\dot{\tilde d})(\tilde a\zeta^\prime
  +\tilde b) \\=(\dot{\tilde a} \tilde c-\dot{\tilde c} \tilde a)\zeta^{\prime ^2}
  +2(\dot{\tilde a} \tilde d-\dot{\tilde c}\tilde b)
  \zeta^\prime+\dot{\tilde b}\tilde d-\dot{\tilde d}\tilde b, 
\end{multline}
while
\begin{equation}
  \label{eq:278}
  Y_{\tilde g^{-1}(u)}(\zeta)=(\dot{\tilde d}\tilde c-\dot{\tilde c}\tilde d)\zeta^2+2(\dot{\tilde a}\tilde d
  -\dot{\tilde b}\tilde c)\zeta+\dot{\tilde b}\tilde a-\dot{\tilde a}\tilde b,
\end{equation}
with $Y^\prime_{\tilde g^{-1}(u)}(\zeta^\prime)=\frac{\partial\zeta^\prime}{\partial\zeta}Y_{\tilde g^{-1}(u)}(\zeta)$. Note also that, with $g$ as in~\eqref{eq:45}, the
coefficient in $Y_{\tilde g^{-1}(u)}(\zeta)$, $Y^\prime_{\tilde g^{-1}(u)}(\zeta^\prime)$ are directly related to the matrix elements of
the right and left invariant Maurer-Cartan forms $d\tilde g {\tilde g}^{-1}$ and
$\tilde g^{-1}d\tilde g$,
\begin{equation}
  \label{eq:296}
  \dot{\tilde g} \tilde g^{-1}
  =\begin{pmatrix} \dot{\tilde a} \tilde d-\dot{\tilde b}\tilde c
    & \dot{\tilde b}\tilde a-\dot{\tilde a}\tilde b
    \\ \dot{\tilde c}\tilde d-\dot{\tilde d}\tilde c
    & \dot{\tilde d}\tilde a -\dot{\tilde c} \tilde b 
  \end{pmatrix},\quad \tilde g^{-1}\dot{\tilde g}
  =\begin{pmatrix} \dot{\tilde a}\tilde d -\dot{\tilde c}\tilde b
    & \dot{\tilde b}\tilde d-\dot{\tilde d}\tilde b \\
    \dot{\tilde c}\tilde a -\dot{\tilde a} \tilde c
    & \dot{\tilde d}\tilde a-\dot{\tilde b}\tilde c 
  \end{pmatrix}.
\end{equation}

The aim is to compute the time dependence of $\varphi^\prime_{\tilde g^{-1}(u)},P^{\prime -2}_{\tilde g^{-1}(u)}$ due to
the time dependent fractional linear transformation. For $\varphi^\prime_{\tilde g^{-1}(u)}$, when
taking account~\eqref{eq:286}, we need to evaluate the term in the squared
parenthesis. When using that
\begin{equation}
    \label{eq:274}
   (\partial\varphi)|={(\tilde c\zeta^\prime +\tilde d)}^2\partial^\prime\varphi^\prime_{\tilde g^{-1}(u)}+\tilde c(\tilde c\zeta^\prime+\tilde d), 
\end{equation}
and taking into account~\eqref{eq:276}, it follows that 
\begin{multline}
  \label{eq:181}
  (\partial \varphi)|\partial_u\zeta(\zeta^\prime)-\partial_u\ln{(\tilde c\zeta^\prime +\tilde d)}=
  \partial^\prime\varphi^\prime_{\tilde g^{-1}(u)} Y^\prime_{\tilde g^{-1}(u)}
  +\frac{\tilde c Y^\prime_{\tilde g^{-1}(u)}-(\dot{\tilde c}\zeta^\prime+\dot{\tilde d})}{\tilde c\zeta^\prime +\tilde d}\\=
  \partial^\prime\varphi^\prime_{\tilde g^{-1}(u)} Y^\prime_{\tilde g^{-1}(u)}+\frac 12 \partial^\prime Y^\prime_{\tilde g^{-1}(u)}=\frac 12 \widebar\eth^\prime_{;\tilde g^{-1}(u)} \widebar{\mathcal{Y}}^\prime_{\tilde g^{-1}(u)}, 
\end{multline}
where the last equality follows from the definition of the fields
$\mathcal{Y}^\prime_{\tilde g^{-1}(u)},\widebar{\mathcal{Y}}^\prime_{\tilde g^{-1}(u)}$ of spin weight $-1,1$ through
\begin{equation}
  \label{eq:283}
  \mathcal Y^\prime_{\tilde g^{-1}(u)}=P^{\prime -1}_{;\tilde g^{-1}(u)}\widebar Y^\prime_{\tilde g^{-1}(u)},\quad \widebar{\mathcal{Y}}^\prime=P^{\prime -1}_{;\tilde g^{-1}(u)} Y^\prime_{\tilde g^{-1}(u)},
\end{equation}
the definition of $\eth^\prime_{;\tilde g^{-1}(u)}$ as the eth operator
involving $P^{\prime}_{;\tilde g^{-1}(u)}$, and using~\eqref{eq:5}. In other
words, if
\begin{equation}
  \label{eq:185}
    \varphi^\prime_{;\tilde g^{-1}(u)}(\zeta^\prime,\widebar\zeta^\prime)
    =\varphi(\zeta,\widebar\zeta)+\ln |-\tilde c\zeta+\tilde a|^2
    =\varphi(\zeta,\widebar\zeta)-\ln |\tilde c\zeta^\prime+\tilde d|^2,
\end{equation}
it follows that
\begin{equation}
  \label{eq:287}
  \partial_u\varphi^\prime_{;\tilde g^{-1}(u)}=\frac 12 (\eth^\prime_{;\tilde g^{-1}(u)}\mathcal Y^\prime_{\tilde g^{-1}(u)}+\widebar\eth^\prime_{;\tilde g^{-1}(u)} \widebar{\mathcal{Y}}^\prime_{\tilde g^{-1}(u)}). 
\end{equation}
In terms of 
\begin{equation}
  \label{eq:288}
  P^\prime_{;\tilde g^{-1}(u)}(\zeta^\prime,\widebar\zeta^\prime)
  =P(\zeta(\zeta^\prime),\widebar\zeta(\widebar\zeta^\prime))|\frac{\partial\zeta^\prime}{\partial \zeta}|, 
\end{equation}
this yields
\begin{equation}
  \label{eq:289}
  \partial_u P^{\prime -2}_{;\tilde g^{-1}(u)}=P^{\prime -2}_{;\tilde g^{-1}(u)}(\eth^\prime_{;\tilde g^{-1}(u)}\mathcal Y^\prime_{\tilde g^{-1}(u)}+\widebar\eth^\prime _{;\tilde g^{-1}(u)}\widebar{\mathcal{Y}}^\prime_{\tilde g^{-1}(u)}). 
\end{equation}
Note that $\mathcal Y^\prime_{\tilde g^{-1}(u)},\widebar{\mathcal{Y}}^\prime_{\tilde g^{-1}(u)}$ satisfy
\begin{equation}
  \label{eq:184}
  \widebar\eth^\prime_{;\tilde g^{-1}(u)} \mathcal Y^\prime_{\tilde g^{-1}(u)}=0=\eth^\prime_{;\tilde g^{-1}(u)}\widebar \widebar\mathcal Y^\prime_{\tilde g^{-1}(u)}, 
\end{equation}
and describe conformal Killing vector fields of
\begin{equation}
  \label{eq:186}
  ds^{\prime 2}=-2P^{\prime -2}_{;\tilde g^{-1}(u)}d\zeta^\prime d\widebar\zeta^\prime, 
\end{equation}
while
\begin{equation}
  \label{eq:279}
  \mathcal Y^\prime_{\tilde g^{-1}(u)\circ}=P^{ -1}_\circ(\zeta^\prime,\widebar\zeta^\prime)\widebar Y^\prime_{;\tilde g^{-1}(u)},\quad \widebar{\mathcal{Y}}^\prime_{\tilde g^{-1}(u)\circ}=P_\circ^{-1}(\zeta^\prime,\widebar\zeta^\prime) Y^\prime_{\tilde g^{-1}(u)},
\end{equation}
satisfy
\begin{equation}
  \label{eq:280}
  \widebar\eth^\prime_\circ \mathcal Y^\prime_{\tilde g^{-1}(u)\circ}=0=\eth^\prime_\circ\widebar{\mathcal Y}^\prime_{\tilde g^{-1}(u)\circ}, 
\end{equation}
and describe conformal Killing vectors of the line element of the unit-sphere,
\begin{equation}
  \label{eq:281}
  ds^2=-2P^{-2}_\circ(\zeta^\prime,\widebar\zeta^\prime) d\zeta^\prime d\widebar\zeta^\prime,
\end{equation}
with
\begin{equation}
  \label{eq:298}
  \frac 12 \widebar\eth^\prime_\circ\widebar{\mathcal Y}^\prime_{\tilde g^{-1}(u)\circ}=\frac{(\dot{\tilde a}\tilde c-\dot{\tilde c}\tilde a)\zeta^\prime+(\dot{\tilde d}\tilde b-\dot{\tilde b}\tilde d)\widebar\zeta^\prime-(\dot{\tilde a} \tilde d-\dot{\tilde c} \tilde b)(|\zeta^\prime|^2-1)}{|\zeta^\prime|^2+1}, 
\end{equation}
while 
\begin{equation}
  \label{eq:297}
  \frac 12 \widebar\eth_\circ\widebar{\mathcal Y}_{\tilde g^{-1}(u)\circ}=\frac{(\dot{\tilde d}\tilde c-\dot{\tilde c}\tilde d)\zeta+(\dot{\tilde b}\tilde a-\dot{\tilde a}\tilde b)\widebar\zeta-(\dot{\tilde a} \tilde d-\dot{\tilde b} \tilde c)(|\zeta|^2-1)}{|\zeta|^2+1}. 
\end{equation}
In terms of $V^\prime_{\circ;g^{-1}(u)}$ defined in~\eqref{eq:256a}, 
the associated infinitesimal conformal factors are related as
\begin{multline}
  \label{eq:282}
  \frac 12(\eth^\prime_{;\tilde g^{-1}(u)}\mathcal Y^\prime_{\tilde g^{-1}(u)}+\widebar\eth^\prime_{;\tilde g^{-1}(u)} \widebar{\mathcal{Y}}^\prime_{;\tilde g^{-1}(u)})=\frac 12(
  \eth^\prime_\circ\mathcal Y^\prime_{\tilde g^{-1}(u)\circ}+\widebar\eth^\prime_\circ \widebar{\mathcal{Y}}^\prime_{\tilde g^{-1}(u)\circ})\\-\mathcal Y^\prime_{\tilde g^{-1}(u)\circ} {(V^\prime_{\circ;\tilde g^{-1}(u)})}^{-1}\eth^\prime_\circ V^\prime_{\circ;\tilde g^{-1}(u)}-\widebar{\mathcal{Y}}^\prime_\circ {(V^\prime_{\circ;\tilde g^{-1}(u)})}^{-1}\widebar\eth^\prime_\circ
  V^\prime_{\circ;\tilde g^{-1}(u)}\\=\frac 12(
  \eth^\prime_\circ\mathcal Y^\prime_{\tilde g^{-1}(u)\circ}+\widebar\eth^\prime_\circ \widebar{\mathcal{Y}}^\prime_{\tilde g^{-1}(u)\circ})
  +\mathcal Y^\prime_{\tilde g^{-1}(u)\circ}
  \eth^\prime_\circ \Phi^\prime_{\circ;\tilde g^{-1}(u)}+\widebar{\mathcal{Y}}^\prime_{\tilde g^{-1}(u)\circ}\widebar\eth^\prime_\circ\Phi^\prime_{\circ;\tilde g^{-1}(u)}.
\end{multline}

Finally, if $\eta^{-1}$ is a field of spin weight $-1$ and $\eta^0$ a real field of spin weight $0$, 
\begin{equation}
  \label{eq:294}
  \eta^{0}_{j=1}=\frac 12(\eth_\circ \eta^{-1}+\widebar\eth_\circ\widebar{\eta^{-1}}) \iff \eta^{-1}=\eta^{-1}_{j=1}=-\widebar\eth_\circ \eta^{0}_{j=1}, 
\end{equation}
so that ${(\eth_\circ \eta^{-1}+\widebar\eth_\circ\widebar{\eta^{-1}})}_{j=1}=0 \iff \eta^{-1}_{j=1}=0$. 

\section{Spin and boost weighted harmonics}\label{sec:spin weighted harmonics}

We collect here some properties of spin and boost weighted spherical harmonics
that we will use.

The action of $\eth_\circ,\widebar\eth_\circ$ on spin weighted spherical harmonics is given by 
\begin{equation}
  \label{eq:304}
  \begin{split}
	& \eth_\circ \,\Yjms{s}{j m} =- \sqrt{\tfrac 12(j-s)(j+s+1)}\; \Yjms{s+1}{j m},\\ &
    \widebar\eth_\circ \,\Yjms{s}{j m}=  \sqrt{\tfrac 12(j+s)(j-s+1)} \;\Yjms{s-1}{j m},                                                                                
  \end{split}
\end{equation}
so that
\begin{equation}
  \label{eq:115}
  \eth_{\circ}\widebar\eth_{\circ}\, {}_{0}Y_{jm}=-\frac{j(j+1)}{2}\, {}_{0}Y_{jm},\quad
  \eth_\circ^2\widebar\eth_\circ^2\,{}_0Y_{jm}=s_j\, {}_{0}Y_{jm},
\end{equation}
with
\begin{equation}
  \label{eq:259}
  s_j=\frac{(j-1)j(j+1)(j+2)}{4}=\frac{1}{4}\frac{(j+2)!}{(j-2)!}.
\end{equation}
Moreover,
\begin{equation}
  \label{eq:303}
  	\eth_\circ^2 \Yjms{0}{j m} = \sqrt{s_j} \; \Yjms{2}{j m} \,,\quad
	\widebar\eth_\circ^2 \Yjms{0}{j m} =\sqrt{s_j} \; \Yjms{-2}{j m},
\end{equation}
while $ \eth_\circ^2 \Yjms{0}{j m} = 0$ for $j=0,1$. It follows that the harmonics at $s=0$, $j=0,1 $ are
in the kernel of $\eth_\circ^2$, $\widebar\eth^2_\circ$ and $\eth_\circ^2\,\widebar\eth_\circ^2 $.

Under complex conjugation\footnote{We thank L.~Szabados for pointing out a
  misprint in (4.15.104) of~\cite{Penrose:1984}.},
\begin{equation}
  \label{eq:251}
  {}_s\widebar Y_{jm}={(-)}^{m-s}{}_{-s}Y_{j,-m}.
\end{equation}
The orthonormality relations are
\begin{equation}
  \label{eq:292}
  \int_{\mathbb S^2} \dd^2\Omega \;\Yjms{s_1}{j_1 m_1}\,\barYjms{s_2}{j_2 m_2}
  =\delta_{s_1 s_2}\delta_{j_1j_2}\delta_{m_1 m_2},
\end{equation}
while
\begin{multline}
  \label{eq:332}
  \int_{\mathbb S^2} \dd^2\Omega \;\Yjms{s_1}{j_1 m_1}\,\Yjms{s_2}{j_2 m_2} \, \Yjms{s}{j m} \\
		\qquad \qquad = \sqrt{\frac{(2j_1+1)(2j_2+1)(2j+1)}{4\pi}} \ThreeJ{j_1}{m_1}{j_2}{m_2}{j}{m} \ThreeJ{j_1}{-s_1}{j_2}{-s_2}{j}{-s},
\end{multline}
in terms of the Wigner 3j-symbols.

The product decomposes as
\begin{equation}
  \label{eq:291}
  {}_{s_1}Y_{j_1m_1}\,{}_{s_2}Y_{j_2m_2}=\mathcal A^j_{s_1j_1m_1;s_2j_2m_2}\,  {}_{s_1+s_2}Y_{j,m_1+m_2}, 
\end{equation}
where the summation over $j$ goes from
$\max(|j_1-j_2|,|s_1+s_2|,|m_1+m_2|)$ to $j_1+j_2$
and
\begin{multline}\label{3Ycoeff}
  A^j_{s_1j_1m_1;s_2j_2m_2}\delta^m_{m_1+m_2}\delta^s_{s_1+s_2}\\
  ={(-1)}^{m-s}\sqrt{\tfrac{(2j_1+1)(2j_2+1)(2j+1)}{4\pi}} \ThreeJ{j_1}{-s_1}{j_2}{-s_2}{j}{s} \ThreeJ{j_1}{m_1}{j_2}{m_2}{j}{-m},
\end{multline}
cf.~\cite{Goldberg1967} and~\cite{Beyer:2013loa}.

In particular,
\begin{multline}
  \label{eq:334}
  4\mathcal A^{6}_{0,5,m_1;0,5,m_2}-
  \mathcal A^{6}_{-2,5,m_1;2,5,m_2}\\={(-)}^{-m_1-m_2}\sqrt{\frac{{(11)}^2 13}{4\pi}}
  \ThreeJ{5}{m_1}{5}{m_2}{6}{-m_1-m_2}
  \ThreeJ{5}{0}{5}{0}{6}{0}
  \big[4+\frac{\ThreeJ{5}{-2}{5}{2}{6}{0}}{\ThreeJ{5}{0}{5}{0}{6}{0}}\big]
  \\={(-)}^{-m_1-m_2}\sqrt{\frac{{(11)}^2 13}{4\pi}}\ThreeJ{5}{m_1}{5}{m_2}{6}{-m_1-m_2}4\sqrt{\frac{5}{7293}}\big[4+\frac{9}{10}\big]\neq 0.
\end{multline}

When needed in order to take into account the appropriate behavior under
fractional linear transformations, one assigns a boost weight to the spin
weighted spherical harmonics
\begin{equation}
  \label{eq:98}
  {}_{sw} Y_{jm}(\zeta,\widebar\zeta)={}_s Y_{jm}(\zeta,\widebar\zeta),
\end{equation}
which transform under~\eqref{eq:138} as fields of spin and boost weights $(s,w)$. In addition,  
\begin{equation}
  \label{eq:379}
  \int_{\mathbb S_2} d^2\Omega\, {}_{s_1w_1}Y_{j_1m_1}\, \widebar{{}_{s_2w_2}Y_{j_2m_2}}
  =\delta_{w_1,-w_2-2}\delta_{s_1 s_2}\delta_{j_1j_2}\delta_{m_1 m_2},
\end{equation}
when taking into account that the integration measure on the sphere
\begin{equation}
  \label{eq:376}
  d^2\Omega=\frac{d\zeta d\widebar\zeta}{\imath P^2_\circ(\zeta,\widebar\zeta)}
\end{equation}
is of spin and boost weights $(0,2)$,
\begin{equation}
  \label{eq:391}
  d^2\Omega^\prime\equiv \frac{d\zeta^\prime d\widebar\zeta^\prime}{\imath P^2_\circ(\zeta^\prime,\widebar\zeta^\prime)}=
  e^{-2E^g_R}(x^\prime)d^2\Omega(x_g(x^\prime)).
\end{equation}
What matters for us here is that, when combining~\eqref{eq:360} with~\eqref{eq:304}, it follows that
\begin{equation}
  \label{eq:363}
  \begin{split}
    \eth^\prime_{1,g} [e^{wE^g_R}\;{}_{sw}Y^\prime_{jm;g^{-1}}]&=
    - \sqrt{\tfrac 12(j-s)(j+s+1)} e^{wE^g_R}\;{}_{s+1,w}Y^\prime_{jm;g^{-1}},\\ 
    \widebar\eth^{\prime}_{1,g} [e^{wE^g_R}\;{}_{sw}Y^\prime_{jm;g^{-1}}]&=  \sqrt{\tfrac 12(j+s)(j-s+1)} e^{wE^g_R}\; {}_{s+1,w}Y^\prime_{jm;g^{-1}} .             
  \end{split}
\end{equation}

\section{STF integrals}\label{app:STF Integrals}
Let \(n^i\) be the unit radial vector and \(\hat n^{L}\equiv n^{\langle i_1}\cdots n^{i_\ell\rangle}\)
the symmetric tracefree (STF) product of order \(\ell\). The goal of this appendix
is to compute integrals of the form
\begin{equation}
	\mathcal{I}_{kL}(u_i)
	=\int_{\mathbb S^2} \frac{d^2\Omega}{4\pi}\,\frac{(-\hat n)_{L}}{{(u_0 -u_i n^i)}^k},
    \quad k\in\mathbb{N},
	\label{eq:ILk_def}
\end{equation}
where $u^a$ denotes a normalized future-directed vector as in~\eqref{eq:237}. Here $u_0=\sqrt{1+|\vec u|^2}>1$. The unusual
sign is due to our choice of adopting the $(+,-,-,-)$ signature for the
Minkowski spacetime metric.

Let us start by giving the integrals $\mathcal I_{k0}$. Using spherical symmetry to perform the integral over the azimuthal angle, one finds
\begin{equation}
  \label{eq:316}
  \mathcal{I}_{k0}(u_0,u_i)=u_0^{-k}\frac 12 \int^1_{-1}d\mu \frac{1}{(1-x\mu)^k},\quad 0\leq  x=\frac{|\vec u|}{u_0}< 1,   
\end{equation}
so that
\begin{equation}
  \label{eq:322}
  \mathcal I_{10}=u_0^{-1}x^{-1}\frac 12 \ln{\big(\frac{1+x}{1-x}\big)}=u_0^{-1}x^{-1}Q_0(x),
\end{equation}
where $Q_0(x)$ is the Legendre function of the second kind on the branch which is analytic on the interval $]-1,1[$. For $k>1$,
\begin{equation}
  \label{eq:323}
  \mathcal I_{k0}=u_0^{-k}x^{-1}\frac{1}{2(k-1)}\big[(1-x)^{1-k}-(1+x)^{1-k}\big]. 
\end{equation}
In particular,
\begin{equation}
  \label{eq:325}
  \begin{split}
    \mathcal I_{20}&=u_0^{-2}x^{-1}\frac{1}{2}\big[(1-x)^{-1}-(1+x)^{-1}]=u_0^{-2}\frac{1}{1-x^2},\\
    \mathcal I_{30}&=u_0^{-3}x^{-1}\frac{1}{4}\big[(1-x)^{-2}-(1+x)^{-2}]=u_0^{-3}\frac{1}{(1-x^2)^2}.
  \end{split}
\end{equation}

In order to get $\mathcal I_{k1}(u_i)$ for $k>1$, one treats $u_0$ and $u_i$ as independent at first, and uses 
\begin{equation}
  \label{eq:324}
  \mathcal I_{k1}(u_0,u_i)=\int_{\mathbb S^2}\frac{d^2\Omega}{4\pi}\frac{(-n_i)}{(u_0-u_in^i)^k}=
  -\frac{1}{k-1}\partial_{u^i} I_{k-10}(u_0,u_i), 
\end{equation}
before setting $u_0=\sqrt{1+|\vec u|^2}$.

More generally, it is convenient to decompose $u_i=|\vec u| \,\hat{u}_i$, where
$|\vec u|$ denotes the norm and $\hat{u}_i$ a unit co-vector. The STF product
$\hat n^L$, and similarly $\hat u_{L}\equiv \hat{u}_{\langle i_1}\cdots \hat{u}_{i_\ell\rangle}$ is an
irreducible representation of rank $\ell=|L|$ of the rotation group. The integral
$\mathcal{I}_{kL}$ transforms as an irreducible representation of rank $\ell$,
which depends only on $|\vec u|, \hat{u}_i$, and thus should take the form
\begin{equation}\label{I vs J}
	\mathcal{I}_{kL}(u_i)=J_{k\ell}(|\vec u|)\,\hat u_{L}.
\end{equation}
where $J_{k\ell}(|\vec u|)$ is invariant under rotations. To this end, we use the
identity (see appendix A.5 of~\cite{Blanchet:1985sp})
\begin{equation}
  \label{eq:248}
  \hat n^{L}{\hat{n}'}_{L}=\frac{\ell!}{(2\ell-1)!!}\,P_\ell(\hat n^i\hat n'_i),
\end{equation}
with $\hat n^i$ a unit vector, $\hat n'_i$ a unit co-vector, and $P_\ell(x)$
Legendre polynomials. If $\hat n'_i=-\hat n_i$, we find the normalization
$\hat n^{L}{(-\hat n)}_{L}=\frac{\ell!}{(2\ell-1)!!}$. Next, contract~\eqref{I vs J} with
\({(- \hat u)}^L\) and use the previous equation to get,
\begin{multline}
	{J}_{k\ell}(|\vec u|)= \frac{(2\ell-1)!!}{\ell!}\,\mathcal{I}_{k}^L\,(-\hat{u})^L 
	=\frac{(2\ell-1)!!}{\ell!}
	\int_{\mathbb S^2}\frac{d^2\Omega}{4\pi} \frac{\hat{n}^L\,\hat{u}_L}{{(u_0-u_i n^i)}^k} 
	\\= \int_{\mathbb S^2}\frac{d^2\Omega}{4\pi} \frac{P_\ell(\mu)}{|\vec u|^k{(z-\mu)}^k},
    \quad \mu\equiv \hat u_i n^i,\quad 1< z\equiv \frac{u_0}{|\vec u|}<\infty, 
  \end{multline}
  where here and in the main text, we work with the convention of the Legendre function which has a branch cut on the interval $[-1,1]$. Rotational invariance of $J_{k\ell}(|\vec u|)$ may be used to put $u_i$
  along the $z$ axis, so that $\mu=\cos\theta$ and azimuthal integration becomes
  trivial,
\begin{equation}
  J_{k\ell}(|\vec u|)=\frac{1}{2|\vec u|^k}\int_{-1}^{1} d\mu\,\frac{P_\ell(\mu)}{{(z-\mu)}^k}
  =\frac{{(-1)}^{k-1}}{2(k-1)!|\vec u|^{k}}\,\partial_{z}^{k-1}\int_{-1}^{1} d\mu\,\frac{P_\ell(\mu)}{z-\mu}.
\end{equation}
Now recall Neumann's integral representation of the Legendre function of the
second kind, $Q_\ell(z) =\frac{1}{2}\int_{-1}^{1}\frac{P_\ell(\mu)}{z-\mu}\,d\mu$, which
implies that
\begin{equation}
	J_{k\ell}(|\vec u|)
	=\frac{{(-1)}^{k-1}}{(k-1)!|\vec u|^{k}}Q_\ell^{(k-1)}(z) \,,
    \quad Q_\ell^{(k)}(z)\equiv\tfrac{\pd^k}{\pd z^k}Q_\ell(z).
\end{equation}
Therefore, the integral evaluates to
\begin{equation}
  \label{eq:242}
  \boxed{\mathcal{I}_{kL}(u_i)
  =\frac{{(-1)}^{k-1}}{(k-1)!|\vec u|^{k}}\,Q_\ell^{(k-1)}(u_0/|\vec u|)\hat{u}_L}.
\end{equation}
One might worry that the limit $|\vec{u}|\to 0$ is singular. However, it can be shown that 
	\begin{align}\label{stat limit}
		\lim_{|\vec{u}|\to 0}J_{k0}=1\,,\qquad \lim_{|\vec{u}|\to 0}J_{k\ell}=0, \ell\geq 1.
	\end{align}
To see this, differentiate Neumann's integral formula $k$ times with respect to $z$, and multiply by $z^{k+1}$ to find
\begin{equation}
	z^{k+1}\frac{d^k Q_\ell(z)}{dz^k}
	=
	\frac{(-1)^k k!}{2}
	\int_{-1}^{1}
	P_\ell(t)
	\left(\frac{z}{z-t}\right)^{k+1}
	\,dt .
\end{equation}
In the stationary limit $z\to \infty$, we thus find 
\begin{equation}
		\lim_{z\to\infty}
		z^{k+1}\frac{d^k Q_\ell(z)}{dz^k}
		=
		(-1)^k k!\,\delta_{\ell 0}.
\end{equation}
which proves \eqref{stat limit}.

We can use the above result to multipole
expand~\cite{Thorne1980,Blanchet:1985sp}
${(w_{1,g})}^{-k}=\frac{1}{{(u_0-u_i n^i)}^{k}}$. For a scalar function $f(n^i)$
on the sphere,
\begin{equation}
  \label{eq:245}
  	f(n^i)=f_L \hat{n}^L,
\end{equation}
where 
\begin{equation}
	f_L=\frac{(2\ell+1)!!}{\ell!}\int_{\mathbb S^2} \frac{d^2\Omega}{4\pi}\, f {(-\hat{n})}_{L},
\end{equation}
by using orthogonality of STF harmonics
\begin{equation}
	\int_{\mathbb S^2} \frac{d^2\Omega}{4\pi}\, \hat n^L(-\hat n)_{L'}=\frac{\ell!}{(2\ell+1)!!}\,\delta^{L}_{L'}.
\end{equation}
Using~\eqref{eq:ILk_def} and~\eqref{eq:242} 
implies 
\begin{equation}
  \label{eq:244}  {(w_{1,g})}^{-k}_L=\frac{{(-1)}^{k-1}}{(k-1)!|\vec u|^{k}}\frac{(2\ell+1)!!}{\ell!}\,
  Q_\ell^{(k-1)}(\frac{u_0}{|\vec u|})\hat{u}_L. 
\end{equation}
When inserted in~\eqref{eq:245} for $f={(w_{1,g})}^{-k}$ and using~\eqref{eq:248} finally yields
\begin{equation}
  \label{eq:246}
  \frac{1}{{(u_0-u_i n^i)}^{k}}={(-1)}^{k-1}\frac{2\ell+1}{(k-1)!|\vec u|^{k}}\,Q_\ell^{(k-1)}(u_0/|\vec u|)P_{\ell}(u_i n^i). 
\end{equation}

\vfill
\newpage

\addcontentsline{toc}{section}{References}

\printbibliography%

\end{document}